\documentclass[10pt]{iopart}

\usepackage{latexsym}
\usepackage{amssymb}
\usepackage{amsfonts}
\usepackage{theorem}
\usepackage{graphicx}
\usepackage{fullpage}
\usepackage[latin1]{inputenc}
\usepackage[T1]{fontenc}
\usepackage[english]{babel}

\newcommand{\pprime}{{\prime\prime}}

\newcommand{\bra}{\langle}
\newcommand{\ket}{\rangle}

\newcommand{\order}{{\mathcal O}}

\newcommand{\bnull}{{\mbox{\boldmath $0$}}}
\newcommand{\bd}{\begin{displaymath}}
\newcommand{\ed}{\end{displaymath}}
\newcommand{\vsp}{\vspace*{3mm}}

\newcommand{\R}{{\rm I\!R}}

\newcommand{\bb}{\ensuremath{\mathbf{b}}}

\newcommand{\bh}{\ensuremath{\mathbf{h}}}

\newcommand{\bM}{\ensuremath{\mathbf{M}}}

\newcommand{\bpsi}{{\mbox{\boldmath $\psi$}}}

\newcommand{\bomega}{{\mbox{\boldmath $\omega$}}}
\newcommand{\bsigma}{{\mbox{\boldmath $\sigma$}}}
\newcommand{\bl}{{\mbox{\boldmath $\ell$}}}
\newcommand{\btau}{{\mbox{\boldmath $\tau$}}}

\newcommand{\s}{\sigma}

\newcommand{\bea}{\begin{eqnarray}}
\newcommand{\eea}{\end{eqnarray}}
\newcommand{\be}{\begin{equation}}
\newcommand{\ee}{\end{equation}}

\newcommand{\Db}{{D(\omega|\beta)}}
\newcommand{\Dbp}{{D(\omega^\prime|\beta)}}
\newcommand{\Dbd}{{D(\omega^\pprime|\beta)}}

\newcommand{\Dpsi}{{D_\psi(\omega|\beta)}}

\newcommand{\Dpsia}{{D_\psi(\omega_\alpha|\beta)}}
\newcommand{\half}{{\frac{1}{2}}}

\newcommand{\pint}{\int\! \{\rmd P \rmd\hat P\}\,}
\newcommand{\Hi}{{\mathcal H}}

\newcommand{\Zset}{{\rm Z\hspace*{-1.2mm}Z}}
\newcommand{\Langle}{{\Bigg\langle}}
\newcommand{\Rangle}{{\Bigg\rangle}}

\begin{document}

\title{Immune networks: multi-tasking capabilities near saturation}

\author{E Agliari$^{1,2}$, A Annibale$^{3,4}$, A Barra$^{5}$, ACC Coolen$^{4,6}$, and D Tantari$^7$}

\address{$^1$ Dipartimento di Fisica, Universit\`{a} degli Studi di Parma, Viale GP Usberti 7/A,  43124 Parma, Italy}
\address{$^2$ INFN, Gruppo Collegato di Parma, Viale Parco Area delle Scienze 7/A,
 43100 Parma, Italy}
\address{$^3$ Department of Mathematics, King's College London,  The Strand, London WC2R 2LS, UK}
\address{$^4$ Institute for Mathematical and Molecular Biomedicine, King's College London, Hodgkin Building, London SE1 1UL, UK}
\address{$^5$ Dipartimento di Fisica, Sapienza Universit\`{a} di Roma, P.le Aldo Moro 2, 00185 Roma, Italy}
\address{$^6$ London Institute for Mathematical Sciences, 35a South St, Mayfair, London W1K 2XF, UK}
\address{$^7$ Dipartimento di Matematica, Sapienza Universit\`{a}  di Roma, P.le Aldo Moro 2, 00185 Roma  Italy}

\begin{abstract}
Pattern-diluted associative networks were introduced recently as models for the immune system,
with nodes representing T-lymphocytes and stored patterns representing signalling protocols between T- and B-lymphocytes.
It was shown earlier that in the regime
of extreme pattern dilution, a system with $N_T$ T-lymphocytes can manage a number $N_B\!=\!\order(N_T^\delta)$ of
B-lymphocytes simultaneously, with $\delta\!<\!1$. Here we study this model in the
extensive load regime $N_B\!=\!\alpha N_T$,
with also a high degree of pattern dilution, in agreement with immunological findings.
We use graph theory and statistical mechanical analysis based on replica methods to show that in the finite-connectivity regime, where each
T-lymphocyte interacts with a finite number of B-lymphocytes as $N_T\to\infty$,
the T-lymphocytes can coordinate effective immune responses to an extensive number of distinct antigen invasions in parallel. As $\alpha$ increases, the system eventually undergoes a second order transition to a phase with clonal cross-talk interference, where the system's performance degrades gracefully.
Mathematically, the model is equivalent to a spin system on a finitely connected graph with many short loops, so one would expect the available analytical methods, which all assume locally tree-like graphs, to fail. Yet it turns out to be solvable.
Our results are supported by numerical simulations.
\end{abstract}


\ead{agliari@fis.unipr.it,alessia.annibale@kcl.ac.uk,adriano.barra@roma1.infn.it,\\ton.coolen@kcl.ac.uk,tantari@mat.uniroma1.it}

\section{Introduction}

After a long period of
dormancy since the pionering paper \cite{parisi},
we have in recent years seen a renewed interest
in statistical mechanical models of the immune system
\cite{BA1,PRE,JTB1,bialek,kosmir1,kosmir2,chakra,PREdeutch2,ton6}. These
 complement the standard approaches to immune system modelling, which are formulated in terms
of dynamical systems \cite{perelson,perelson2,din1,din2}.
However, to make further progress,
we need quantitative tools that are able to handle the complexity of
the immune system's intricate signalling patterns.
Fortunately, over the last decades a powerful arsenal of
statistical mechanical techniques was developed in the disordered system community
to deal with heterogeneous many-variable systems on complex topologies
\cite{barabasi,prlTon,ton3,ton4,smallworld}.
In the present paper we exploit these new techniques to model
the multitasking capabilities of the (adaptive) immune network, where
effector branches (B-cells) and coordinator branches (T-cells) interact
via (eliciting and suppressive) signaling proteins called cytokines.
From a theoretical physics perspective, a network of interacting B- and
T-cells resembles a bi-partite spin glass.
It was recently shown that such a bi-partite spin-glass
is thermodynamically equivalent to a Hopfield-like
neural network with effective Hebbian interactions
\cite{BGG,hotelNN}.

The analogy between immune and neural networks was noted already decades ago: both networks
are able to learn (e.g. how to fight new antigens), memorize (e.g. previously encountered antigens) and `think'  (e.g. select the best strategy to cope with pathogens).
However, their architectures are very different.
Models with fully connected topology, mathematically convenient simplifications of
biological reality, are tolerable for neural networks, where each
neuron is known to have a huge number of connections with others \cite{NNbook}.
In immune networks, however, interactions among lymphocytes are much
more specific and signalled via chemical messengers, leading to network topologies
that display finite connectivity. This difference is not purely
formal, but plays also a crucial operational role.
Neural networks are designed to perform high-resolution serial
information processing, with neurons interacting with many others to retrieve collectively a
{\em single} pattern at a time. The immune system, in contrast,
must simultaneously recall multiple patterns
(i.e. defense strategies), since
many antigens will normally attack the host at the same time.
Remarkably, diluting interactions
in the underlying bi-partite spin-glass
causes a switch from serial to parallel processing
(i.e. to simultaneous pattern recall) of the thermodynamically
equivalent Hopfield network\footnote{
In contrast, diluting the bonds in a Hopfield network
does not affect pattern retrieval qualitatively
\cite{ton3,ton4,sompo1,ton0,ton1,sompo2}: the system would still recall only one pattern at a time, but simply have a lower storage capacity. }
\cite{alps2_lett,alps2_lungo}.

The inextricable link between retrieval and toplogical features of such systems
requires a combination of techniques from statistical mechanics and graph
theory, which will be
the focus of the present paper, which is organized as follows. In Section 2 we describe a minimal biological scenario for the immune system, based on the
analogy with neural networks, and define our model.
Section 3 gives a comprehensive analysis  of the topological properties of the network in the finite connectivity and high load regime,
which is the one assumed throughout our paper. Section 4 is dedicated to the statistical mechanical analysis of the system, focusing on simultaneous pattern recall of the network.
In Section 5 we use a population dynamics
algorithm to inspect numerically different regions of the phase diagram.
We end with a summary of our main findings.

\section{Statistical mechanical modelling of the adaptive immune system}

\subsection{The underlying biology}

All mammals have an innate (broad range) immunity, managed by macrophages, neutrophils, etc., and an adaptive immune response.
We refer to the excellent books \cite{janaway,abbas} for comprehensive  reviews of the immune system, and to a selection of papers
\cite{BA1,PRE,JTB1,alps2_lett,alps2_lungo,BA2}
for theoretical modeling inspired by biological reality.
Our prime interest is in B-cells and in T-cells; in particular, among T-cells, in the subgroups of so-called `helpers' and `suppressors'.
B-cells produce antibodies which
are able to recognize and bind pathogens, and those that produce the same antibody are said to form a clone. The human immune repertoire consists of $\mathcal{O}(10^8-10^9)$ clones.
The size of a clone, i.e. the number of identical B-cells, may vary strongly. A clone at rest may contain some $\mathcal{O}(10^3-10^4)$ cells, but when it
undergoes clonal expansion its size may increase by several orders of magnitude, up to $\mathcal{O}(10^6-10^7)$. Beyond this size the state of the immune system
would be pathological, and is referred to as lymphocytosis.


When an antigen enters the body, several antibodies produced by different clones may be able to bind to it, making it chemically inert and biologically inoffensive.
In this case, conditional on authorization by T-helpers (mediated via cytokines), the binding clones undergo clonal expansion and start
releasing high quantities of soluble antibodies to inhibit the enemy. After the antigen has been deleted,
B-cells are instructed by T-suppressors, again via cytokines, to stop producing antibodies and undergo apoptosis. In this way the clones reduce their sizes, and
order is restored.
Thus, two signals are required within a small time interval for B-cells to
start clonal expansion:
the first is binding to antigen, the second is a `consensus' signal, in the
form of an eliciting cytokine \cite{chitochine,chitochinebook}
secreted by T-helpers.
This mechanism, known as the `two-signal model' \cite{goodnow1,goodnow2,goodnow3,anergy3},
prevents abnormal reactions, such as autoimmune
manifestations\footnote{Through a phenomenon called `cross-linking', a B-cell can also have the ability to bind a self-peptide, and may accidentally start
duplication and antibody release, which is a dangerous unwanted outcome.}.
The focus of this study is to understand, from a statistical mechanics perspective, the ability of helpers and suppressors to coordinate and
manage an extensive ensemble of B-clones {\em simultaneously}.

\subsection{A minimal model} \label{sec:formal}

\begin{figure}[t]
\unitlength=0.12mm
\centering
\includegraphics[width=270\unitlength,height=290\unitlength]{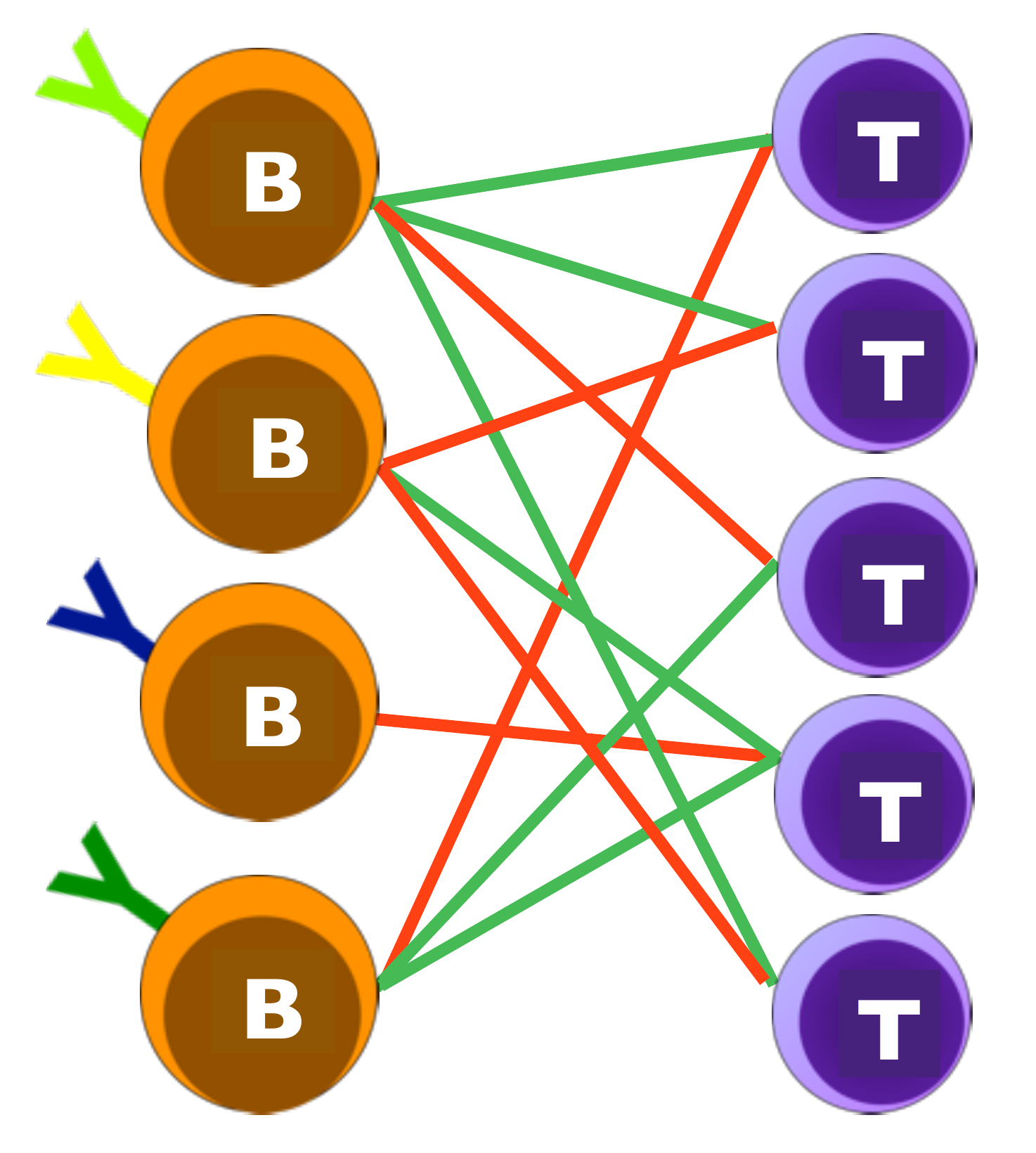}
~~~
\includegraphics[width=280\unitlength,height=270\unitlength]{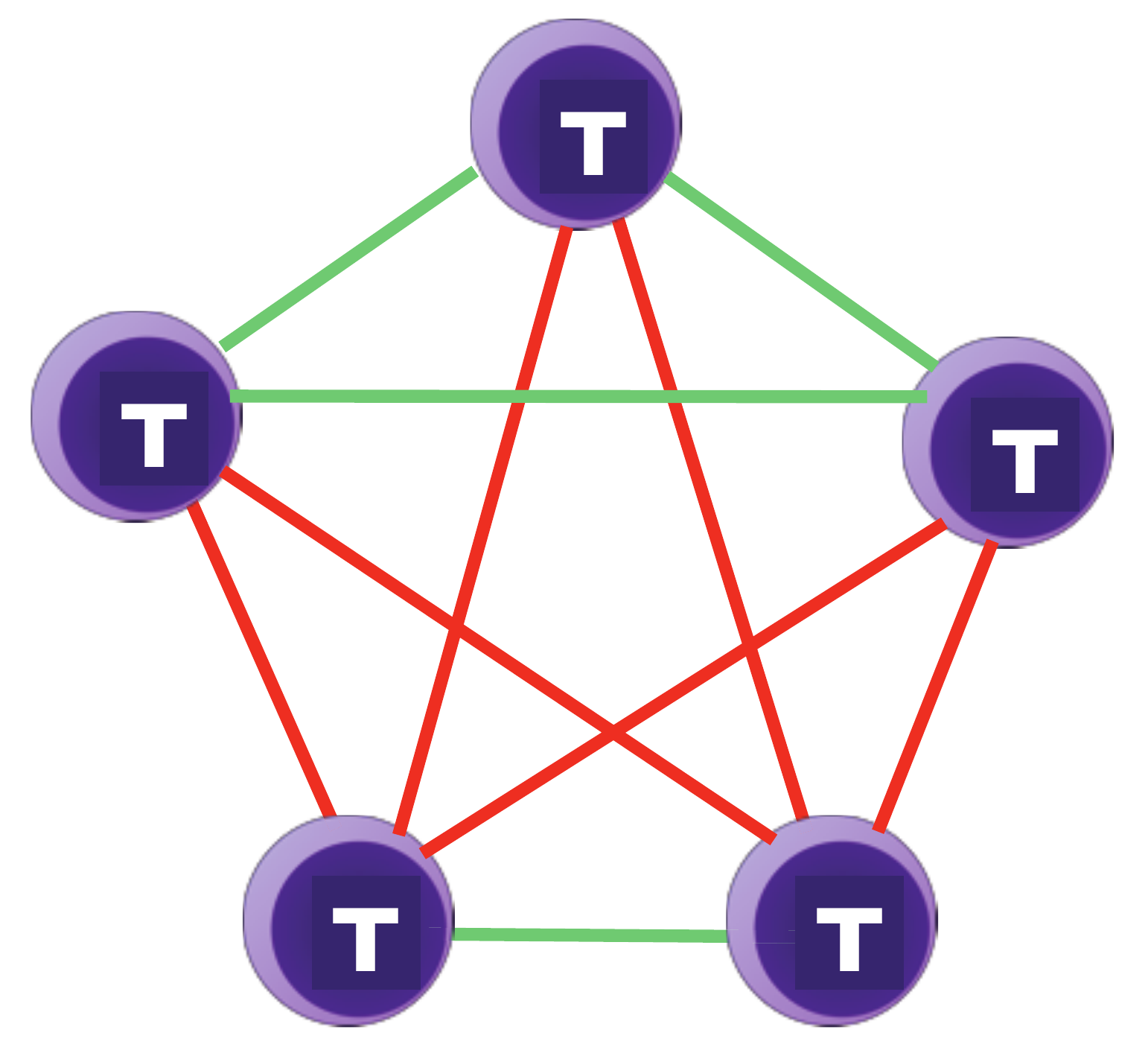}
\vspace*{1.5mm}

\caption{Left: the bi-partite spin-glass which
models the interaction between B- and T-cells through cytokines.
Green (red) links represent stimulatory (inhibitory) cytokines.
Note that the network is diluted. Right: the equivalent associative
multitasking network consisting of T-cells only, obtained by integrating out
the B-cells. This network is also diluted, with links given by the Hebbian
prescription.}
\label{fig:map}
\end{figure}

We consider an immune repertoire of $N_B$ different clones, labelled by $\mu \in \{1,...,N_B\}$. The size of  clone $\mu$ is $b_{\mu}$.
In the absence of interactions with helpers, we take the clone sizes to be Gaussian distributed; without loss of generality we may take the mean to be zero  and unit width, so $b_{\mu} \sim \mathcal{N}(0,1)$. A value $b_{\mu}\gg 0$ now implies that clone $\mu$ has expanded (relative to the typical clonal size), while $b_{\mu}\ll 0$ implies  inhibition.
The Gaussian clone size distribution is supported both by experiments and by theoretical arguments \cite{JTB1}.
Similarly, we imagine having $N_T$ helper clones, labelled by $i \in \{1,...,N_T\}$.
The state of helper clone $i$ is denoted by $\s_i$. For simplicity, helpers are assumed to be in only two possible states: secreting cytokines ($\s_i=+1$) or quiescent ($\s_i=-1$).  Both the clone sizes $b_\mu$ and the helper states $\s_i$ are dynamical variables.
We will abbreviate $\bsigma=(\s_1,\ldots,\s_{N_T})\in\{-1,1\}^{N_T}$, and $\bb=(b_1,\ldots,b_{N_B})\in\R^{N_B}$.

The interaction between the helpers and B-clones is implemented by cytokines. These are taken to be frozen (quenched) discrete  variables.
The effect of a  cytokine secreted by helper $i$ and detected by clone $\mu$ can be nonexistent ($\xi_i^\mu=0$), excitatory ($\xi_i^{\mu} = 1$), or inhibitory
 ($\xi_i^{\mu} = -1$). To achieve a Hamiltonian formulation of the system, and  thereby enable equilibrium statistical mechanical analysis, we have to impose symmetry of the cytokine interactions. So, in addition to the B-clones being influenced by cytokine signals from helpers, the helpers will similarly feel a signal from the B-clones.  This symmetry assumption can be viewed as a necessary first step, to be relaxed in future investigations, similar in spirit to the early formulation of symmetric spin-glass models for neural networks \cite{DGZ,ton5}.  We are then led to  a Hamiltonian $\hat{\mathcal{H}}(\bb,\bsigma|\xi)$ for the combined system of the following form (modulo trivial multiplicative factors):
\be
\hat{\mathcal{H}}(\bb,\bsigma|\xi) = -\sum_{i=1}^{N_T}\sum_{\mu=1}^{N_B}\xi_i^{\mu}\s_i b_{\mu} +\frac{1}{2\sqrt{\beta}}\sum_{\mu=1}^{N_B}b_\mu^2.
\ee
In the language of disordered systems, this is a bi-partite spin-glass. We can integrate out the variables $b_\mu$, and map our system to a model with helper-helper interactions only.
The partition function $Z_{N_T}(\beta,\xi)$, at inverse clone size noise level $\sqrt{\beta}$ (which is the level consistent with our assumption
$b_{\mu} \sim \mathcal{N}(0,1)$) follows straightforwardly,  and reveals the mathematical equivalence with an associative attractor network:
\begin{eqnarray}
Z_{N_T}(\beta,\xi) &=& \sum_{\bsigma}\int\!db_1\ldots db_{N_B}\exp[-\sqrt{\beta}~\hat{\mathcal{H}}(\bb,\bsigma|\xi) ]
\nonumber\\
&=& \sum_{\bsigma}\exp [-\beta \mathcal{H}(\bsigma|\xi)],
\label{equivalence}
\end{eqnarray}
in which, apart from an irrelevant additive constant,
\begin{eqnarray}
\mathcal{H}(\bsigma|\xi)
&=&-\frac{1}{2}\sum_{ij=1}^{N_T}\s_i J_{ij}\s_j,~~~~~~J_{ij}=\sum_{\mu=1}^{N_B}\xi_i^{\mu}\xi_j^{\mu}.
\label{eq:hopfield}
\end{eqnarray}
Thus, the system with Hamiltonian $\hat{\mathcal{H}}(\bb,\bsigma|\xi)$, where helpers and B-clones interact stochastically through cytokines, is thermodynamically equivalent to a Hopfield-type associative network represented by $\mathcal{H}(\bsigma|\xi)$, in which helpers mutually interact through an effective Hebbian coupling
(see Fig. \ref{fig:map}). Learning a pattern in this model means adding a new B-clone with an associated string of new cytokine variables.

\begin{figure}[t]
\unitlength=0.12mm
\centering
\includegraphics[width=270\unitlength,height=270\unitlength]{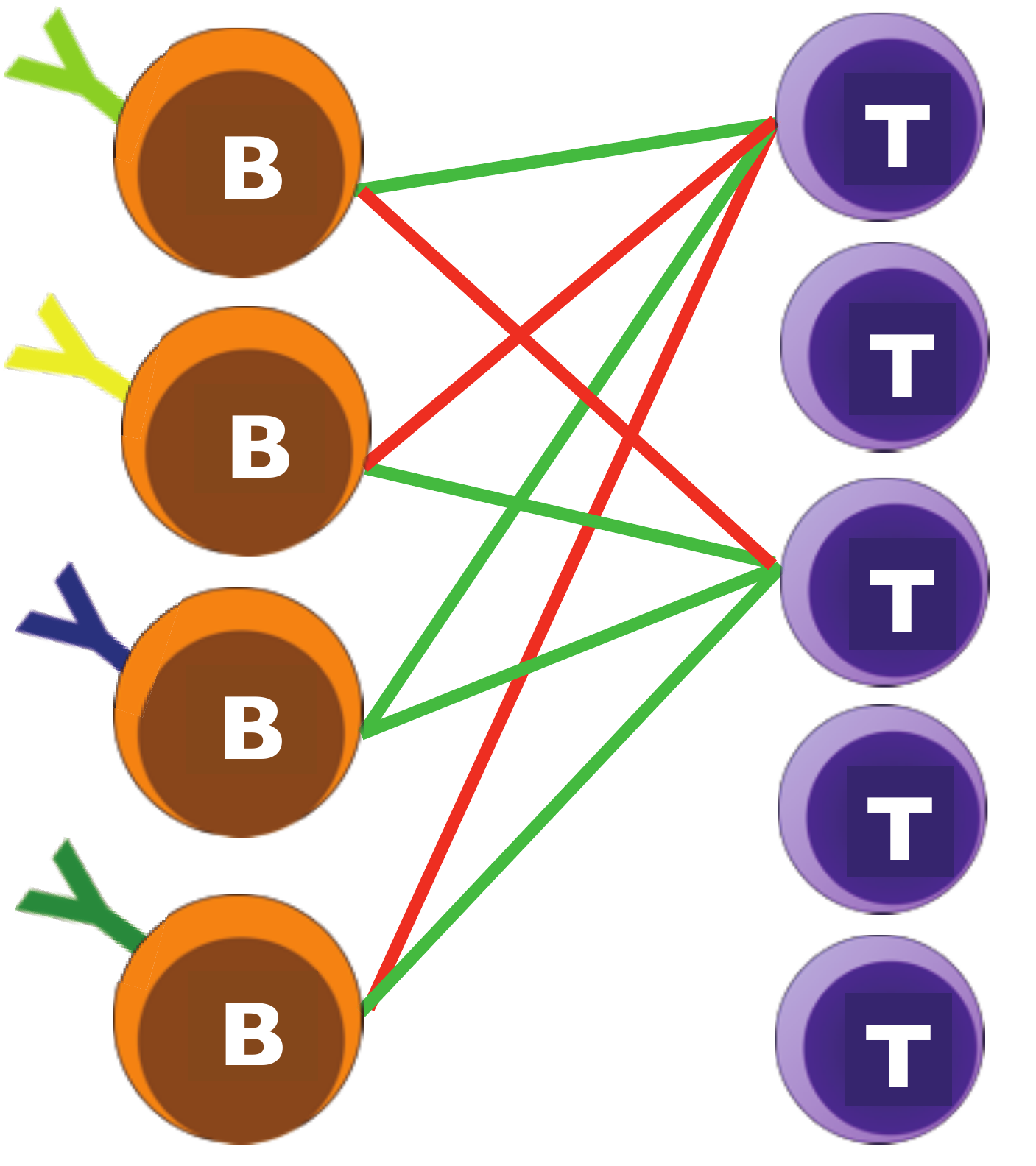}
\vspace*{1.5mm}

\caption{The specific  T-cell configuration that would give the strongest possible positive signal
to the first clone. Upward arrows indicate cytokine secreting T-cells, downward arrows indicate quiescent ones.
Eliciting and suppressive cytokines are
represented by green and red links, respectively.
}
\label{fig:Signal}
\end{figure}

If  all $\{\xi_i^\mu\}$ are nonzero, the system characterized by (\ref{eq:hopfield}) is well known in the information processing systems community.  It is able to retrieve each of the $N_B$ `patterns'
$(\xi_1^\mu,\ldots,\xi_{N_T}^\mu)$, provided these are sufficiently uncorrelated, and both the ratio $\alpha=N_B/N_T$ and the noise level $1/\beta$ are sufficiently small  \cite{JTB1,ton0,amit,ton2}.
Retrieval quality can be quantified by introducing $N_B$ suitable order parameters, the so-called Mattis magnetizations
$m_{\mu}(\bsigma)= N_T^{-1}\sum_i \xi_i^{\mu}\s_i$, in terms of which  we can write (\ref{eq:hopfield})  as
\begin{equation} \label{eq:intemedio}
\mathcal{H}(\bsigma|\xi)= -\frac{N_T}{2}\sum_{\mu=1}^{N_T} m_{\mu}^2(\bsigma).
\end{equation}
If $\alpha$ is sufficiently small, the minimum energy configurations of the system are those where  $m_{\mu}(\bsigma)=1$ for some $\mu$ (`pure states'),  which implies that $\bsigma=(\xi_1^\mu,\ldots,\xi_{N_T}^\mu)$ and  pattern $\mu$ is said to be retrieved perfectly.
In our immunological context this means the following. If $m_\mu(\bsigma)=1$, all the helpers are `aligned' with their coupled cytokines: those $i$ that inhibit clone $\mu$ (i.e. secrete $\xi_{i}^\mu=-1$)  will be quiescent ($\s_{i}=-1$), and
those $i$ that excite clone $\mu$ (i.e. secrete $\xi_{i}^\mu=1$)  will be active ($\s_{i}=1$) and release the eliciting cytokine. As a result the B-clone $\mu$ receives the strongest possible positive signal (i.e. the random environment becomes a `staggered magnetic field'), hence it is forced to expand
(see Fig.
\ref{fig:Signal}).
Conversely, for $m_\mu(\bsigma)=-1$, clone $\mu$ receives the strongest
possible negative signal and it is forced to contract.
However, in this scenario of $\xi_i^\mu\in\{-1,1\}$ for all $(i,\mu)$ (where the bi-partite network is fully connected) only one B-clone at a time can expand (apart from minor spurious mixture states). This would be a disaster for the immune system.

We need the dilution in the bi-partite B-H network that is caused by having also  $\xi_i^\mu=0$ (i.e. no  signalling between helper $i$ and clone $\mu$), to  enable multiple clonal expansions. In this case,
the network (\ref{eq:hopfield}) stores patterns that also have blank
entries,
and retrieving a pattern no longer employs all spins $\sigma_i$:
those corresponding to null entries can be used to recall other patterns.
This is energetically favorable since the energy is quadratic  in the
magnetizations $m_{\mu}(\bsigma)$.
Conceptually, this is only a redefinition of the network's recall task: no  theoretical bound for information content is violated, and  global retrieval is still performed through $N_B$ bits. However, the perspective is shifted: the system no longer requires a sharp resolution in information exchange between a helper clone and a B-clone\footnote{In fact, the high-resolution analysis is performed in the antigenic recognition on the B-cell surface, which is based on a sharp key-and-lock mechanism \cite{BA1}.}. It suffices that a B-clone receives an attack signal, which could be encoded  even by a single bit. In a diluted bi-partite B-H system the associative capabilities of the helper network are distributed, in order  to simultaneously manage the whole ensemble of B-cells.
The analysis of these immunologically relevant pattern-diluted versions of associative networks
has so far been carried out in the low storage case $N_B \!\sim\! \log N_T$ \cite{alps2_lett,alps2_lungo} and the medium storage case
$N_B\! \sim\! N_T^{\delta},~0\!<\!\delta\!<\!1$,
where the system indeed performs as a multitasking associative memory \cite{medio}.
The focus of this paper is to analyse the ability of the network to retrieve
simultaneously an extensive number of patterns, i.e. $N_B=\alpha N_T$ with $\alpha>0$ fixed and $N_T\to\infty$,  while in addition
implementing a higher degree of dilution  for the B-H system, in agreement with immunological findings \cite{janaway,abbas}.


\section{Topological properties of the emergent network}
\label{sec:topo}

\subsection{Definitions}

The system composed of $N_T$ T-lymphocytes that interact with $N_B$ B-lymphocytes, via cytokines, can be described as a bi-partite graph $\mathcal{B}$, in which  the nodes,
belonging to the sets $\mathcal{V}_T$ and $\mathcal{V}_B$, of cardinality $|\mathcal{V}_T| = N_T$ and $|\mathcal{V}_B| = N_B$, respectively,
are pairwise connected via undirected links.
We assign to the link between T-lymphocyte $i$ and B-lymphocyte $\mu$ a variable
$\xi_i^\mu$, which takes values $1$ if the cytokines produced by T-lymphocyte
$i$ triggers expansion of B-clone $\mu$, $-1$ if it triggers contraction and
$0$ if  $i$ and $\mu$ don't interact.
We assume that the $\{\xi_i^{\mu}\}$ are identically and independently distributed random variables,
drawn from
\begin{equation}
P(\xi_i^{\mu}|d) = \frac{1-d}{2} \delta_{\xi_i^{\mu}-1,0} +
\frac{1-d}{2} \delta_{\xi_i^{\mu}+1,0} + d \, \delta_{\xi_i^{\mu},0},
\end{equation}
where $\delta_{x,0}$ is the Kronecker delta symbol.
$P(\xi_i^{\mu}|d)$ implicitly accounts for bond dilution within the graph $\mathcal{B}$.

It is experimentally well established that although helpers are much more numerous than B-cells, their relative sizes are still comparable in a statistical mechanical sense, hence we will assume
\begin{equation} \label{eq:HD}
N_B=\alpha N_T,\quad 0\!<\!\alpha\!<\!1.
\end{equation}
We have shown in the previous section how the signaling process of the B- and T-cells on this bi-partite graph
can be mapped to a thermodynamically equivalent process on a new
graph, $\mathcal{G}$, built only of the $N_T$ nodes in $\mathcal{V}_T$, and  occupied by spins $\sigma_i$ that
interact pairwise through the coupling matrix
\be \label{eq:coupling}
J_{ij} = \sum_{\mu = 1}^{N_B} \xi_i^{\mu} \xi_j^{\mu}.
\ee
The topology of the (weighted,
monopartite) graph $\mathcal{G}$
can range from fully-connected to sparse, as $d$ is tuned.
Our interest is in the ability of this system to perform
as a multitasking associative memory such that the maximum number of
pathogens can be fought simultaneously.
A recent study \cite{medio} suggested that in order to bypass the spin-glass structure of phase space at the load level (\ref{eq:HD}),
a finite-connectivity topology
is required:
\begin{equation} \label{eq:d}
d = 1 - c/N_T.
\end{equation}
Remarkably, the finite-connectivity topology
is also in agreement with the biological picture of highly-selective touch-interactions among B and T cells.


\subsection{Simple characteristics of graph $\mathcal{B}$}

Let us now describe in more detail the topology of the graph $\mathcal{B}$ under condition (\ref{eq:d}). We denote with $k_{i}$ the degree of node $i \in \mathcal{V}_T$ (the number of links stemming from $i$), and with $\kappa_{\mu}$ the degree of node $\mu \in \mathcal{V}_B$ (the number of links stemming from $\mu$):
\begin{eqnarray}
k_{i} = \sum_{\mu \in \mathcal{V}_B} |\xi_i^{\mu}| \in [0,N_B],\quad\quad\quad
\kappa_{\mu} = \sum_{i \in \mathcal{V}_T} |\xi_i^{\mu}| \in [0,N_T].
\end{eqnarray}
Since links in $\mathcal{B}$ are independently and identically drawn, $k$ and $\kappa$ both follow a binomial distribution
\begin{eqnarray}
\label{eq:binomial_rho}
P_T(k| d, N_B) = \Big(\!\begin{array}{c}N_B\\k\end{array}\!\Big)
 (1-d)^{k} d^{N_B-k},\quad\quad\quad
P_B(\kappa| d, N_T) =
\Big(\!\begin{array}{c}N_T\\ \kappa\end{array}\!\Big)
(1-d)^{\kappa} d^{N_T-\kappa},
\end{eqnarray}
hence we have $\mathbb{E}_T (k) \equiv \sum_{k} P_T(k| d, N_B) k  = (1-d)N_B = c \alpha$, $~\mathbb{E}_T (k^2) - [\mathbb{E}_T (k)]^2 = (1-d)dN_B = c \alpha (1- c /N_T)$, and $\mathbb{E}_B (\kappa) \equiv \sum_{\kappa} P_B(\kappa| d, N_T) \kappa  = (1-d)N_T = c $, $~\mathbb{E}_B (\kappa^2) - [\mathbb{E}_B (\kappa)]^2 = (1-d)dN_T=c(1-c/N_T)$.

\begin{figure}[t]
\vspace*{-4mm}
\unitlength=0.12mm
\hspace*{-4mm}
\includegraphics[width=410\unitlength]{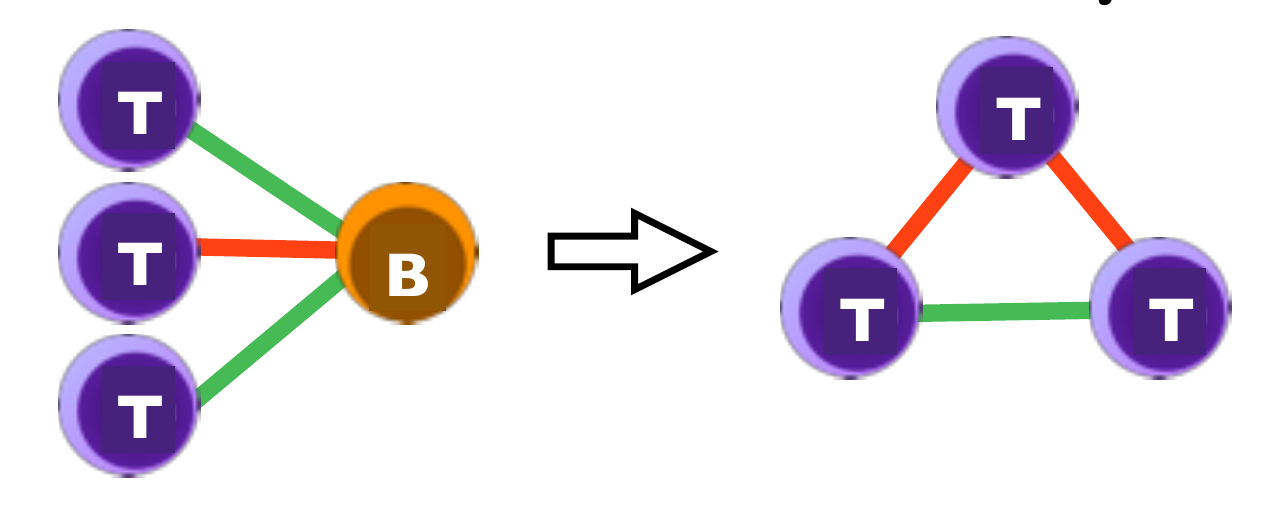}\hspace*{10mm}
\includegraphics[width=410\unitlength]{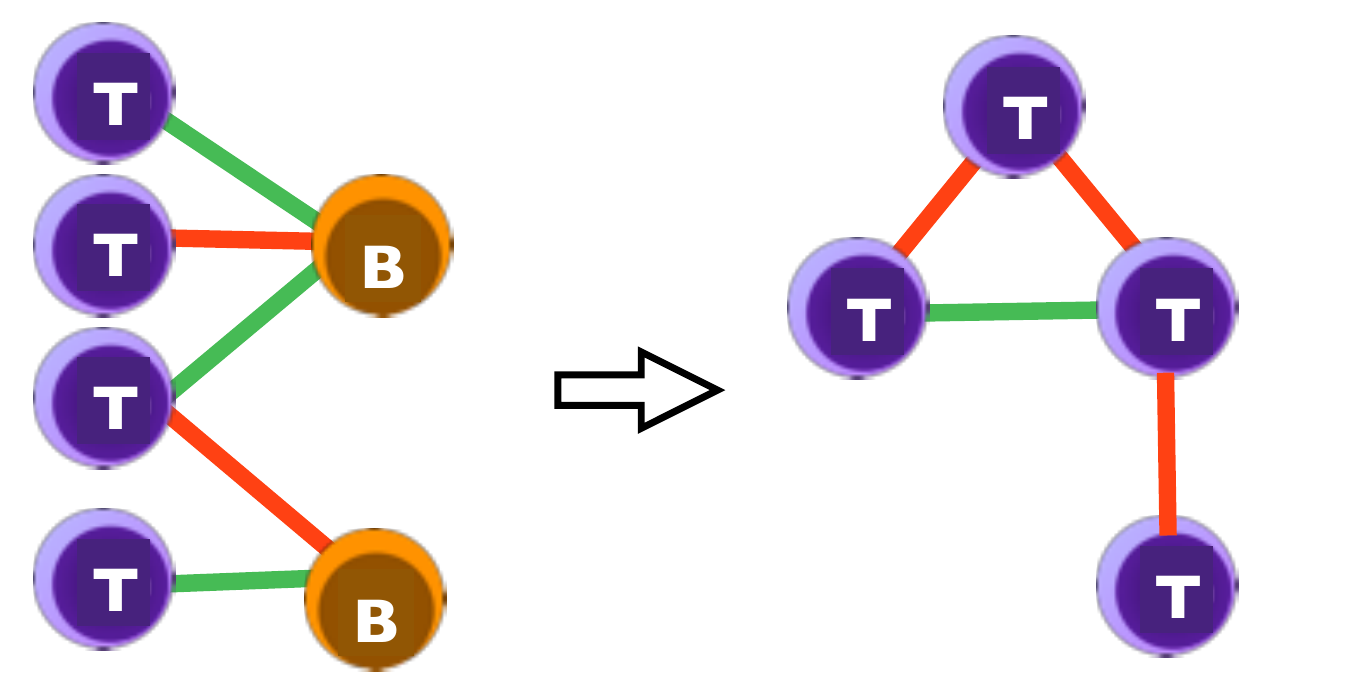}\hspace*{10mm}
\includegraphics[width=410\unitlength]{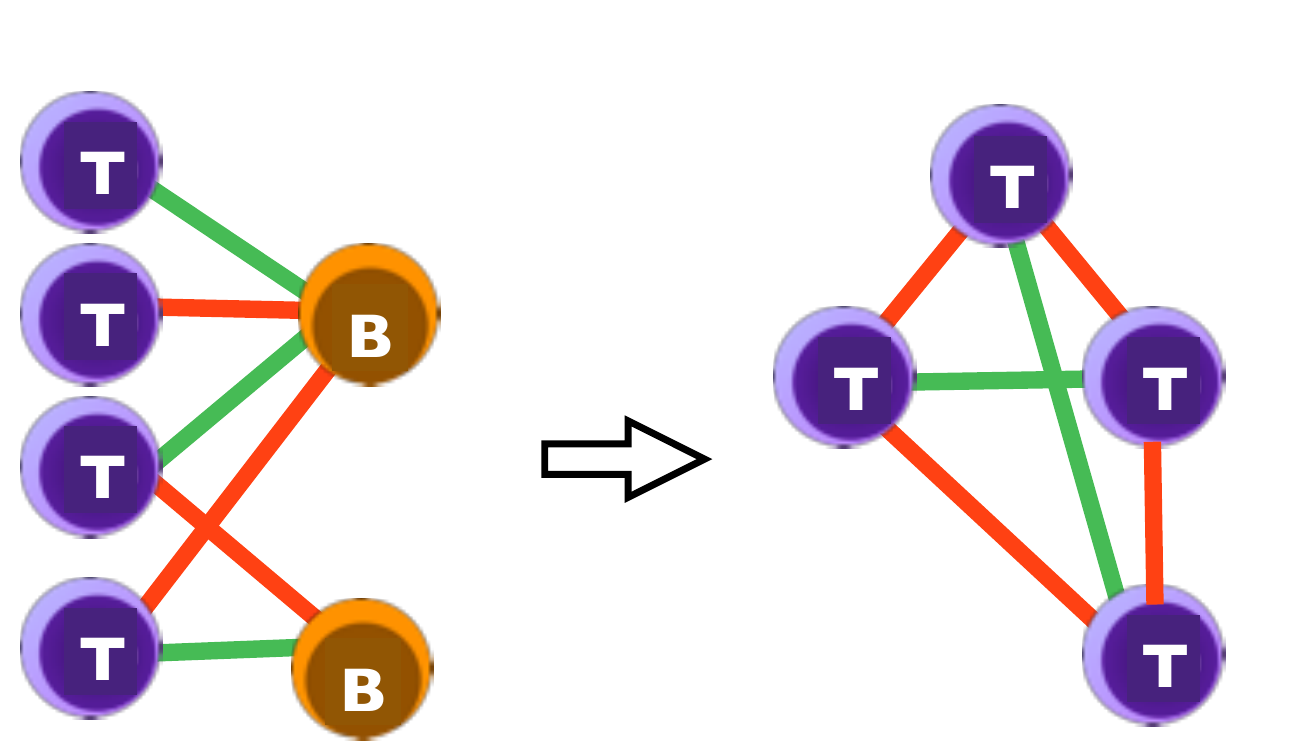}
\vspace*{-2mm}

\caption{\label{fig:casi} Examples showing how different components within $\mathcal{B}$
are mapped into $\mathcal{G}$ upon marginalization. Left: any star $S_n$ in $\mathcal{B}$  with a node in $\mathcal{V}_B$ at its center and $n$ leaves $i_1, i_2, ..., i_n \in \mathcal{V}_T$ corresponds in $\mathcal{G}$ to a complete graph $K_n$, where each link $J_{ij}$ has unit magnitude. Middle: two stars $S_n$ and $S_m$ in $\mathcal{B}$  that share a leaf correspond to two connected complete graphs in $\mathcal{G}$, $K_n$ and $K_m$ respectively, that have a common node. Again, each link $J_{ij}$ has unit weight. Right: when a loop of length $4$ is present in the bi-partite graph $\mathcal{B}$, the corresponding nodes in $\mathcal{G}$ may be connected by a link with
weight larger (in modulus) than one.}
\end{figure}

Due to the finite connectivity, we expect an extensive fraction of nodes $i \in \mathcal{V}_T$ and $\mu \in \mathcal{V}_B$ to be isolated. In the thermodynamic limit, the fraction of isolated nodes in $\mathcal{V}_T$ and
$\mathcal{V}_B$ is
$d^{N_B} = (1-c/N_T)^{\alpha N_T} \rightarrow e^{- c \, \alpha}$
and $d^{N_T} = (1-c/N_T)^{N_T} \rightarrow e^{- c}$, respectively. In order to have a low number of non-signalled B cells, one should therefore choose a  relatively large value of $c$.
Moreover, as will be shown below, by reducing $\alpha$ and/or $c$ we can break $\mathcal{B}$ into small components, each yielding, upon marginalization (\ref{equivalence}), a distinct component within $\mathcal{G}$ (see Fig.~\ref{fig:casi}).
This fragmentation is crucial to allow parallel pattern retrieval. In general, a macroscopic component emerges when the link probability $(1-d)$ exceeds the percolation threshold $1/\sqrt{N_T N_B}$ \cite{medio,Newman02}, which, recalling equations (\ref{eq:HD}) and (\ref{eq:d}),
can be translated into
\be
c > 1/\sqrt{\alpha}.
\label{eq:percolation}
\ee

\noindent
\subsection{Analysis of graph $\mathcal{G}$}

Due to the finite connectivity of $\mathcal{B}$, we expect that also $\mathcal{G}$ will have a macroscopic fraction $f(\alpha,c)$ of isolated nodes, which will be larger than $e^{- c \, \alpha}$ (the fraction of nodes that are isolated in $\mathcal{B}$). In fact, a node $i \in \mathcal{V}_T$, which in $\mathcal{B}$ has a number of neighbors $\mu \in \mathcal{V}_B$, but does not share any of these with any other node $j \in \mathcal{V}_T$, remains isolated upon marginalization. Put differently, whenever $i$ is the centre of a star $S_n$ in $\mathcal{B}$, with $n=0, 1, ..., N_B$, it will be isolated in $\mathcal{G}$. We recall that a star $S_n$ is a tree with a central node and $n$ leaves; this includes isolated nodes ($n=0$), dimers ($n=1$), etc. The larger $n$, the less likely the occurrence of the component $S_n$ in $\mathcal{B}$ and the smaller the related contribution $f_n(\alpha,c)$ to $f(\alpha,c)$. On average, one will have
\be
f_n(\alpha,c)=(1-d)^n d^{N_B-n}
\Big(\!\begin{array}{c}N_B\\[-0.5mm] n\end{array}\!\Big),
\ee
so that, overall, the fraction of isolated nodes in $\mathcal{G}$ is roughly $e^{-c \alpha} + c \alpha e^{-c \alpha}$.
In the following subsections we will inspect the architecture of $\mathcal{G}$ in more detail.


\subsubsection{Coupling distribution}

Let us introduce the probability distribution $P(J | N_B, N_T, c)$
that an arbitrary link $J_{ij}$,
as given by (\ref{eq:coupling}), has weight $J$. The average link
probability is then $1-P(J=0 | N_B, N_T, c) \equiv 1-p$.
This distribution $P(J | N_B, N_T, c)$ can be viewed as the probability distribution for the end-to-end distance of a one-dimensional random walk endowed with a waiting probability $p_w$, which corresponds to the probability that a term $\xi_i^{\mu} \xi_j^{\mu}$ is null, and equal probabilities $p_l=p_r$ of moving left or right, respectively:
\begin{eqnarray}
p_w = d(2-d) = 1 - \Big( \frac{c}{N_T} \Big)^2,\\
p_l = p_r = \frac{1-p_w}{2} = \frac{1}{2} (1-d)^2 =  \frac{1}{2} \Big( \frac{c}{N_T} \Big)^2.
\end{eqnarray}
Therefore, we can write
\be \label{eq:RW}
P(J | N_B, N_T, c) =\sum_{S=0}^{N_B-J}\hspace*{-2mm}^\prime~ \frac{N_B!} {S! \left( \frac{N_B-S-J}{2} \right)! \left( \frac{N_B-S+J}{2} \right)!} \, p_w^S \, p_r^{(N_B-S+J)/2} \, p_l^{(N_B-S-J)/2},
\ee
where the primed sum means that only values of $S$ with the same parity as
$(N_B \pm J)$ are taken into account.
The distribution (\ref{eq:RW}) can easily be generalized to the case of a biased random walk, i.e. biased distribution for weights \cite{Amit-PRA1987}.
The couplings among links have (in the limit of large $N_T$) the following average values \cite{medio}
\begin{eqnarray}\label{eq:coupling_estimate1}
\langle{J} \rangle &=& 0\\
\label{eq:coupling_estimate2}
\langle{J^2} \rangle &=&  \alpha \, c^2/N_T,
\end{eqnarray}
\vspace*{-3mm}
and for $J=0$ one has
\begin{eqnarray} \label{eq:link_MF}
P(0 | N_B, N_T,  c) =\sum_{S=0}^{N_B}\hspace*{-1mm}^\prime~ \frac{N_B!} {S! \Big[ \Big( \frac{N_B-S}{2} \Big)\Big]^2 } \, p_w^S \, p_r^{(N_B-S)}
=\sum_{S=0}^{N_B}\hspace*{-1mm}^\prime~ \Big(\!\begin{array}{c}N_B\\ S\end{array}\!\Big)
\Big(\!\begin{array}{c}N_B-S\\ \frac{N_B-S}{2}\end{array}\!\Big)
 p_w^S p_r^{N_B-S},
\end{eqnarray}
where now $S$ must have the same parity as $N_B$.
Assuming $N_B$ even, we can write
\begin{eqnarray}
P(0 | N_B, N_T, c)  &=& p_r^{N_B}
\Big(\!\begin{array}{c}N_B\\ N_B/2\end{array}\!\Big)
\, {_{2}F}_1\Big(-\frac{N_B}{2},-\frac{N_B}{2},\frac{1}{2}
, \frac{p_w^2}{4p_r^2} \Big) \\
&\approx& \Big( \frac{c \alpha}{N_B} \Big)^{\!2N_B} \sqrt{\frac{2}{\pi N_B}}   \, {_{2}F}_1\Big(-\frac{N_B}{2},-\frac{N_B}{2},\frac{1}{2} ,\Big[\Big(\frac{N_B}{c \alpha}\Big)^2-1\Big]^2\Big),
\end{eqnarray}
where in the last step we used $N_B \gg1$.
Hence, upon expanding the hypergeometric function we get
\be \label{eq:PJ0}
P(0 | N_B, N_T,  c)  \approx \Big[ 1- \Big( \frac{c}{N_T} \Big)^2 \Big]^{\alpha N_T} \left[ 1 + \mathcal{O} (N_T^{-2}) \right] \sim \rme^{-c^2 \alpha /N_T}.
\ee
Following a mean-field approach, we can estimate the degree distribution $P_{\textrm{MF}}(z| N_B, N_T, c)$ in $\mathcal{G}$ by means of a binomial, in which the link probability is simply $p \equiv 1-P(J=0 | N_B, N_T, c)$, namely
\be
P_{\footnotesize\textrm{MF}}(z| N_B, N_T, c) =
\Big(\!\begin{array}{c}N_T\\z\end{array}\!\Big)
 (1 - p)^z p^{N_T-z},
\ee
in which the average degree and the variance are
\begin{eqnarray}
\label{eq:z_MF}
\langle z \rangle_{\footnotesize\textrm{MF}} &=& (1-p) N_T \sim (1 - e^{-c^2 \alpha /N_T}) N_T \sim c^2 \alpha, \\
\label{eq:z2_MF}
\langle z^2 \rangle_{\footnotesize\textrm{MF}} - \langle z \rangle^2_{\footnotesize\textrm{MF}} &=&  (1- p) p N_T \sim c^2 \alpha.
\end{eqnarray}
Due to the homogeneity assumption intrinsic to the mean-field approach,
we expect our estimate to be accurate only for the first moment,
while fluctuations are underestimated \cite{medio}.
In order to account for the topological inhomogeneity characteristic of $\mathcal{G}$ we need to return to analysis of $\mathcal{B}$.

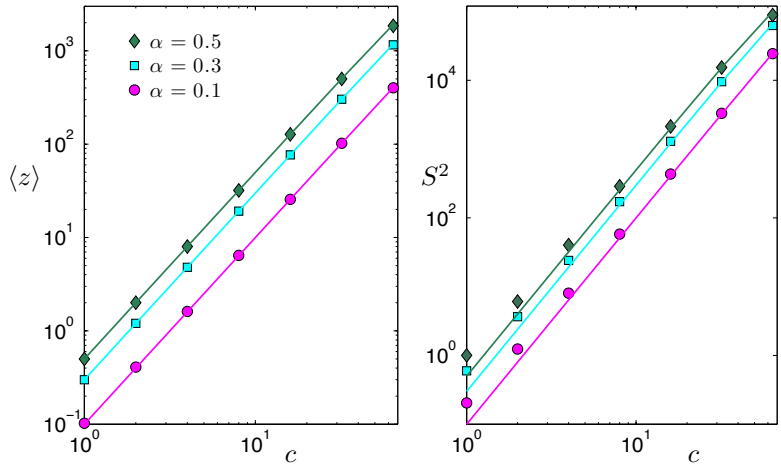
\begin{figure}[t]

\unitlength=0.67mm
 \hspace*{30mm}\begin{picture}(150,90)
\put(0,0){\includegraphics[width=150\unitlength]{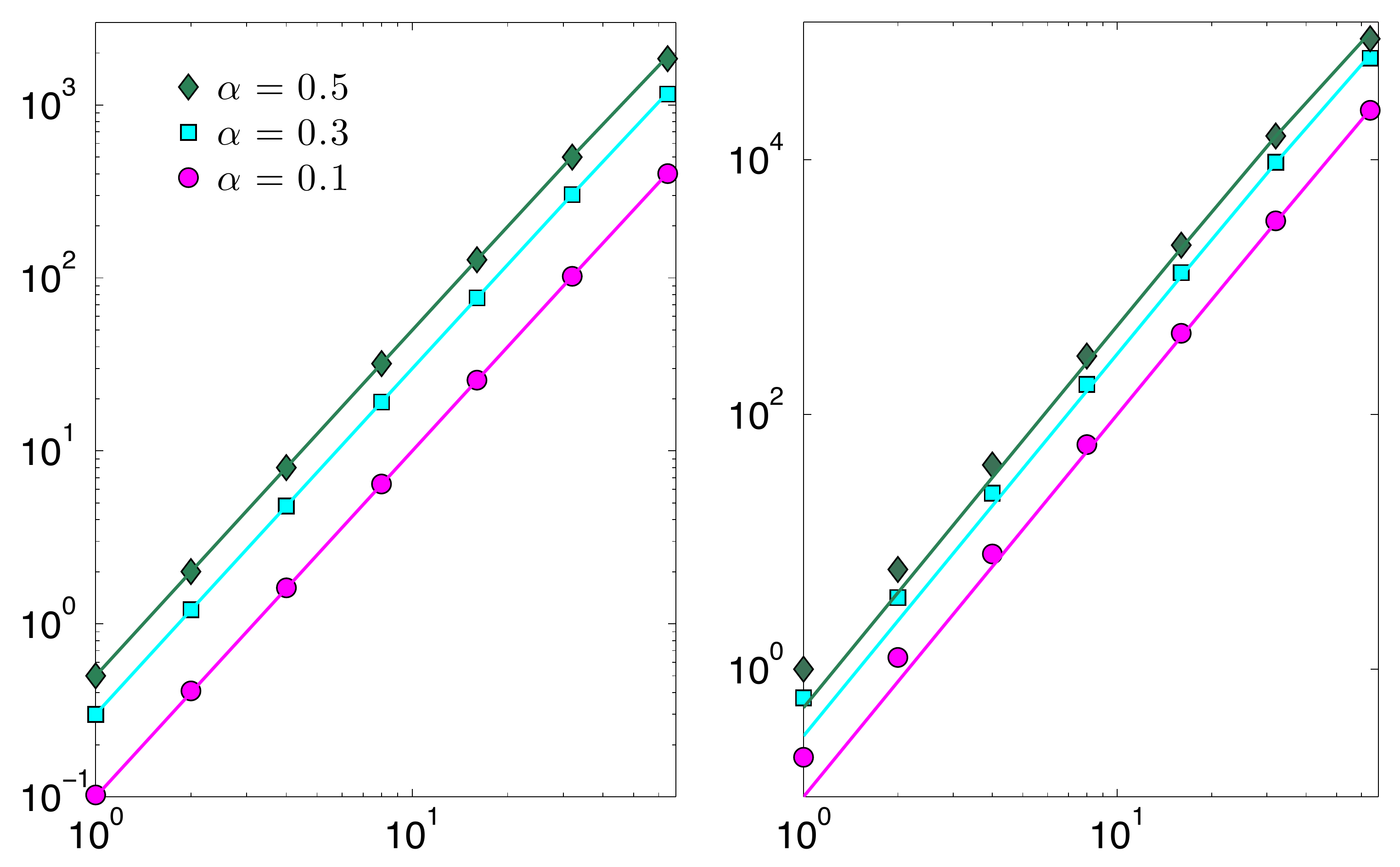}}
\put(50,0){$c$}
\put(130,0){$c$}
\put(-5,55){$\langle z\rangle$}
\put(77,55){$S^2$}
\end{picture}

\vspace*{1mm}

\caption{Average degree $\langle z \rangle$ (left) and its fluctuations $S^2=\langle z^2 \rangle - \langle z^2 \rangle$ (right) in the bi-partite graph ${\mathcal B}$, as a function of $c$ and for different values of $\alpha$ as shown in the legend. Data from numerical simulations (symbols) performed on systems  of fixed size $N_T=1.5 \times 10^3$ are compared with the analytical predictions (solid lines)
given by (\ref{eq:z}) and (\ref{eq:z2}).}
\label{fig:z}
\end{figure}

In the bi-partite graph, given two nodes $i, j \in \mathcal{V}_T$, with $k_i$ and $k_j$ nearest neighbours respectively, the number $\ell$ of shared nearest-neighbors corresponds to the number of non-null matchings between the related strings, $(\xi_i^1,\ldots,\xi_i^{N_B})$ and $(\xi_j^1,\ldots,\xi_j^{N_B})$,  which is distributed according to
\be \label{eq:elle}
P(\ell | k_i, k_j, N_B) = \frac{N_B!}{(N_B+ \ell - k_i - k_j)! (k_i - \ell)! ( k_j - \ell)! \ell!} \left[
\Big(\!\begin{array}{c}N_B\\ k_i\end{array}\!\Big)
\Big(\!\begin{array}{c}N_B\\ k_j\end{array}\!\Big)
 \right]^{-1}.
\ee
Note that the number $\ell$ also provides an upper bound for $J_{ij}$. From
(\ref{eq:elle}) the average $\langle \ell \rangle_{k_i,k_j}$ is found to be
\be
\langle \ell \rangle_{k_i,k_j} = k_i k_j/N_B.
\ee
We evaluate the typical environment for node $i$ by averaging $P(\ell | k_i, k_j, N_B)$ over $P_T(k_j, c, N_B)$, as given by
(\ref{eq:binomial_rho}), and get
\be
\hspace*{-10mm}
P(\ell | k_i, c, N_B) =  \sum_{k_j=0}^{N_B} \Big(\!\begin{array}{c}N_B\\ k_j\end{array}\!\Big)
\Big(\frac{c}{N_T} \Big)^{k_j} \Big(1 \!-\! \frac{c}{N_T} \Big)^{N_B - k_j} \! P(\ell | k_i, k_j, N_B) ~= \frac{d^{k_i - \ell} (1\!-\!d)^{\ell} k_i!}{\ell! (k_i - \ell)!}.
\ee
In particular, the probability for $i$ to be connected to an arbitrary node $j$, can be estimated as $p_{k_i} \equiv 1 - P(\ell =0 | k_i, c, N_B)$, with which the average degree of node $i$ can be written as
\begin{eqnarray}
\langle z \rangle_{k_i} = p_{k_i} N_T =  (1 - d^{k_i}) N_T.
\end{eqnarray}
Upon averaging $\langle z \rangle_{k_i}$ and $\langle z \rangle_{k_i}^2$ over $P_T(k_i | c, N_B)$, we get estimates for the average degree and its variance:
\begin{eqnarray}
\label{eq:z}
\langle z \rangle &=& \Big\{ 1-[ d(2\!-\!d)]^{N_B} \Big\} N_T \sim (1-e^{ \alpha c^2/N_T}) N_T \sim \alpha c^2, \\
\label{eq:z2}
\langle z^2 \rangle - \langle z \rangle^2 &=& \Big\{1 - 2[d(2\!-\!d)]^{N_B} + d^{N_B} (1\!+\!d\!-\!d^2)^{N_B} - [1 \!-\! (d(2\!-\!d))^{N_B}]^2  \Big\}N_T^2  \sim \alpha c^3,
\end{eqnarray}
where the last aproximation holds when $c/N_T$ is small.
As expected, we indeed recover the average degree predicted by the mean-field approach
(\ref{eq:z_MF}), while the fluctuations display an additional factor $c$ (see (\ref{eq:z2_MF})). The analytical results (\ref{eq:z}) and (\ref{eq:z2}) are compared with numerical data in Fig.~\ref{fig:z}.
The agreement is very good, especially for large $c$ where the number of bonds is larger and hence the statistics more sound.

\subsubsection{Growth and robustness} \label{sec:robust}

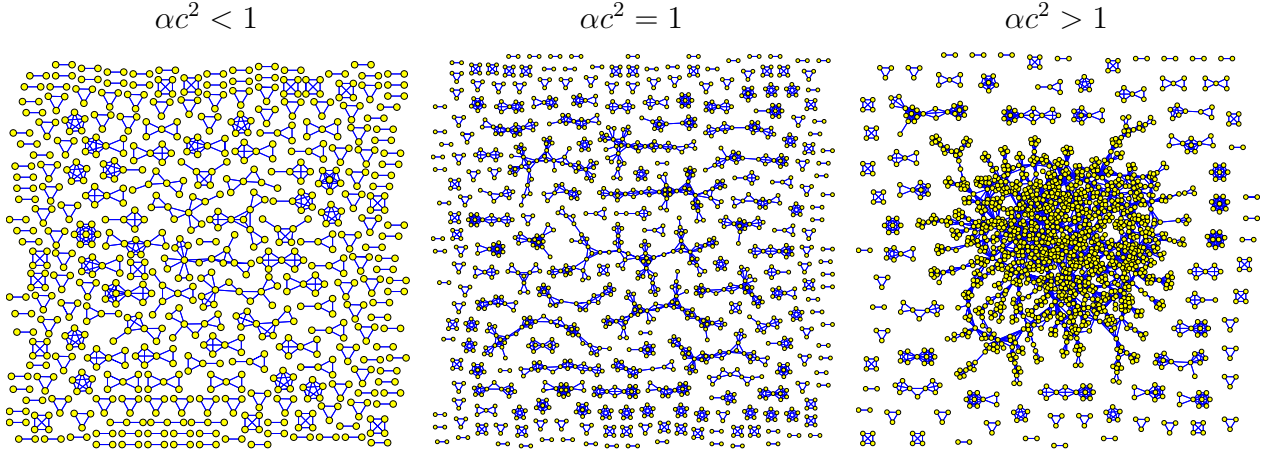
\begin{figure}[t]
\unitlength=0.565mm
 \begin{picture}(300,102)
\put(0,0){\includegraphics[width=95\unitlength]{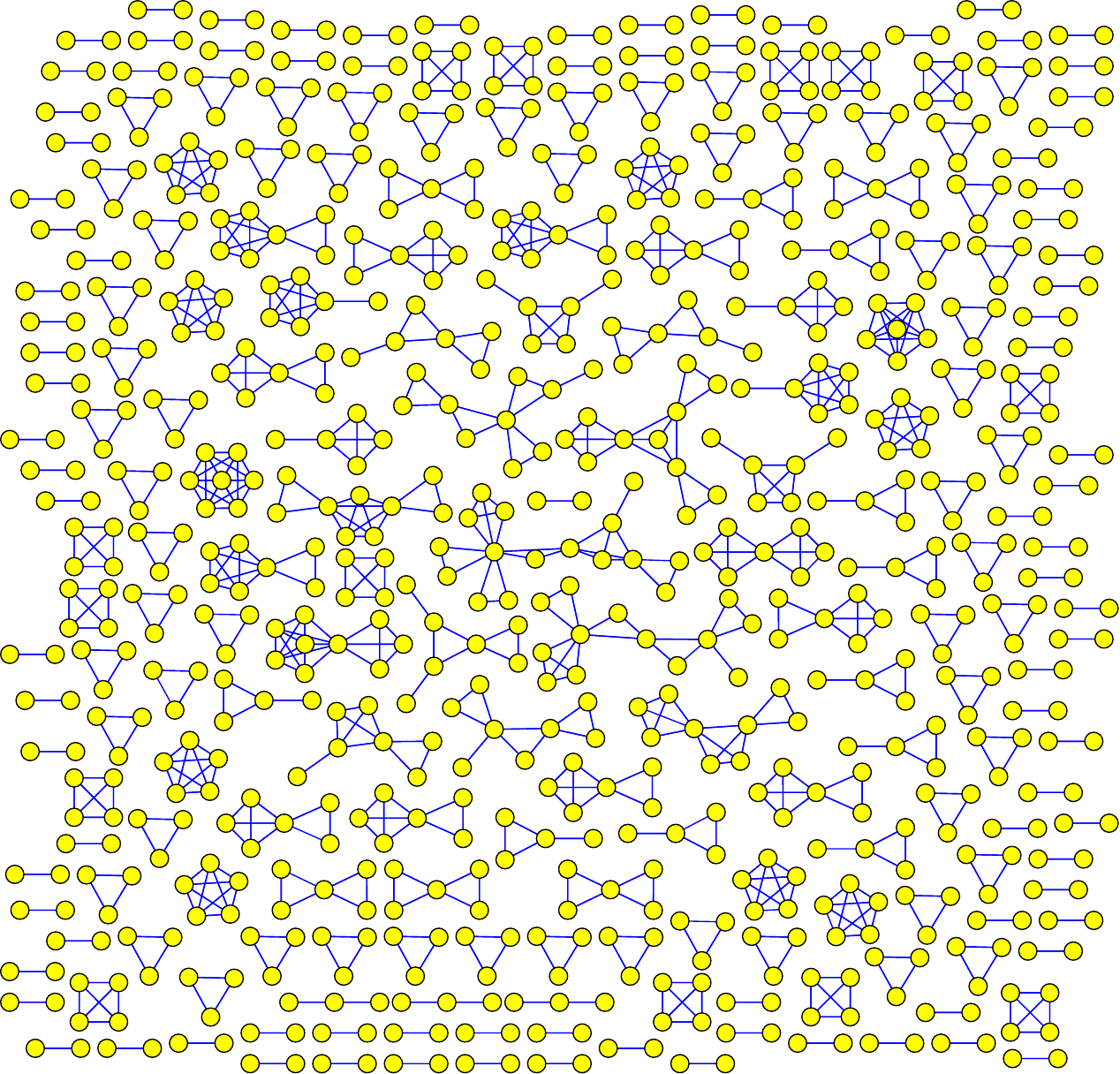}}
\put(100,0){\includegraphics[width=95\unitlength]{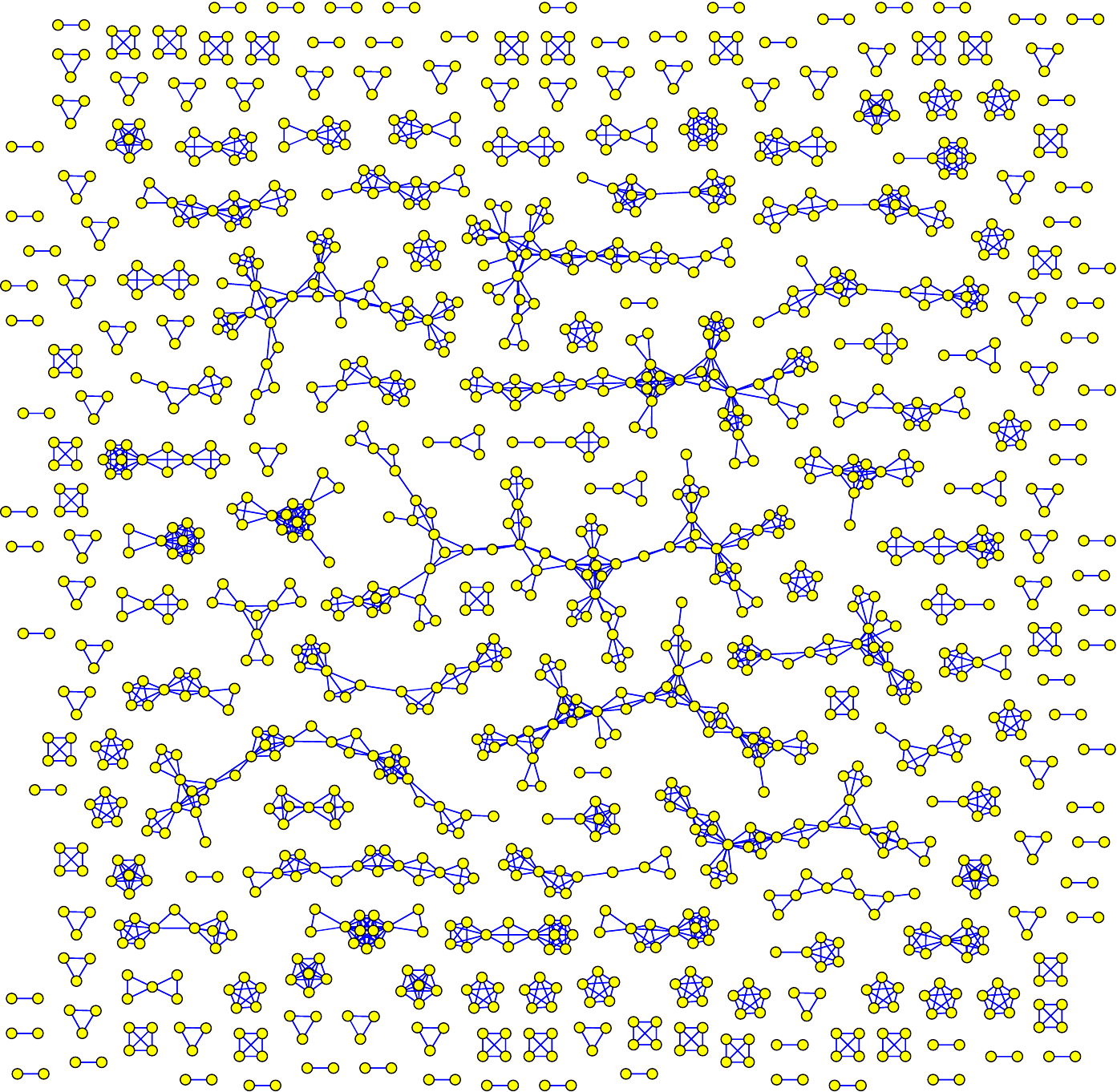}}
\put(200,0){\includegraphics[width=95\unitlength]{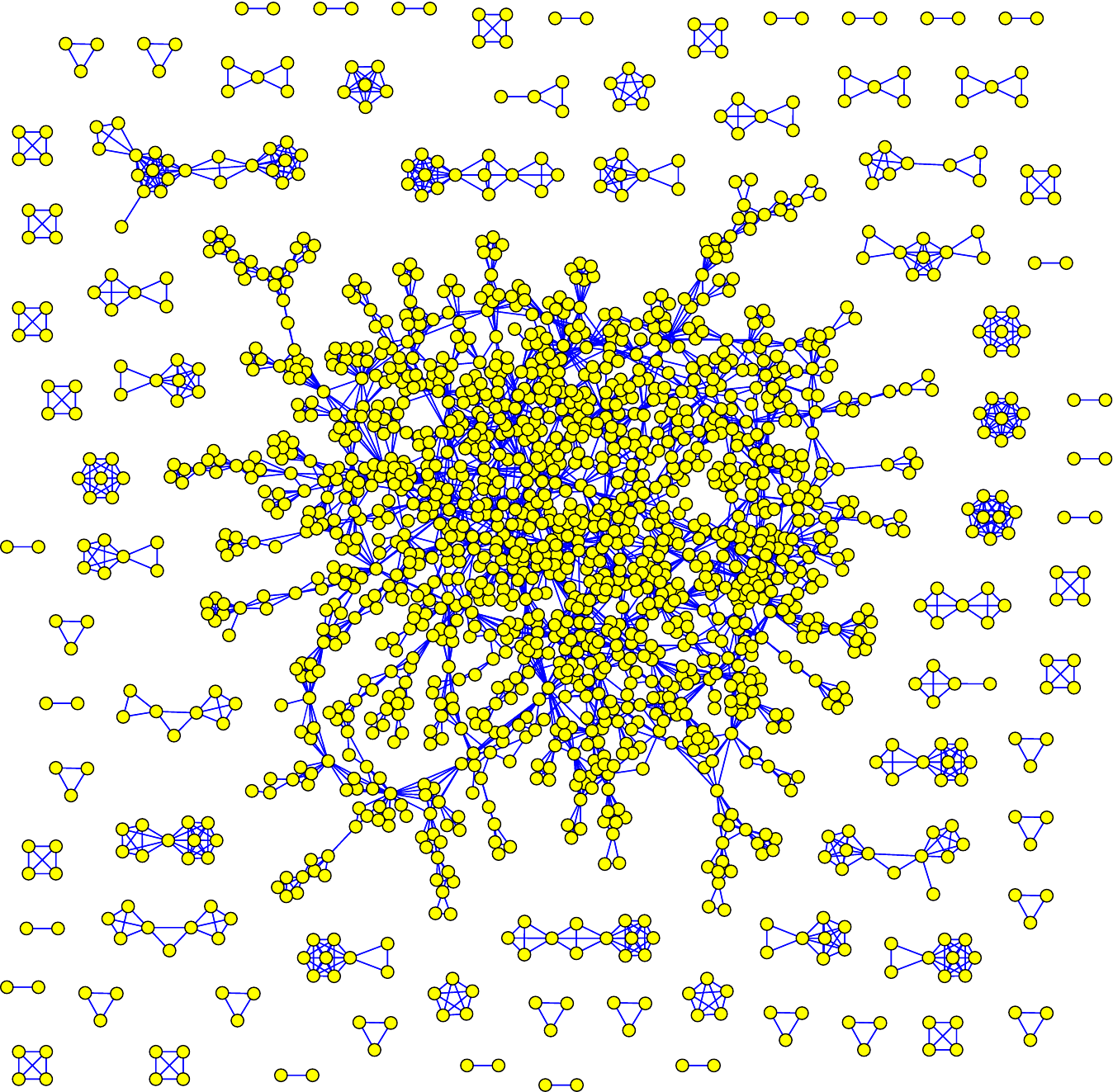}}
\put(35,99){\large $\alpha c^2 < 1$}
\put(135,99){\large $\alpha c^2 =1$}
\put(235,99){\large $\alpha c^2  > 1$}
\end{picture}
\vspace*{-1.5mm}

\caption{\label{fig:grafi2} Examples of typical graphs $\mathcal{G}$ obtained for different values of $c$, while $\gamma=1, \delta=1$, $N_T=5.10^3$ and $\alpha=0.1$ are kept fixed. Left:  the under-percolated regime. Middle: the percolation threshold. Right: the
over-percolated regime.
Isolated nodes, amounting to $4229$, $3664$ and
$3243$, respectively,
 are not shown here. As expected, although many short loops are already present for low connectivity, non trivial (longer) loops start to
occur at the percolation threshold $\alpha c^2=1$.}
\end{figure}

\begin{figure}[t]
\vspace*{1mm}
 \begin{center}
\includegraphics[width=90mm]{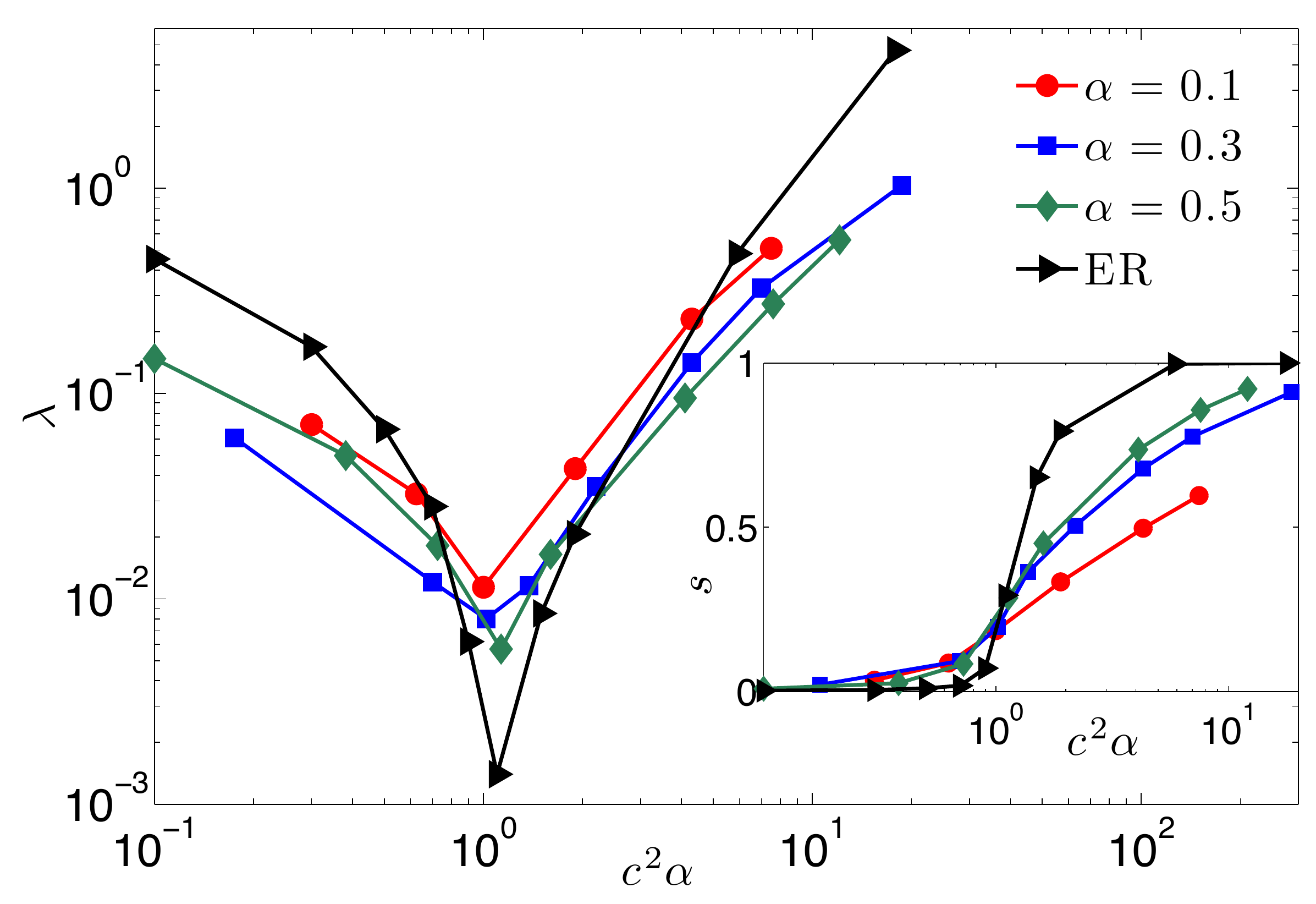}
\caption{\label{fig:Laplacian} Main plot: algebraic connectivity $\lambda$ versus $c^2 \alpha$, measured on the largest components  of graphs of size $N_T=5000$, with different values of $\alpha$ (see legend). Similar results for Erd\"{o}s-R\'{e}nyi (ER) graphs are shown for comparison; here $\lambda$ is plotted versus the link probability times $N_T$, which represents the mean coordination number over the whole network.  Inset: size $S$ of the largest component for $\alpha=0.1$ versus $\alpha c^2$. }
\end{center}
\end{figure}

As anticipated, the point where $ \alpha c^2 = 1$ defines the percolation threshold for the bi-partite graph $\mathcal{B}$:
when $c < 1/ \sqrt{\alpha}$ the graph is fragmented into a number of components with sub-extensive size, while for $c > 1/ \sqrt{\alpha}$ a giant (i.e. extensive) component emerges.
This phenomenology is mirrored in the monoportite graph $\mathcal{G}$.
In particular, we will show that
for $c < 1/ \sqrt{\alpha}$ there is a large number of disconnected components in $\mathcal{G}$
with finite size and  a high degree of modularity,
while for $c > 1/ \sqrt{\alpha}$ bridges appear between these components,
modularity progressively decays, and again a giant component emerges
(see Fig.~\ref{fig:grafi2}).
The transition across the percolation threshold
is rather smooth, as it stems from a main component which encompasses, as
$\alpha c^2$ is increased, more
and more isolated nodes and small-sized components.
This contrasts sharply with the situation in explosive percolation processes
\cite{Achlioptas-Science2009}, where a number of components develop and their
merging at the percolation threshold gives rise to a steep growth in the size
$s$ of the largest component.
Here $s$ grows smoothly and, even at relatively large values of $c^2 \alpha$, a significant fraction of nodes remain isolated or form small-size components (see Fig.~\ref{fig:Laplacian}, inset).

Moreover, the largest component exhibits high levels of modularity and clustering (see \cite{medio} for more details). This can be understood. For $\alpha <1$, any set $\mathcal{C}_{\mu}$ such that all nodes $i \in \mathcal{C}_{\mu} \subseteq  \mathcal{V}_T$ share at least one neighbor $\mu \in \mathcal{V}_B$ will, upon marginalization, result in a clique in $\mathcal{G}$. Hence, $\mathcal{G}$ is relatively compact and redundant and, due to its smooth growth, will remain so even around $\alpha c^2  =1$. One can check this by measuring the algebraic connectivity, i.e. the spectral gap, $\lambda$ of the largest component; results are shown in Fig.~\ref{fig:Laplacian}, main plot
\footnote{The algebraic connectivity $\lambda$, or `spectral gap',
i.e. the second smallest eigenvalue of the Laplacian matrix of a graph, is regarded as a useful quantifier in the analysis of various robustness-related
problems. For instance, $\lambda$ is a lower bound on both the node and the
link connectivity. More precisely, a small algebraic connectivity means that it is relatively easy to disconnect the graph, i.e. to cut it
 into independent components. This means that there exist `bottle-necks', i.e. one can identify subgraphs that are connected only via  a small number of `bridges'. A small algebraic connectivity is also known to
influence transport processes on the graph itself and to favour instability of synchronized states (synchronizability) \cite{Donetti-JSTAT2006}.}.
A graph with a small $\lambda$ has a relatively clean bisection, while large $\lambda$ values characterise
non-structured networks, in which a simple clear-cut separation
into subgraphs is not possible.
As shown in Fig.~\ref{fig:Laplacian}, the minimum of $\lambda$ provides a consistent  signature of percolation, since the possible coalescence of different components is likely to yield the formation of bridges. Moreover, by comparing data for $\mathcal{G}$ and for an Erd\"{o}s-R\'{e}nyi graph we see that when $ \alpha c^2 \approx 1$, where the related largest-size components are comparable, the former displays a larger $\lambda$ and it is hence more structured.

\subsubsection{Component size distribution and retrieval} \label{sec:components}

\begin{figure}[t]
\vspace*{-3mm}
 \begin{center}
\includegraphics[width=.60\textwidth]{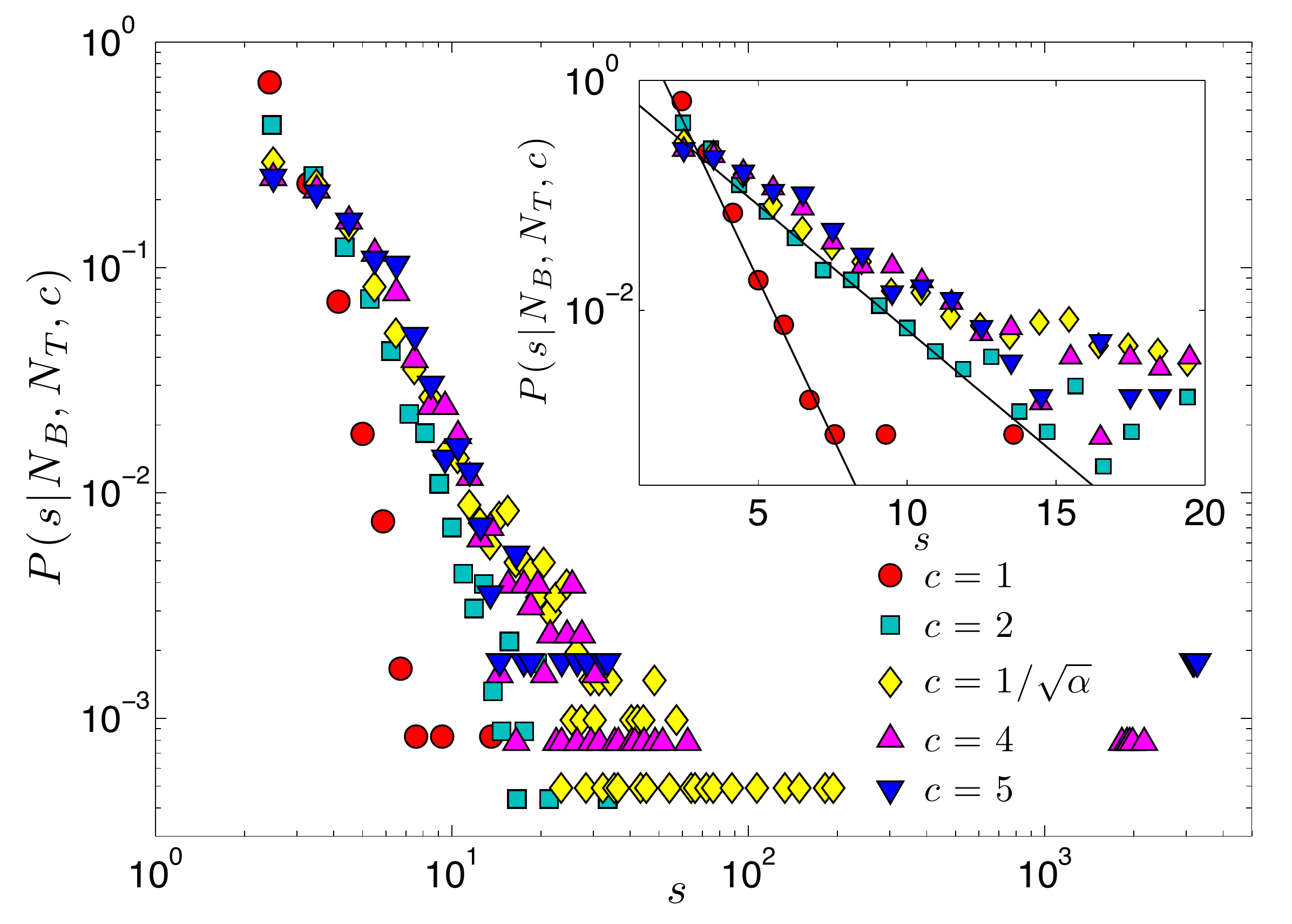}
\vspace*{-1mm}

\caption{\label{fig:compo} Size distribution of the connected components
in $\mathcal{G}$ for $\alpha=0.1$ and different values of $c$ (see legend).
Data are obtained from numerical simulations of systems with
$N_T=\frac{3}{2}.10^3$.
For $c\!<\!1/\sqrt{\alpha}$ (in the under-percolated regime,
{\large$\circ$},$\square$) the decay is exponential with a finite cut-off,
for $c\!=\!1/\sqrt{\alpha}$ ($\lozenge$) the exponential decay broadens,
while for $c\!>\!1/\sqrt{\alpha}$ (in the over-percolated regime, $\vartriangle$,$\triangledown$)
large components appear.
}
\end{center}
\end{figure}

We next analyse the structure of the small components in $\mathcal{G}$, as they are strongly related to retrieval properties, starting with the \emph{underpercolated regime}. Here the typical components in $\mathcal{B}$ are stars $S_n$ centered in a node $\mu \in \mathcal{V}_B$ (because $N_T > N_B$), possibly with arms of length larger than $1$, or combination of stars. In all these cases, two nodes $i, j \in \mathcal{V}_T$ share at most one neighbor $\mu \in \mathcal{V}_B$, so the spins  $\sigma_i$ and $\sigma_j$ can communicate non-conflicting signals to $\mu$.
More precisely, such components allow for spin configurations with
nonzero Mattis magnetizations for {\em all} the patterns involved in the component  (see  Figure \ref{fig:casi}).
This scenario is therefore compatible with parallel retrieval.

Parallel retrieval can be jeopardized by loops in $\mathcal{B}$, which can create disruptive feed-back mechanisms between spins which prevent the complete and simultaneous retrieval of all patterns within the component (see the image on the right in Fig.~\ref{fig:casi}).
We can estimate the probability that a loop involving two nodes $i, j \in \mathcal{V}_T$ occurs in $\mathcal{B}$: since the graph is bi-partite, the minimum length for loops is $4$, which requires that $i$ and $j$ share a number $\ell \geq 2$ of neighbours in $\mathcal{B}$.
We can write $P(\ell \leq 2 | k_i, k_j, N_B) = P(\ell =0 | k_i, k_j, N_B) + P(\ell =1 | k_i, k_j, N_B)$. By replacing $\ell =0$ and $\ell=1$ in (\ref{eq:elle}), we get, respectively,
\begin{eqnarray}
P(\ell =0 | k_i, k_j, N_B) &=&
\Big(\!\begin{array}{c}N_B-k_i\\ k_j\end{array}\!\Big)
\Big(\!\begin{array}{c}N_B\\ k_j\end{array}\!\Big)^{-1}, \\
P(\ell =1 | k_i, k_j, N_B) &=& k_j
\Big(\!\begin{array}{c}N_B-k_i\\ k_j-1\end{array}\!\Big)\Big(\!\begin{array}{c}N_B\\ k_j\end{array}\!\Big)^{-1}.
\end{eqnarray}
By averaging over the distribution $P(k_j | d, N_B)$  (\ref{eq:binomial_rho}) of
$k_j$, we obtain the typical behaviour for an arbitrary node $i$
\begin{equation}
P(\ell \leq 2 | k_i, d, N_B) = d^{k_i} +  d^{k_i-1} (1\!-\!d) k_i=  \Big( 1\!-\! \frac{c}{N_T} \Big)^{k_i} \Big(1\!+\! \frac{c}{N_T} k_i \Big)
\sim  1 - \left( \frac{c k_i}{N_T} \right)^2,
\end{equation}
where in the last step we used $k_i\ll N_T$.
In particular, when $c$ is relatively large ($c \alpha >1$), the approximation $k_i \approx \langle k \rangle$ is valid, and we see that the number of node pairs sharing at least two neighbours in $\mathcal{B}$ scales as $(\alpha c^2)^2$.
Hence, in the underpercolated regime $\alpha c^2 < 1$,
the graph $\mathcal{B}$ is devoid of loops, which is a necessary
condition for straightforward error-free parallel retrieval.
As mentioned before, in this regime the typical components in $\mathcal{B}$ are stars $S_n$ centered in a node $\mu \in \mathcal{V}_B$, possibly presenting arms of length larger than $1$, or combination of stars.
Upon marginalization, these arrangements give rise to complete graphs $K_n$, with nodes possibly linked to small trees, or combinations of complete graphs, respectively (see Fig.~\ref{fig:grafi2}, left panel).
Hence, in the underpercolated regime the typical components in $\mathcal{G}$ are of finite size, and form cliques. The typical size of these cliques decays exponentially with $s$, as shown in Fig.~\ref{fig:compo} (see also \cite{medio}).

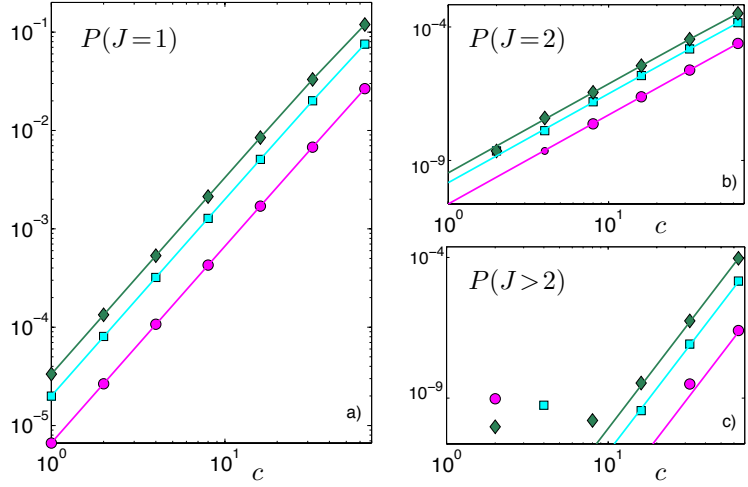
\begin{figure}[t]
\unitlength=0.67mm
 \hspace*{30mm}\begin{picture}(150,90)
\put(0,0){\includegraphics[width=150\unitlength]{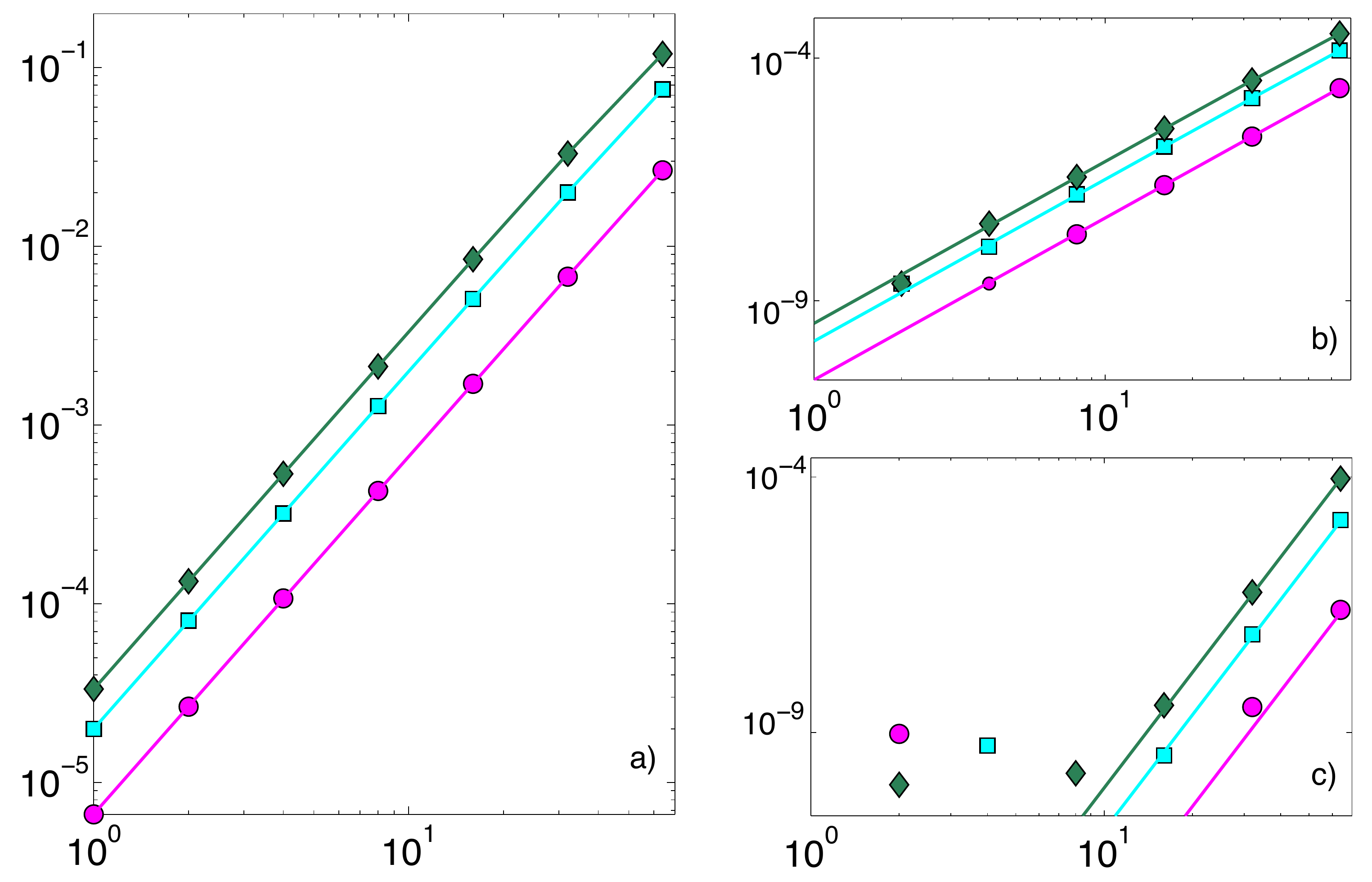}}
\put(50,0){$c$}
\put(130,0){$c$}
\put(130,49){$c$}
\put(16,86){$P(J\!=\!1)$}
\put(93,86){$P(J\!=\!2)$}
\put(93,38){$P(J\!>\!2)$}
\end{picture}

\vspace*{1mm}

\caption{\label{fig:J1J2} These plots show the average probability for an arbitrary link in $\mathcal{G}$ to have weight $J=1$, $J=2$ or $J>2$, respectively, as a function of $c$, for several values of $\alpha$ (being
$\alpha=0.1$, $\alpha=0.3$ and $\alpha=0.5$, with markers as  in the legend of Fig.~\ref{fig:z}).
Numerical simulations (symbols)
were carried out for systems with $N_T=\frac{3}{2}.10^3$ nodes, and  are compared
with the analytical estimates provided by (\ref{eq:J1_1}) and (\ref{eq:J1_2}).
}
\end{figure}

At the percolation threshold, larger loops start to appear in $\mathcal{B}$. For the graphs $\mathcal{G}$ this implies that two cliques can share not only nodes, but even links, and that two nodes $i,j$ can display a coupling $|J_{ij}| \geq 2$ (see Fig.~\ref{fig:grafi2}, middle panel). As a result, the simultaneous retrieval of all patterns within the same component is no longer ensured, and
the distribution of component sizes will broaden.

In the over-percolated regime, a giant component of size $\mathcal{O}(N_T)$ emerges, while many isolated nodes and finite-size components will remain still. Now the average coordination number in the whole graph $\mathcal{G}$ is approximately $c^2 \alpha$ (see (\ref{eq:z})), but will be larger on the giant component. It is worth focusing on the
macroscopic component to find out how it is organized, and how
it compares to a random structure such as the Erd\"{o}s-R\`{e}nyi graph.
We note that even for the giant component the distribution $P(J|N_B, N_T, c)$ has a finite variance, and is concentrated on small values of $J$. To see this we calculate from (\ref{eq:RW}) $P(J=1|N_B, N_T,c)$, $P(J=2|N_B, N_T,c)$ and infer the asymptotic behavior for $P(J|N_B, N_T,c)$.
From (\ref{eq:RW}) we get
\begin{eqnarray}
\label{eq:J1_1}
\nonumber
\hspace*{-23mm}
P(J\!=\!1|N_B, N_T, c) &=& \sum_{S=1}^{N_B-1}\!\!^{\prime}~\frac{N_B!}{S! \Big( \frac{N_B-S+1}{2} \Big)! \Big( \frac{N_B-S-1}{2} \Big)! } p_w^S p_r^{N_B-S}\\
\nonumber
&=& \frac{(N_B-\frac{1}{2})!}{\sqrt{\pi}(N_B-1)!}  \Big(\frac{c \alpha}{N} \Big)^{2N_B-2} \Big[ 1 \!-\! \Big( \frac{c \alpha}{N} \Big)^{2}  \Big]
{_{2}F}_1\Big(\!-\frac{N_B}{2}, 1\!-\! \frac{N_B}{2}, \frac{3}{2}, \Big(\frac{c \alpha}{N_B}\Big)^4 \Big[1 \!-\!\Big(\frac{c \alpha}{N_B}\Big)^2 \Big]^2 \Big)
\nonumber
\\
&\approx&  N_B \Big(\frac{c \alpha}{N_B}\Big)^2 \Big[1 \!-\!\Big(\frac{c \alpha}{N_B}\Big)^2 \Big]^{N_B-1} \sim \Big(\frac{c^2 \alpha^2}{N_B}\Big) \rme^{-c^2 \alpha^2/N_B},
\end{eqnarray}
where the prime restricts the sum to values of $S$ with different parity
from $N_B$ (here assumed even), and where we used
the isotropy ($p_r = p_l$) of the random walk. The
asymptotic form obtained in the last step  applies to $N_B \gg1$.
Similarly, for the case $J=2$ we find
\begin{eqnarray}
\label{eq:J1_2}
\nonumber
\hspace*{-10mm}
P(J=2|N_B, N_T, c) &=& \sum_{S=0}^{N_B-2}\!\!^{\prime} \frac{N_B!}{S! \left( \frac{N_B-S+2}{2} \right)! \left( \frac{N_B-S-2}{2} \right)! }~ p_w^S p_r^{N_B-S}\\
\nonumber
&=& \frac{( \frac{c \alpha}{N_B}) N_B! }{(\frac{N_B}{2}\!+\!1)! (\frac{N_B}{2}\!-\!1)! 2^{N_B}} {_{2}F}_1\Big(\!-1\!-\frac{N_B}{2}, 1\!-\! \frac{N_B}{2}, \frac{1}{2}, \Big(\frac{N_B}{c \alpha}\Big)^4 \Big[1 \!-\!\Big(\frac{c \alpha}{N_B}\Big)^2 \Big]^2 \Big)\\
&\approx&   \Big(\frac{c^2 \alpha^2}{N_B}\Big)^2 \Big[1\! -\!\Big(\frac{c \alpha}{N_B}\Big)^2 \Big]^{N_B-2} \sqrt{1\!-\!\frac{2}{N}} \frac{1}{8}~\sim \Big(\frac{c^2 \alpha^2}{N_B}\Big)^2 \rme^{-c^2 \alpha^2/N_B}.
\end{eqnarray}
Hence we expect the leading terms to scale as
$P(J) \approx (c^2 \alpha^2/N_B)^J [ 1 \!- \!(c \alpha/N_B
)^2]^{N_B-J} $.
These results are confirmed by numerical simulations, with different choices for the parameters $c$ and $\alpha$, see Figure \ref{fig:J1J2}.

\section{Equilibrium analysis in the regine of high storage and finite connectivity}

\subsection{Definitions}

We now turn to a statistical mechanics analysis and
consider the effective network consisting solely of T-cells,
with effective interactions
 described by the following Hamiltonian (rescaled by a factor $c$ relative to (\ref{eq:hopfield})):
\be
\Hi(\bsigma|\xi)=-\frac{1}{2c}\sum_{i,j}^{N_T}\sum_{\mu}^{N_B}\xi^{\mu}_i\xi^{\mu}_j\s_i\s_j.
\label{eq:hopfield_again}
\ee
It is not a priori obvious that solving this model analytically will be possible. Most methods for spins systems on finitely connected heterogeneous graphs rely (explicitly or implicitly) on these being locally tree-like; due to the pattern dilution, the underlying topology of the  system (\ref{eq:hopfield_again}) is a heterogeneous graph with many short loops.
From now on we will write $N_T$ simply as $N$, with $N_B=\alpha N$ and $N\to\infty$.
The cytokine components $\xi_i^{\mu}\in \{-1,0,1\}$ are quenched random variables, identically and independently distributed according to
\be
P(\xi^{\mu}_i=1)=P(\xi^{\mu}_i=-1)=\frac{c}{2N},\quad
\quad P(\xi^{\mu}_i=0)=1-\frac{c}{N},
\ee
with $c$ finite.
The Hamiltonian is  normalized correctly: since the term $\sum_{i=1}^{N}\xi^{\mu}_i\s_i$ is $\order{(1)}$ both for condensed and non condensed patterns \cite{medio},
 (\ref{eq:hopfield_again})  is indeed extensive in  $N$.
The aim of this section is to compute the disorder-averaged free energy $\overline{f}$, at inverse temperature $\beta=T^{-1}$, where
$\overline{\cdots}$ denotes averaging over the $\alpha N^2$ variables $\{\xi_i^\mu\}$ and
\be
f=-\lim_{N\to\infty}\frac{1}{\beta N}\log Z_N(\beta,\xi),
\label{eq:f}
\ee
where $Z_N(\beta,\xi)$ is the partition function
\be
Z_N(\beta,\xi)=\sum_{\bsigma\in\{-1,1\}^N}\rme^{\frac{\beta}{2c}\sum_{\mu=1}^{\alpha N}\left(\sum_{i=1}^N\xi^{\mu}_i\s_i\right)^2}.
\ee

The state of the system can be characterized in terms of the $\alpha N$ (non-normalised)
Mattis magnetizations, i.e. the overlaps between the system configuration
and each cytokine pattern
\begin{eqnarray}
M_\mu(\bsigma)=\sum_{i=1}^N\xi_i^\mu \sigma_i.
\end{eqnarray}
However, since in the high load regime the number of overlaps is extensive,
it is more convenient to work with the overlap distribution
\be
P(M|\bsigma)=\frac{1}{\alpha N} \sum_{\mu=1}^{\alpha N} \delta_{M_{\mu}(\bsigma),M}.
\label{eq:marginal}
\ee
Although $M_{\mu}(\bsigma)$ can take (discrete) values in the whole range $\left\{-N, -N+1,\cdots, N\right\}$, we expect that, due to dilution,
the number of values that the $M_{\mu}(\bsigma)$ assume remains effectively finite for
large $N$, so that (\ref{eq:marginal}) represents an effective finite  number of order parameters.
In order to probe responses of the system to selected perturbations or triggering of clones
we introduce external fields $\{\psi_\mu\}$ coupled to the overlaps
$\{M_\mu(\bsigma)\}$, so we consider the extended Hamiltonian
\be
\Hi(\s,\xi)=-\frac{1}{2c}\sum_{i,j}^{N}\sum_{\mu}^{\alpha N}\xi^{\mu}_i\xi^{\mu}_j\s_i\s_j -\sum_{\mu=1}^{\alpha N} \psi_\mu M_\mu(\bsigma).
\label{eq:H}
\ee
We also define the field distribution $P(\psi)$ and   the joint distribution $P(M,\psi|\bsigma)$ of magnetizations and fields, which is the most informative observable from a biological point of view (and of which $P(\psi)$ is a marginal):
\begin{eqnarray}
P(\psi)=\frac{1}{\alpha N}\sum_{\mu=1}^{\alpha N} \delta(\psi-\psi_\mu),~~~~~~~~
P(M,\psi|\bsigma)=\frac{1}{\alpha N}\sum_{\mu=1}^{\alpha N} \delta_{M,M_\mu(\bsigma)}\delta(\psi-\psi_\mu).
\label{eq:joint}
\end{eqnarray}

\subsection{The free energy}

The free energy per spin (\ref{eq:f}) for the Hamiltonian (\ref{eq:H}) can be writen as
\bea
f&=&  -\lim_{N\to\infty}\frac{1}{\beta N}\log \sum_\bsigma
\rme^{\frac{\beta}{2c}\sum_{\mu=1}^{\alpha N} M_{\mu}^2(\bsigma)+
\beta
\sum_{\mu=1}^{\alpha N} \psi_\mu M_{\mu}(\bsigma)}.
\label{eq:f_M}
\eea
We insert the following integrals of delta-functions written in Fourier representation
\begin{eqnarray}
1&=&\prod_{M}\prod_{\psi}\int\!\rmd P(M,\psi)~\delta\Big[P(M,\bpsi)-\frac{1}{\alpha N}\sum_{\mu=1}^{\alpha N}\delta_{M,M_\mu(\bsigma)}\delta(\psi-\psi_\mu)\Big]
\nonumber
\\
&=& \prod_{M}\prod_{\psi}\int\!\frac{\rmd P(M,\psi)\rmd\hat{P}(M,\psi)}{2\pi/\Delta N}
~\rme^{\rmi N\Delta\hat{P}(M,\psi)[P(M,\psi)-\frac{1}{\alpha N}\sum_{\mu=1}^{\alpha N}\delta_{M,M_\mu(\bsigma)}\delta(\psi-\psi_\mu)]}.
\end{eqnarray}
In the limit $\Delta\to 0$ we use $\Delta\sum_\psi\ldots \to \int\! \rmd\psi\ldots$, and we define the path integral measure $\{\rmd P\rmd \hat{P}\}=\lim_{\Delta\to 0}\rmd P(M,\psi)\rmd\hat{P}(M,\psi)\Delta N/2\pi$. This gives us
\begin{eqnarray}
1&=&\int\!\{\rmd P\rmd\hat{P}\}
~\rme^{\rmi N\int\!\rmd\psi\sum_M\hat{P}(M,\psi)P(M,\psi)-\frac{i}{\alpha}\sum_{\mu=1}^{\alpha N}\hat{P}(M_\mu(\bsigma),\psi_\mu)}.
\end{eqnarray}
Insertion into (\ref{eq:f_M}) leads us to an expression for $f$ involving the density of states $\Omega[\hat P]$:
\bea
\hspace*{-3mm}
f&=&-\lim_{N\to\infty}\frac{1}{\beta N}\log\pint \rme^{N\big\{\rmi\int\! \rmd\psi\, \sum_M P(M,\psi)\hat P(M,\psi) +\beta \alpha \int\! \rmd\psi\, \sum_M P(M,\psi)
\left(\frac{M^2}{2c}+M\psi\right)+\Omega[\hat P]\big\}}
\nonumber
\\[-2mm]
&&
\\
\hspace*{-8mm}\Omega[\hat P]&=&\lim_{N\to\infty}\frac{1}{N}\log \sum_\bsigma
\rme^{-\frac{\rmi}{\alpha}\sum_\mu \hat P(M_\mu(\bsigma),\psi_\mu)}.
\label{eq:Omega}
\eea
Hence via steepest descent integration for $N\to\infty$, and after avering the result over the disorder,  we obtain:
\bea
\hspace*{-5mm}
\overline{f}=-\frac{1}{\beta}{\rm extr}_{\{P,\hat P\}}\Big\{\rmi\!\int\! \rmd\psi\, \sum_M P(M,\psi)\hat P(M,\psi) +\beta \alpha \int\! \rmd\psi\, \sum_M P(M,\psi)(\frac{M^2}{2c}+M\psi)+\overline{\Omega}[\hat P]\Big\},
\nonumber
\\[-2mm]&&
\label{eq:barf}
\\[-9mm]&&\nonumber
\eea
with
\bea
\overline{\Omega}[\hat P]&=&\lim_{N\to\infty}\frac{1}{N}
~\overline{\log \sum_{\bsigma}\rme^{-\frac{\rmi}{\alpha}\sum_\mu \hat P(M_\mu(\bsigma),\psi_\mu)}}.
\label{eq:log_density}
\eea
Working out the functional saddle-point equations that define the extremum in (\ref{eq:barf}) gives
\bea
\hat P(M,\psi)=\rmi\alpha \beta \Big(\frac{M^2}{2c}+M\psi\Big), \quad\quad P(M,\psi)=\rmi\frac{\delta \overline{\Omega}[\hat P]}{\delta\hat P(M,\psi)},
\label{eq:P_hatP}
\eea
and inserting the first of these equations into (\ref{eq:barf}) leads us to
\bea
\overline{f}=-\frac{1}{\beta}\overline{\Omega}[\hat P]\Big|_{\hat P(M,\psi)=\rmi\alpha \beta(\frac{M^2}{2c}+M\psi)}.
\label{eq:barf2}
\eea
Hence calculating the disorder-averaged free-energy boils down to calculating
(\ref{eq:log_density}).  This can be done using the replica method, which is
based on the identity $\overline{\log Z}=\lim_{n\to 0}n^{-1}\log \overline{Z^n}$,
yielding
\bea
\overline{\Omega}[\hat P]
&=&  \lim_{N\to\infty}\lim_{n\to 0}\frac{1}{Nn}\log \sum_{\bsigma^1\ldots\bsigma^n}\overline{\rme^{-\frac{\rmi}{\alpha}\sum_{\alpha=1}^n\sum_{\mu=1}^{\alpha N} \hat P(M_\mu(\bsigma^\alpha),\psi_\mu)}}.
\label{eq:replica_density}
\eea
The free energy (\ref{eq:barf2})
could also have been calculated directly from (\ref{eq:f_M}), by taking the
average over disorder and using the replica identity.
The advantage of working with the log-density of states is that,
working out $\overline{\Omega}[\hat P]$ first for arbitrary functions
$\hat P$ gives us via (\ref{eq:P_hatP}) a formula for the distribution $P(M,\psi)$,
from which we can obtain useful information on the system retrieval
phases and response to external perturbations.
Finally we set $\hat P(M,\psi)=\rmi\alpha \beta\chi(M,\psi)$
with a real-valued function $\chi$, to compactify our equations,  with which we can write our problem as follows
\bea
\hspace*{-1cm}&&\overline{f}=
f[\chi]\Big|_{\chi(M,\psi)=\frac{M^2}{2c}+M\psi}\quad\quad
f[\chi]=-\lim_{N\to\infty}\lim_{n\to 0}\frac{1}{\beta Nn}\log \sum_{\bsigma^1\ldots\bsigma^n}\overline{\rme^{\beta\sum_{\alpha=1}^n\sum_{\mu=1}^{\alpha N} \chi(M_\mu(\bsigma^\alpha),\psi_\mu)}},
\label{eq:free_energy}
\\
\hspace*{-1cm}&&
P(M,\psi)=-\frac{1}{\alpha}\frac{\delta f[\chi]}{\delta \chi}\Big|_{\chi(M,\psi)=\frac{M^2}{2c}+M\psi}.
\label{eq:chi_derivative}
\eea
For simple tests of (\ref{eq:free_energy}) and
(\ref{eq:chi_derivative}) in special limits see \ref{app:Simple_cases}.

\subsection{Derivation of saddle-point equations}

From now on, unless indicated otherwise, all  summations and products over $\alpha$, $i$, and $\mu$ will be understood to imply $\alpha=1\ldots n$, $i=1\ldots N$,  and $\mu=1\ldots \alpha N$, respectively.  We next need to introduce order parameters that allow us to carry out the disorder average in (\ref{eq:free_energy}).
The simplest choice is to  isolate the overlaps themselves by inserting
\begin{eqnarray}
1&=& \prod_{\alpha\mu}\Big[\sum_{M_{\alpha\mu}=-N}^{N}\delta_{M_{\alpha\mu},\sum_i\xi_i^\mu \sigma_i^\alpha}\Big]=
\prod_{\alpha\mu}\Big[\sum_{M_{\alpha\mu}=-N}^{N}\int_{-\pi}^{\pi}\!\frac{\rmd\omega_{\alpha\mu}}{2\pi}\rme^{\rmi\omega_{\alpha\mu}(M_{\alpha\mu}-\sum_i\xi_i^\mu \sigma^\alpha_i)}
\Big].
\label{eq:order}
\end{eqnarray}
This gives
\begin{eqnarray}
f[\chi]&=&   -\lim_{N\to\infty}\lim_{n\to 0}\frac{1}{\beta Nn}\log\Big\{
\prod_{\alpha\mu}\Big[\sum_{M_{\alpha\mu}=-\infty}^{\infty}\int_{-\pi}^{\pi}\!\frac{\rmd\omega_{\alpha\mu}}{2\pi}\Big]
\rme^{\rmi\sum_{\alpha\mu}\omega_{\alpha\mu} M_{\alpha\mu}+\sum_{\alpha\mu}\beta \chi(M_\mu^\alpha,\psi_\mu)}
\nonumber
\\
&&\hspace*{60mm} \times
\sum_{\bsigma^1\ldots\bsigma^n}\!\!\overline{
\rme^{-\rmi\sum_i\sum_{\alpha\mu}\omega_{\alpha\mu}\xi_i^\mu \sigma^\alpha_i}}
\Big\}.
\label{eq:f1}
\end{eqnarray}
We can carry out the disorder average
\begin{eqnarray}
\overline{
\rme^{-\rmi\sum_i\sum_{\alpha\mu}\omega_{\alpha\mu}\xi_i^\mu \sigma^\alpha_i}}&=&
\prod_{i\mu}\Big\{
1-\frac{c}{N}
+\frac{c}{2N}\Big(
\rme^{\rmi\sum_{\alpha}\omega_{\alpha\mu}\sigma^\alpha_i}
\!+\!
\rme^{-\rmi\sum_{\alpha}\omega_{\alpha\mu}\sigma^\alpha_i}
\Big)
\Big\}
\nonumber
\\
&=&  \rme^{\frac{c}{N}\sum_{i\mu}\big[
 \cos(\sum_{\alpha}\omega_{\alpha\mu}\sigma^\alpha_i)-1\big]+\order(N^0)},
\end{eqnarray}
which leads us to
\begin{eqnarray}
f[\chi]&=&    -\lim_{N\to\infty}\lim_{n\to 0}\frac{1}{\beta Nn}\log\Big\{
\prod_{\alpha\mu}\Big[\sum_{M_{\alpha\mu}}\int_{-\pi}^{\pi}\!\frac{\rmd\omega_{\alpha\mu}}{2\pi}\Big].
\rme^{\rmi\sum_{\alpha\mu}\omega_{\alpha\mu} M_{\alpha\mu}+\sum_{\alpha\mu}\beta \chi(M_\mu^\alpha,\psi_\mu)}
\nonumber
\\&&
\hspace*{50mm}
\times \Big[\sum_{\sigma_1\ldots\sigma_n}\rme^{\frac{c}{N}\sum_{\mu}\big[\cos(\sum_{\alpha}\omega_{\alpha\mu}\sigma_\alpha)-1\big]}
\Big]^N\Big\}.
\label{eq:f2}
\end{eqnarray}
We next introduce $n$-dimensional vectors: $\bsigma=(\sigma_1,\ldots,\sigma_n)\in\{-1,1\}^n$, $\bM^{\mu}=(M_{1\mu},\ldots,M_{n\mu})\in
{\Zset}^n$ and $\bomega^\mu=(\omega_{1\mu},\ldots,\omega_{n\mu})\in[-\pi,\pi]^n$.
This allows us to write (\ref{eq:f2}) as
\begin{eqnarray}
f[\chi]&=&   -\lim_{N\to\infty}\lim_{n\to 0}\frac{1}{\beta Nn}\log\Big\{
\prod_{\mu}\Big[\sum_{\bM^\mu}\int_{-\pi}^{\pi}\!\frac{\rmd\bomega^\mu}{(2\pi)^n}\Big]\cdot
\rme^{\rmi\sum_{\mu}\bomega^\mu\cdot \bM^{\mu}+\sum_{\mu}\beta \chi(\bM^\mu,\psi_\mu)}
\nonumber
\\&&
\hspace*{50mm}\times \Big[\sum_{\bsigma}\rme^{\frac{c}{N}\sum_{\mu}[\cos(\bomega^\mu\cdot\bsigma)-1]}
\Big]^N\Big\}.
\label{eq:f3}
\end{eqnarray}
This last expression invites us to  introduce the distribution $P(\bomega)=(\alpha N)^{-1}\sum_\mu \delta(\bomega-\bomega^\mu)$, for $\bomega\in[-\pi,\pi]^n$, via path integrals. We therefore insert
\begin{eqnarray}
1&=& \prod_{\bomega}\int\!\rmd P(\bomega)~\delta\Big[P(\bomega)-\frac{1}{\alpha N}\sum_\mu \delta(\bomega-\bomega^\mu)
\Big]
\nonumber
\\
&=& \prod_{\bomega}\int\!\frac{\rmd P(\bomega)\rmd\hat{P}(\bomega)}{2\pi/\Delta N}\rme^{\rmi N\Delta\hat{P}(\bomega)\Big[P(\bomega)-\frac{1}{\alpha N}\sum_\mu \delta(\bomega-\bomega^\mu)
\Big]}.
\end{eqnarray}
In the limit $\Delta\to 0$ we use $\Delta\sum_{\bomega} \ldots \to \int\!\rmd\bomega \ldots$, and we define the usual path integral measure $\{\rmd P\rmd\hat{P}\}=\lim_{\Delta\to 0}\rmd P(\bomega)\rmd\hat{P}(\bomega)\Delta N/2\pi$. This converts the above
to
\begin{eqnarray}
1&=&
\int\!\{\rmd P\rmd\hat{P}\}~\rme^{\rmi N\int\!\rmd\bomega~\hat{P}(\bomega)P(\bomega)
-(i/\alpha)\sum_\mu\hat{P}(\bomega^\mu)
}.
\end{eqnarray}
and upon insertion into (\ref{eq:f3}) we get
\begin{eqnarray}
f[\chi]&=&   -\lim_{N\to\infty}\lim_{n\to 0}\frac{1}{\beta Nn}\log
\int\!\{\rmd P\rmd\hat{P}\}~\rme^{\rmi N\!\int_{-\pi}^\pi\!\rmd\bomega~\hat{P}(\bomega)P(\bomega)}
\Big[\sum_{\bsigma}\rme^{\alpha c\int\! \rmd\bomega~P(\bomega)[\cos(\bomega\cdot\bsigma)-1]}
\Big]^{\!N}
\nonumber
\\&&
\hspace*{20mm}
\times
\prod_{\mu=1}^{\alpha N}\Big(\sum_{\bM}\int_{-\pi}^{\pi}\!\frac{\rmd\bomega}{(2\pi)^n}
\rme^{\rmi\bomega\cdot \bM+\sum_\alpha\beta \chi(M^\alpha,\psi_\mu)
-\frac{\rmi}{\alpha}\hat{P}(\bomega)
} \Big).
\label{eq:f4}
\end{eqnarray}
In the limit $N\to\infty$, evaluation of the integrals by steepest descent leads to
\begin{eqnarray}
f[\chi]&=& -\lim_{n\to 0}\frac{1}{\beta n}{\rm extr}_{\{P,\hat{P}\}}~\Psi_n[\{P,\hat{P}\}],
\label{eq:f_psi_n}
\\
\Psi_n[\{P,\hat{P}\}]&=& \rmi\int_{-\pi}^\pi\!\rmd\bomega~\hat{P}(\bomega)P(\bomega)
+\alpha \Big\bra \log \Big(\sum_{\bM}\int_{-\pi}^{\pi}\!\frac{\rmd\bomega}{(2\pi)^n}
\rme^{\rmi\bomega\cdot \bM+\sum_\alpha \beta \chi(M^\alpha,\psi)-\frac{i}{\alpha}\hat{P}(\bomega)
} \Big)\Big\ket_\psi
\nonumber
\\
&& +\log\Big(\sum_{\bsigma}\rme^{\alpha c\int_{-\pi}^{\pi}\! \rmd\bomega~P(\bomega)[\cos(\bomega\cdot\bsigma)-1]}
\Big),
\label{eq:psi_n}
\end{eqnarray}
in which $\bra \ldots \ket_\psi=\int\! \rmd\psi\, P(\psi)\ldots$. We mostly
write $\bra \ldots \ket$ in what follows, when there is no risk of ambiguities.
The saddle-point equations are found by functional variation of $\Psi_n$ with respect to $P$ and $\hat P$, leading to
\begin{eqnarray}
\hat P(\bomega)&=&
\rmi c\alpha ~
\frac{\sum_{\bsigma}\big[\cos(\bomega\cdot\bsigma)-1\big]\rme^{\alpha c\int_{-\pi}^{\pi}\! \rmd\bomega^\prime~P(\bomega^\prime)[\cos(\bomega^\prime\cdot\bsigma)-1]}}
{\sum_{\bsigma}\rme^{\alpha c\int_{-\pi}^{\pi}\! \rmd\bomega^\prime~P(\bomega^\prime)[\cos(\bomega^\prime\cdot\bsigma)-1]}},
\\
P(\bomega)&=&
\Langle \frac{\sum_{\bM}
\rme^{\rmi\bomega\cdot \bM+\sum_\alpha \beta \chi(M^\alpha,\psi)
-\frac{\rmi}{\alpha}\hat P(\bomega)}}
 {\sum_{\bM}\int_{-\pi}^{\pi}\!\rmd\bomega^\prime~
\rme^{\rmi\bomega^\prime\cdot \bM+\sum_\alpha \beta \chi(M^\alpha,\psi)
-\frac{\rmi}{\alpha}\hat P(\bomega^\prime)}}\Rangle.
\end{eqnarray}
The joint distribution of fields and magnetizations now follows directly from
(\ref{eq:chi_derivative}) and  (\ref{eq:f_psi_n},~\ref{eq:psi_n}), and is seen to require only knowledge of the conjugate order parameters $\hat{P}(\bomega)$:
\bea
\frac{P(M,\psi)}{P(\psi)}&=& \lim_{n\to 0}\frac{\sum_\bM \Big(\frac{1}{n}\sum_\gamma \delta_{M,M_\gamma}\Big)
\int_{-\pi}^\pi\! \rmd\bomega~\rme^{\rmi\bomega\cdot \bM + \beta\sum_\alpha \chi(M^\alpha,\psi)
-\frac{\rmi}{\alpha}\hat P(\bomega)}}
{\sum_\bM
\int_{-\pi}^\pi\! \rmd\bomega~\rme^{\rmi\bomega\cdot \bM + \beta \sum_\alpha \chi(M^\alpha,\psi)
-\frac{\rmi}{\alpha}\hat P(\bomega)}}\Bigg|_{\chi=\frac{M^2}{2c}+\psi M}.
\label{eq:PM_general}
\eea
Thus the right-hand side is an expression for $P(M|\psi).$
A last simple transformation $F(\bomega)=-\frac{\rmi}{c\alpha}\hat{P}(\bomega)+1$ converts the saddle point equations into
\begin{eqnarray}
F(\bomega)&=&
\frac{\sum_{\bsigma}\cos(\bomega\cdot\bsigma)\rme^{\alpha c\int_{-\pi}^{\pi}\! \rmd\bomega^\prime~P(\bomega^\prime)\cos(\bomega^\prime\cdot\bsigma)}}
{\sum_{\bsigma}\rme^{\alpha c\int_{-\pi}^{\pi}\! \rmd\bomega^\prime~P(\bomega^\prime)\cos(\bomega^\prime\cdot\bsigma)}},
\label{eq:SP2}
\\
P(\bomega)&=&
\Langle\frac{
\rme^{cF(\bomega)}\prod_\alpha \Dpsia}
{\int_{-\pi}^\pi\! \rmd\bomega^\prime~\rme^{cF(\bomega^\prime)}
\prod_\alpha \Dpsia}\Rangle,
\label{eq:SP1}
\end{eqnarray}
where we have introduced
\begin{eqnarray}
\Dpsi&=& \frac{1}{2\pi}\sum_{M\in\Zset}
\rme^{\rmi\omega M+\beta \chi(M,\psi)}.
\label{eq:D}
\end{eqnarray}
Similarly,
(\ref{eq:PM_general}) and (\ref{eq:f_psi_n}) can now be expressed as, respectively,
\bea
\hspace*{-0.8cm}P(M|\psi)=\lim_{n\to 0}
\frac{\sum_\bM \Big(\frac{1}{n}\sum_\gamma \delta_{M,M_\gamma}\Big)
\int_{-\pi}^\pi\! \rmd\bomega~\rme^{\rmi\bomega \cdot\bM + \beta\sum_\alpha \chi(M^\alpha,\psi)
+cF(\bomega)}}
{\sum_\bM
\int_{-\pi}^\pi\! \rmd\bomega~\rme^{\rmi\bomega\cdot \bM +\beta \sum_\alpha \chi(M^\alpha,\psi)
+cF(\bomega)}} \Bigg|_{\chi=M^2/2c+M\psi},
\label{eq:PM_F}
\eea
and
\bea
f[\chi]&=&
-\lim_{n\to 0}\frac{1}{\beta n}
\Big\{
-c\alpha \int_{-\pi}^\pi\!\rmd\bomega~ F(\bomega)P(\bomega)
+\log\Big(\sum_{\bsigma}\rme^{\alpha c\int_{-\pi}^{\pi}\! \rmd\bomega~P(\bomega)[\cos(\bomega\cdot\bsigma)-1]}
\Big)
\nonumber
\\
&&
\hspace*{30mm}
+\alpha \Big\bra \log \Big(\sum_{\bM}\int_{-\pi}^{\pi}\!\frac{\rmd\bomega}{(2\pi)^n}
\rme^{\rmi\bomega\cdot \bM+\sum_\alpha \beta \chi(M^\alpha,\psi)+cF(\bomega)
} \Big)\Big\ket
\Big\}.
\label{eq:f_F}
\eea
We note that the saddle-point equations guarantee that $P(\bomega)$ is normalised correctly on $[-\pi,\pi]^n$, while for $F(\bomega)$
we have (see \ref{app:normalization})
\bea
\int_{-\pi}^{\pi}\! \rmd\bomega\, F(\bomega)=0.
\label{eq:normalization}
\eea
We observe that
in the absence of external fields, i.e. for $\psi=0$, the function (\ref{eq:D}) is real and symmetric:
\begin{eqnarray}
&& D_0(\omega|\beta)=\frac{1}{2\pi}\sum_{M\in\Zset}
\rme^{\rmi\omega M+\frac{\beta}{2c} M^2} \in \R,
\quad\quad
\forall\omega\in[-\pi,\pi]:~~ D_0(-\omega|\beta)=D_0(\omega|\beta).
\label{eq:symmetry_0}
\end{eqnarray}
The introduction of external fields breaks the symmetry of $D_\psi(\omega|\beta)$ under the transformation $\omega \to -\omega$.

\subsection{The RS ansatz -- route I}

To solve the saddle point equations for $n\to 0$ we need to make
an ansatz on the form of the order parameter functions $P(\bomega)$ and $F(\bomega)$.
Since the conditioned overlap distribution (\ref{eq:PM_F}) depends on $F(\bomega)$ only, a first route to proceed is
eliminating the order function $P(\bomega)$ from our equations and making a
replica-symmetric (RS) ansatz for $F(\bomega)$.
Since $\bomega\in[-\pi,\pi]^n$ is continuous, the RS ansatz for $F(\bomega)$ reads:
\begin{eqnarray}
F(\bomega)&=& \int\!\{\rmd\pi\}~W[\{\pi\}]\prod_{\alpha=1}^n \pi(\omega_\alpha),
\label{eq:RS_F}
\end{eqnarray}
where $W[\ldots]$ is a measure over functions, normalised according to
$\int\!\{\rmd\pi\}~W[\{\pi\}]=1$ and nonzero (in view of (\ref{eq:normalization})) only for functions $\pi(\ldots)$ that are real
and obey
$\int_{-\pi}^{\pi}\! \rmd\omega~\pi(\omega) =0$.
The RS ansatz (\ref{eq:RS_F}) is to be inserted into the saddle point equations.
Insertion into (\ref{eq:SP1}) gives, with a normalization factor $C_n(\psi)$,
\bea
P(\bomega)&=&\Big\bra C_n^{-1}(\psi)
\prod_\alpha \Dpsia~ \rme^{c\int \{ d\pi\} \,W[\{\pi\}]\prod_\alpha
\pi(\omega_\alpha)}\Big\ket
\nonumber\\
&=&\Big\bra C_n^{-1}(\psi)
\prod_\alpha \Dpsia \sum_{k\geq 0}\frac{c^k}{k!}
\Big[\int \{\rmd\pi\} \,W[\{\pi\}]\prod_\alpha
\pi(\omega_\alpha)\Big]^k\Big\ket
\nonumber\\
&=&\Big\bra C_n^{-1}(\psi)
\sum_{k\geq 0}\frac{c^k}{k!} \int \prod_{\ell=1}^k \big[\{\rmd\pi_\ell\} W[\{\pi_\ell\}]\big]\prod_\alpha R_k(\omega_\alpha)\Big\ket,
\eea
with
\be
R_k(\omega)=\Dpsi \prod_{\ell=1}^k
\pi_\ell(\omega).
\ee
%
Next we turn to (\ref{eq:SP2}).
We first work out for $\bsigma\in\{-1,1\}^n$ the quantity
\bea
L(\bsigma)&=& \alpha c\int_{-\pi}^{\pi}\! \rmd\bomega~P(\bomega)\cos(
\bomega\cdot\bsigma)
\nonumber
\\
&=& \alpha c\bra C^{-1}_n(\psi)
\sum_{k\geq 0}\frac{c^k}{k!} \int \prod_{\ell=1}^k \big[\{\rmd\pi_\ell\} W[\{\pi_\ell\}]\big]
\nonumber
\\
&&\times
\Big[
\half \prod_\alpha \int_{-\pi}^\pi\! \rmd\omega_\alpha\,R_k(\omega_\alpha) \rme^{\rmi\omega_\alpha \sigma^\alpha}\!+\!
\half \prod_\alpha \int_{-\pi}^\pi\! \rmd\omega_\alpha\,R_k(\omega_\alpha) \rme^{-\rmi\omega_\alpha \sigma^\alpha}
\Big]\ket,
\nonumber\\
\label{eq:L}
\eea
with $\int_{-\pi}^\pi\! \rmd\bomega\, P(\bomega)=1$ requiring $L({\bf 0})=\alpha c$.
For Ising spins one can use the general identity
\bea
\tilde R_k(\sigma)=\int_{-\pi}^\pi\! \rmd\omega\, R_k(\omega)\rme^{\rmi\omega \sigma}
=B(\{R_k\})\rme^{\rmi A(\{R_k\})\sigma},
\eea
where $B$ and $A$ are, respectively, the absolute value and the argument
of the complex function $\tilde R_k$ evaluated at the point $1$,
$\tilde R_k(1)=|\tilde R_k(1)|\,\rme^{\rmi\phi_{\tilde R(1)}}$, i.e.
\be
B(\{R_k\})=|\tilde R_k(1)|, \quad\quad A(\{R_k\})=\phi_{\tilde R(1)}=
\arctan\Big(\frac{{\rm Im}[\tilde R_k(1)]}{{\rm Re}[\tilde R_k(1)]}\Big).
\label{eq:A}
\ee
This simplifies (\ref{eq:L}) to
\bea
L(\bsigma)=\alpha c\bra C^{-1}_n(\psi)
\sum_{k\geq 0}\frac{c^k}{k!} \int \prod_{\ell=1}^k \Big[\{\rmd\pi_\ell\} W(\{\pi_\ell\})\Big]
B^n(\{R_k\}) \cos\Big[A(\{R_k\}) \sum_\alpha
\sigma^\alpha\Big]\ket.
\eea
In order to have $L({\bf 0})=\alpha c$ in the limit $n\to 0$, one must have
$C_0(\psi)=e^c~\forall \psi$.
Inserting $L(\bsigma)$ into (\ref{eq:SP2}) gives
\bea
K_n F(\bomega)=\sum_\bsigma \cos(\bomega \!\cdot\! \bsigma)
\rme^{c\alpha
\bra C_n^{-1}(\psi)
\sum_{k\geq 0}\frac{c^k}{k!} \int \prod_{\ell=1}^k \big[\{\rmd\pi_\ell\} W[\{\pi_\ell\}]\big]
B^n(\{R_k\}) \cos\big[A(\{R_k\}) \sum_\alpha
\sigma^\alpha\big]\ket,
}\nonumber
\\[-4mm]
&&
\\[-11mm]\nonumber
\eea
with
\be
K_n =\sum_\bsigma
\rme^{c\alpha
\bra C^{-1}_n(\psi)
\sum_{k\geq 0}\frac{c^k}{k!} \int \prod_{\ell=1}^k
\big[\{\rmd\pi_\ell\} W[\{\pi_\ell\}]\big]
B^n(\{R_k\}) \cos\big[A(\{R_k\}) \sum_\alpha
\sigma^\alpha\big]\ket}.
\ee
Upon isolating the term $\sum_\alpha \sigma^\alpha$ via $\sum_m \int_{-\pi}^\pi \frac{\rmd\theta}{2\pi}
\rme^{\rm im\theta-\rmi\theta \sum_\alpha \sigma^\alpha}=1$ we obtain
\bea
K_n F(\bomega)&=&\sum_m \int_{-\pi}^\pi\!\frac{\rmd\theta}{2\pi}~\rme^{\rmi m\theta+c\alpha
\bra C^{-1}_n(\psi)
\sum_{k\geq 0}\frac{c^k}{k!} \int \prod_{\ell=1}^k
\big[\{\rmd\pi_\ell\} W[\{\pi_\ell\}]\big]
B^n(\{R_k\})\cos[A(\{R_k\})m]\ket}
\nonumber\\[-2mm]
&&\hspace*{30mm}\times
\sum_\bsigma \rme^{-\rmi\theta \sum_\alpha \sigma^\alpha}
\Big(\frac{1}{2}\rme^{\rmi\sum_\alpha \sigma^\alpha \omega^\alpha}
\!+\!\frac{1}{2}\rme^{-\rmi\sum_\alpha \sigma^\alpha \omega^\alpha}
\Big)
\nonumber\\
&=&2^{n-1}\sum_m \int_{-\pi}^\pi\!\frac{\rmd\theta}{2\pi}\rme^{\rmi m\theta+c\alpha
\bra C^{-1}_n(\psi)
\sum_{k\geq 0}\frac{c^k}{k!} \int \prod_{\ell=1}^k
\big[\{\rmd\pi_\ell\} W[\{\pi_\ell\}]\big]
B^n(\{R_k\})\cos[A(\{R_k\})m]\ket}
\nonumber\\[-2mm]
&&\hspace*{30mm}\times
\Big[
\prod_\alpha \cos(\omega^\alpha\!-\!\theta)+\prod_\alpha
\cos(\omega^\alpha\!+\!\theta)
\Big].
\eea
The two terms inside the
square brackets in the last line yield identical contributions to the $\theta$-integral, so
\bea
\hspace*{-19mm}
K_n F(\bomega)&=&2^n\sum_m \!\int_{-\pi}^\pi\!\frac{\rmd\theta}{2\pi}~\rme^{\rmi m\theta+c\alpha
\bra C^{-1}_n(\psi)
\sum_{k\geq 0}\frac{c^k}{k!} \int \prod_{\ell=1}^k
\big[\{d\pi_\ell\} W[\{\pi_\ell\}]\big]
B^n(\{R_k\})\cos[A(\{R_k\})m]\ket}\prod_\alpha \cos(\omega^\alpha\!\!-\!\theta),
\nonumber
\\[-3mm]
\hspace*{-15mm}&&
\eea
with $K_0$ simply following from the demand  $F(\bomega={\bf 0})=1$, as required by (\ref{eq:SP2}).
Next we insert
\be
1=\int\! \{\rmd\pi\}\, \prod_\omega \delta\Big[
\pi(\theta)-\cos(\omega\!-\!\theta)\Big],
\ee
where we have used the symbolic notation
$\prod_\omega \delta[\pi(\omega)-f(\omega)]$
for the functional version of the $\delta$-distribution, as defined by the
identity $\int \{\rmd\pi\}\, G[\{\pi\}] \prod_\omega \delta[\pi(\omega)-f(\omega)]
=G[\{f\}]$. This leads us to
\bea
K_n F(\bomega)&=&2^n
\sum_m \int_{-\pi}^\pi\!\frac{\rmd\theta}{2\pi}\rme^{\rmi m\theta+c\alpha
\bra C_n^{-1}(\psi)
\sum_{k\geq 0}\frac{c^k}{k!} \int \prod_{\ell=1}^k
\big[\{\rmd\pi_\ell\} W[\{\pi_\ell\}]\big]
B^n(\{R_k\})\cos[A(\{R_k\})m]\ket}
\nonumber\\[-2mm]
&&\hspace*{30mm}\times \int\! \{\rmd\pi\}\,
\prod_\omega \delta\Big[
\pi(\theta)\!-\!\cos(\omega\!-\!\theta)
\Big]\prod_\alpha \pi(\omega^\alpha).
\eea
Substituting (\ref{eq:RS_F}) for $F(\bomega)$ in the left-hand side of this last equation shows that
in the replica limit $n\to 0$, our RS ansatz indeed generates a saddle point if
\bea
W[\{\pi\}]&=&
\int_{-\pi}^\pi\!\frac{\rmd\theta}{2\pi}~
\lambda(\theta|W)
\prod_\omega \delta\Big[\pi(\omega)-\cos(\omega\!-\!\theta)\Big],
\label{eq:W}
\eea
with the short-hand
\be
\lambda(\theta|W)=K_0^{-1}\sum_{m\in\Zset}\rme^{\rmi m\theta+c\alpha
\sum_{k\geq 0}\frac{c^k e^{-c}}{k!}
\bra
\int \prod_{\ell=1}^k
\big[\{\rmd\pi_\ell\} W[\{\pi_\ell\}]\big]
\cos[A(\{R_k\})m]\ket}.
\ee
The constant $K_0$ follows simply from normalisation, which now takes the form $\int_{-\pi}^\pi\!\frac{\rmd\theta}{2\pi}~
\lambda(\theta|W)=1$, giving
\bea
K_0&=&\int\!\frac{\rmd\theta}{2\pi} \sum_{m\in\Zset}\rme^{\rmi m\theta+c\alpha
\sum_{k\geq 0}\frac{c^k e^{-c}}{k!} \bra\int \prod_{\ell=1}^k
\big[\{\rmd\pi_\ell\} W[\{\pi_\ell\}]\big]
\cos[A(\{R_k\})m]\ket}
\nonumber\\
&=&\sum_{m\in \Zset}\delta_{m,0}~\rme^{c\alpha
\sum_{k\geq 0}\frac{c^k e^{-c}}{k!} \bra \int \prod_{\ell=1}^k
\big[\{\rmd\pi_\ell\} W[\{\pi_\ell\}]\big]
\cos[A(\{R_k\})m]\ket} =\rme^{c\alpha}.
\eea
We then arrive at
\be
\lambda(\theta|W)=\sum_{m\in\Zset} \rme^{\rmi m\theta+c\alpha
\sum_{k\geq 0}\frac{c^k e^{-c}}{k!} \bra \int \prod_{\ell=1}^k
\big[\{\rmd\pi_\ell\} W[\{\pi_\ell\}]\big]
\big[\cos[A(\{R_k\})m]-1\big]\ket}.
\label{eq:lambda}
\ee
It is convenient to write $D(\omega|\beta)=D^\prime(\omega|\beta)\!+\!
\rmi D^{\pprime}(\omega|\beta)$,
with $D^\prime(\omega|\beta)={\rm Re}[D(\omega|\beta)]$ and
$D^\pprime(\omega|\beta)={\rm Im}[D(\omega|\beta)]$. Similarly, we write
$R_k(\omega)=R_k^\prime(\omega)\!+\!\rmi R_k^\pprime(\omega)$.
We note that for $\chi(M,\psi)=M^2/2c\!+\!M\psi$ the function
$D_\psi(\omega|\beta)$ defined in (\ref{eq:D})
has several useful
properties, e.g.
\begin{eqnarray}
&& \forall\omega\in[-\pi,\pi]:~~~~~ D_\psi^\prime(-\omega|\beta)=D_\psi^\prime(\omega|\beta),
\quad\quad
D_\psi^\pprime(-\omega|x)=-D_\psi^\pprime(\omega|x),
\label{eq:symmetry}
\\[2mm]
&& \int_{-\pi}^{\pi}\! \rmd\omega~D_\psi(\omega|\beta)= \sum_{M\in\Zset}
\rme^{\beta\chi(M,\psi)}\int_{-\pi}^{\pi}\! \frac{\rmd\omega}{2\pi}~\rme^{\rmi\omega M}=
\sum_{M\in\Zset}
\rme^{\beta \chi(M,\psi)}\delta_{M,0}=1,
\label{eq:initial_cdt}
\\[1mm]
&&  D_\psi(\omega|0)=
\frac{1}{2\pi}\sum_{M\in\Zset}
\rme^{\rmi\omega M}=\delta(\omega) ~~~{\rm for}~\omega\in[-\pi,\pi].
\end{eqnarray}
From (\ref{eq:A}) we have
\be
A(\{R_k\})=\arctan\Big[\frac{{\rm Im}[\tilde R_k(1)]}{{\rm Re}[\tilde R_k(1)]}\Big]=
\arctan\Big[\frac{\int_{-\pi}^\pi\! \rmd\omega\, [R_k'(\omega) \sin\omega
+R_k''(\omega) \cos\omega]
}{\int_{-\pi}^\pi\! \rmd\omega\, [R_k'(\omega) \cos \omega-R_k''(\omega) \sin\omega]}
\Big],
\ee
and insertion in (\ref{eq:lambda}) gives
\be
\lambda(\theta|W)=\sum_{m\in\Zset} \rme^{\rmi m\theta+c\alpha
\sum_{k\geq 0}\frac{c^k \rme^{-c}}{k!} \int \prod_{\ell=1}^k \big[\{\rmd\pi_\ell\}
W[\{\pi_\ell\}]\big] \big\{\cos[m \arctan f_k(\{\pi_1,\ldots,\pi_k\})]
-1\big\}},
\label{eq:hlambda}
\ee
with\vspace*{-3mm}
\be
f_k(\{\pi_1,\ldots,\pi_k\})=
\frac{\int_{-\pi}^\pi\! \rmd\omega\, [D'(\omega|\beta)\sin\omega + D''(\omega|\beta)\cos\omega]
\prod_{\ell=1}^k \pi_\ell(\omega)}
{\int_{-\pi}^\pi\! \rmd\omega\, [ D'(\omega|\beta)\cos\omega- D''(\omega|\beta)\sin\omega]
\prod_{\ell=1}^k \pi_\ell(\omega)}.
\label{eq:fk}
\ee
For high temperatures $D'(\omega|0)=\delta(\omega)$ and $D''(\omega|0)=0$,
so
$f_k(\{\pi_1,\ldots,\pi_k\})=0$
and
$\lambda(\theta|W)=\delta(\theta)$. Hence
\bea
\beta=0:&& ~~~
W[\{\pi\}]=\prod_\omega \delta\Big[\pi(\omega)-
\cos(\omega)\Big].
\label{eq:para}
\eea
We note that for any symmetric set of functions $\{\pi_1,\ldots,\pi_k\}$
one has, from (\ref{eq:fk}), $f_k(\{\pi_1,\ldots,\pi_k\})=0$
due to the symmetry properties (\ref{eq:symmetry}) of $D_\psi$,
and thus $\lambda(\theta|W)=\delta(\theta)$. Hence, (\ref{eq:para}) is a solution of (\ref{eq:W}) for {\em all} temperatures, and
the only solution at infinite temperature.

\subsection{Conditioned distribution of overlaps}

In order to give a physical interpretation to the RS
solution (\ref{eq:RS_F},\ref{eq:para}), we consider the conditioned overlap distribution  (\ref{eq:PM_F}).
Insertion of (\ref{eq:para}) into (\ref{eq:RS_F}) gives
$$
F(\bomega)=\int \!\{\rmd\pi\}~W[\{\pi\}] \prod_\alpha \pi(\omega_\alpha)=\prod_\alpha
\cos(\omega_\alpha),
$$
and subsequent insertion into (\ref{eq:PM_F}) leads to, with
$C_n$ and $\tilde{C}_n$ representing normalization constants,
\bea
P(M|\psi)&=&\lim_{n\to 0} C_n^{-1}
\sum_\bM \Big(\frac{1}{n}\sum_{\gamma=1}^n \delta_{M,M_\gamma}\Big)
\int_{-\pi}^\pi\! \rmd\bomega~\rme^{\rmi\bomega\cdot \bM+\beta \sum_\alpha \chi(M_\alpha,\psi)}
\sum_{k\geq 0}\frac{c^k}{k!}\prod_\alpha \cos^k(\omega_\alpha)
\nonumber\\
&=&
\lim_{n\to 0} \frac{\tilde{C}_n^{-1}}{n}\sum_{k\geq 0}\frac{c^k}{k!}\int_{-\pi}^\pi\!\rmd\bomega\prod_\alpha
\cos^k(\omega_\alpha)
\int_{-\pi}^\pi\!\rmd\lambda~\rme^{\rmi\lambda M}\sum_{\gamma=1}^n \sum_{M_\gamma\in\Zset}\!
\rme^{\rmi(\omega_\gamma-\lambda) M_\gamma+\chi(M_\gamma,\psi)}
\nonumber\\
&&\hspace*{40mm}\times
\prod_{\alpha\neq \gamma}\sum_{M_\alpha}\rme^{\rmi \omega_\alpha M_\alpha
+\chi(M^\alpha,\psi)}
\nonumber\\
&=&
\lim_{n\to 0} \frac{C_n^{-1}}{n}\sum_{k\geq 0}\frac{c^k}{k!}
\int_{-\pi}^\pi\! \rmd\lambda~\rme^{\rmi\lambda M}\sum_{\gamma=1}^n
\int_{-\pi}^\pi\! \rmd\omega_\gamma~ \cos^k(\omega_\gamma)D_\psi(\omega_\gamma\!-\!\lambda|\beta)
\nonumber
\\
&&\hspace*{40mm}\times
\prod_{\alpha\neq \gamma}\int_{-\pi}^\pi\! \rmd\omega_\alpha~\cos^k(\omega_\alpha)
D_\psi(\omega_\alpha|\beta)
\nonumber\\
&=&\lim_{n\to 0} C_n^{-1}\sum_{k\geq 0}\frac{c^k}{k!}
\int_{-\pi}^\pi\! \rmd\lambda~\rme^{\rmi\lambda M}I_k(\lambda,\beta)
I_k^{n-1}(0,\beta),
\label{eq:PM}
\eea
with
\bea
\hspace*{-5mm}
I_k(\lambda,\beta)&=&\int_{-\pi}^\pi\! \rmd\omega~\cos^k(\omega)D_\psi(\omega\!-\!\lambda|\beta)
= \frac{1}{2^k}\sum_{n=0}^k
\Big(\!\!
\begin{array}{l} k\\ n\end{array}
\!\!\Big)
\int_{-\pi}^\pi\! \rmd\omega~\rme^{-\rmi\omega(k-2n)}
\!\sum_{m\in\Zset} \rme^{\rmi(\omega-\lambda) m +\beta\chi(m,\psi)}
\nonumber\\
&=& \frac{1}{2^k}\sum_{n=0}^k \Big(\!\!
\begin{array}{l}
k\\
n
\end{array}
\!\!\Big)
\rme^{-\rmi\lambda (k-2n) +\beta\chi(k-2n,\psi)}
= \frac{1}{2^k}\sum_{m=-k}^k \Big(\!\!
\begin{array}{c}
k\\
\frac{k-m}{2}
\end{array}\!\!
\Big)
\rme^{-\rmi\lambda m +\beta \chi(m,\psi)}.
\eea
We can now work out
\bea
\int_{-\pi}^\pi\! \rmd\lambda~\rme^{\rmi\lambda M}I_k(\lambda,\beta)
=\left\{\begin{array}{ll}
2^{-k}
\Big(\!\!
\begin{array}{c}
k\\
(k\!-\!M)/2
\end{array}
\!\!\Big)
\rme^{\beta\chi(M,\psi)} & {\rm if}~|M|\leq k
\\
0 & {\rm if}~|M|>k
\end{array}\right.,
\eea
and  obtain our desired formula for $P(M|\psi)$ corresponding to the saddle-point (\ref{eq:para}), in which the
normalisation constant comes out as $C_0=\rme^{c}$. The result then is
\bea
P(M|\psi)=
\sum_{k\geq |M|}\rme^{-c}
\frac{c^k}{k!} ~
\frac{
\Big(\!\!
\begin{array}{c}
k\\
(k\!-\!M)/2
\end{array}
\!\!
\Big)
\rme^{\beta \chi(M,\psi)}}{\sum_{m=-k}^k
\Big(\!\!
\begin{array}{c}
k\\
(k\!-\!m)/2
\end{array}\!\!
\Big)
\rme^{\beta\chi(m,\psi)}}.
\label{eq:PM}
\eea
We can rewrite this result, with the short-hand $p_c(k)=\rme^{-c}c^k/k!$, in the more intuitive form
\bea
P(M|\psi)&=&\sum_{k\geq 0} p_c(k)P(M|k,\psi),
\label{eq:conditional_PM}
\\
P(M|k,\psi)&=& \theta(k\!-\!|M|\!+\!\frac{1}{2})~\frac{
\Big(\!\!\begin{array}{c}
k\\
(k\!-\!M)/2
\end{array}
\!\!
\Big)
\rme^{\beta \chi(M,\psi)}}{\sum_{m=-k}^k
\Big(\!\!
\begin{array}{c}
k\\
(k\!-\!m)/2
\end{array}
\!\!
\Big)
\rme^{\beta\chi(m,\psi)}}.
\label{eq:PM_beta}
\eea
We recognise that $p_c(k)$ is the asymptotic probability that any cytokine pattern $(\xi_1^\mu,\ldots,\xi_N^\mu)$ has $k$ non-zero entries; since   each pattern has $N$ independent entries with probability $c/N$ to be nonzero, $k$ will for $N\to\infty$ indeed be a Poissonian random variable with average $c$.
Hence, $P(M|k,\psi)$
is the conditional probability to have an overlap of value $M$,
given the cytokine pattern concerned has $k$ non-zero entries and is triggered by an external field $\psi$.
We have apparently mapped the neural network with $N$ neurons and $N_B=\alpha N$ diluted stored
patterns to a system of $k$ neurons with a single undiluted binary pattern. We will see that
this is due to the fact that in the regime where replica-symmetric theory holds
one is always able, as a consequence of the dilution, to decompose the original
system into an extensive number of independent finite-sized subsystems, each
recalling one particular pattern.

The solution (\ref{eq:para}), leading to (\ref{eq:PM_beta}), is a saddle-point for any temperature. At infinite temperatures it is the only solution, and simplifies further. For $\beta=0$ expression  (\ref{eq:PM_beta})  gives
\bea
P(M|k,\psi)= 2^{-k}
\Big(\!\!\begin{array}{c}
k\\
(k\!-\!M)/2
\end{array}
\!\!
\Big),
\label{eq:Pi_high}
\eea
which is  the
probability that a system of $k$ spins has an overlap $M$ with an undiluted
stored pattern, if each spin behaves completely randomly.
This describes, as expected, an immune network
behaving as a paramagnet, i.e. unable to retrieve stored strategies.
For the distribution of overlaps we find
\be
P(M)=\rme^{-c}\sum_{k\geq 0}\frac{(\frac{1}{2}c)^k}{k!}
\Big(\!\!
\begin{array}{c}
k\\
(k\!-\!M)/2
\end{array}\!\!
\Big).
\label{eq:PM_high}
\ee

In the limit $\beta\to \infty$, the sum in the denominator of
(\ref{eq:PM_beta}) is dominated by the value of $m$ which maximises
$\chi(M,\psi)=m^2/2c \!+\! \psi m$, being $m\!=\!k~{\rm sgn}(\psi)$ if $\psi\neq 0$ and $m\!=\!\pm k$ for $\psi=0$. In either case we obtain
\bea
\sum_{m=-k}^k
\Big(\!\!
\begin{array}{c}
k\\
(k\!-\!m)/2
\end{array}
\!\!
\Big)
\rme^{\beta\chi(m,\psi)}&\sim&
\rme^{\beta(k^2/2c+k|\psi|)}.
\label{eq:denominator}
\eea
Substitution into (\ref{eq:PM_beta}) and (\ref{eq:conditional_PM}) subsequently gives
\bea
\lim_{\beta\to\infty}
P(M|\psi)&=& \lim_{\beta\to\infty} \rme^{-c}\sum_{k\geq |M|} \frac{c^k}{k!}\Big(\!\!\begin{array}{c}
k\\
(k\!-\!M)/2
\end{array}
\!\!
\Big)
\rme^{-\beta(k^2-M^2)/2c-\beta |\psi|(k- {\rm sgn}(\psi)M)}
\nonumber
\\
&=& \left\{
\begin{array}{ll}
\rme^{-c}\sum_{k\geq |M|} \frac{c^k}{k!}\Big(\!\!\begin{array}{c}
k\\
(k\!-\!M)/2
\end{array}
\!\!
\Big)
\rme^{-\beta(k^2-M^2)/2c-\beta |\psi|(k- {\rm sgn}(\psi)M)}
& {\rm if}~\psi\neq 0
\\
\rme^{-c}\sum_{k\geq |M|} \frac{c^k}{k!}\Big(\!\!\begin{array}{c}
k\\
(k\!-\!M)/2
\end{array}
\!\!
\Big)
\rme^{-\beta(k^2-M^2)/2c}
& {\rm if}~\psi=0
\end{array}
\right.
\nonumber
\\
&=& \left\{
\begin{array}{ll}
\theta(M\psi)~\rme^{-c} c^{|M|}/|M|!
& {\rm if}~\psi\neq 0,~M\neq 0
\\
\rme^{-c}
& {\rm if}~\psi\neq 0,~M=0
\\
\rme^{-c} c^{|M|}/|M|!
& {\rm if}~\psi=0
\end{array}.
\right.
\eea
Similarly we have
\bea
\psi\neq 0:&~~~&
P(M|k,\psi)=\delta_{|M|,k}\Big(\delta_{M,0}+\theta(\psi M)(1\!-\!\delta_{M,0})\Big),
\label{eq:PM_low}
\\
\psi=0:&~~~&
P(M|k,\psi)=\delta_{|M|,k}\Big(\delta_{M,0}+\frac{1}{2}(1\!-\!\delta_{M,0})\Big).
\eea
For $k>0$ this describes error-free activation or inhibition
of a stored strategy with $k$ nonzero entries.

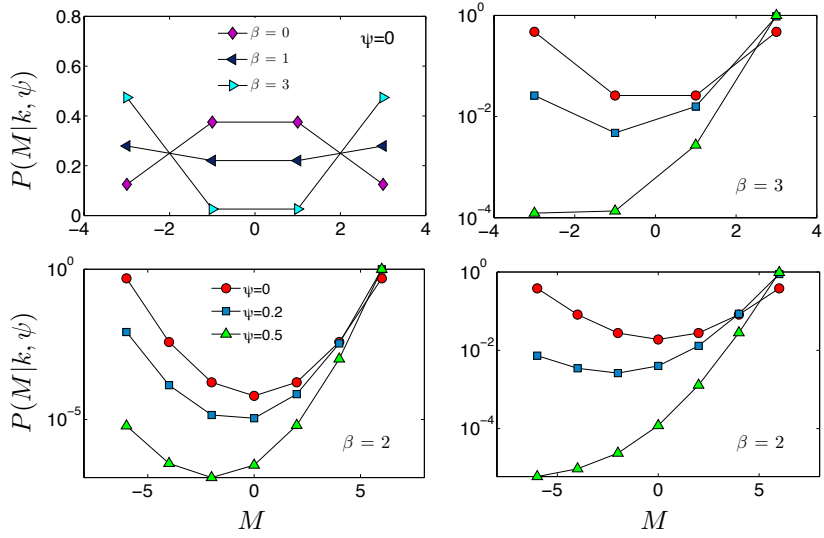
\begin{figure}[t]
\unitlength=0.67mm
 \hspace*{30mm}\begin{picture}(160,100)
\put(0,0){\includegraphics[width=160\unitlength]{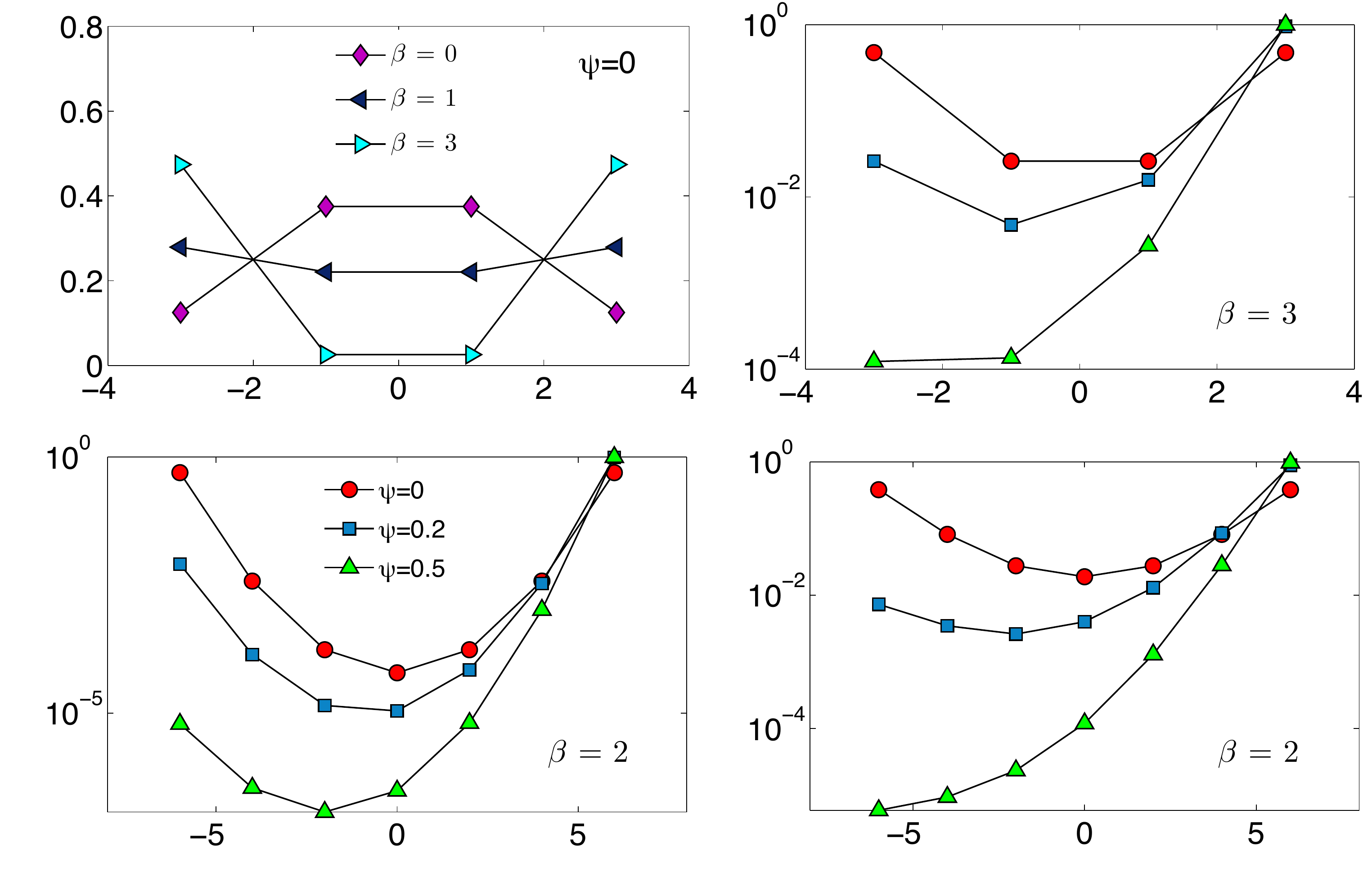}}
\put(43,-2){$M$}
\put(123,-2){$M$}
\put(-2,17){\rotatebox{90}{$P(M|k,\psi)$}}
\put(-2,65){\rotatebox{90}{$P(M|k,\psi)$}}

\end{picture}
\vspace*{2mm}

\caption{Conditioned overlap distribution $P(M|k,\psi)$ corresponding to the state (\ref{eq:RS_F},\ref{eq:para}), as given by formula (\ref{eq:PM_beta}).
Top panels refer to $k=c=3$. Left: $\beta=0, 1, 3$ and $\psi=0$; Right:
$\beta=3$ and $\psi=0, 0.2, 0.5$.
Bottom panels refer to  $\psi=0, 0.2, 0.5$ and $\beta=2$.
Left: $c=3, k=6$. Right: $c=k=6$. Note that $M\in\{-k,-k+1,\ldots,k-1,k\}$, so that the lines connecting markers are only guides to the eye.
}
\label{fig:PMk}
\end{figure}

For intermediate temperatures  a plot of (\ref{eq:PM_beta}) shows that
without  external fields, $P(M|0)$ acquires  two symmetric peaks at large
overlaps (in absolute value), as $\beta$ is increased from $\beta=0$; see Fig. \ref{fig:PMk}, top left panel.
Unlike typical magnetic systems in the thermodynamic limit, there is no spontaneous ergodicity breaking at $\psi=0$;
 the system acts effectively as an extensive number of independent finite subsystems, each devoted to a single B-clone. Each
size-$k$ subsystem oscillates randomly between the the two peaks in $P(M|0)$, with a characteristic switching timescale $t_k\sim e^{\beta k^2/2c}$,
which grows with the size $k$ of the subsystem
and remains finite at finite temperature.

Introducing a field $\psi$
 reduces the overlap peak at $M$ values opposite in sign  to the field; this peak will
eventually disappear for sufficiently strong fields (Fig. \ref{fig:PMk}, top right panel).
The field-induced asymmetry in the height of the two peaks increases
at smaller temperatures and
larger sizes (Fig. \ref{fig:PMk}, bottom panels).
Thus, external fields  trigger the system
towards either activation or inhibition of a strategy
(e.g. clonal expansion versus contraction), whereas in their absence the system
oscillates stochastically between the two.

Beyond the multiple clonal expansions, achieved in the present model through activation signalling
from the T-cells to B-cells via appropriately diluted cytokine patterns, the  apparent emergence of regular
inhibitory signals sent to the B-clones  that are not expanding (in the absence of external fields triggering those clones)
is a biologically fundamental feature for homeostasis.
 B-cells that are not receiving a significant number of signals undergo a process called `anergy'  \cite{goodnow1, goodnow2}, and will eventually die. Thus, the ability to support fast switching between positive and
negative signals to multiple clones in parallel,
which is achieved in a rather natural way in the present multitasking network,
has further welcome implications.

\subsection{Simplification of the RSB theory}

The approach developed in the previous section
led to transparent formulae for the distribution of
overlaps in the RS state (\ref{eq:para}), and even allows us to derive analytically the condition defining the (continuous) phase transition where (\ref{eq:para}) ceases to hold (see \ref{app:transition}). However, the states
beyond the transition point are better described within an alternative  (but mathematically equivalent)
formulation of the theory.
This alternative approach  is
based on formulating our equations first in terms of the following quantities:
\be
L(\bsigma)=\alpha c\int_{-\pi}^\pi\! \rmd\bomega~ P(\bomega) \cos(\bomega\cdot \bsigma),\quad\quad
Q(\bomega)=\rme^{cF(\bomega)}.
\label{eq:newOPs}
\ee
Both $P(\bomega)$ and $Q(\bomega)$ are only defined for
$\bomega\in[-\pi,\pi]^n$.
In terms of (\ref{eq:newOPs}) we can write our earlier saddle point equations
(\ref{eq:SP1}, \ref{eq:SP2}) as
\begin{eqnarray}
P(\bomega)&=&
\Big\bra\frac{
Q(\bomega)\sum_{\bM\in\Zset^n} \rme^{\rmi\bomega \cdot\bM+\sum_\alpha \chi(M_\alpha,\psi)}}
{\int_{\-\pi}^\pi\! \rmd\bomega^\prime \,Q(\bomega^\prime)\sum_{\bM\in\Zset^n}
\rme^{\rmi\bomega^\prime \cdot\bM+\sum_\alpha \chi(M_\alpha,\psi)}}\Big\ket_\psi,
\label{eq:PinQ}
\\
\log Q(\bomega)&=&
c~\frac{\sum_{\bsigma\in\{-1,1\}^n}\cos(\bomega\cdot\bsigma)\rme^{L(\bsigma)}}
{\sum_{\bsigma\in\{-1,1\}^n}\rme^{L(\bsigma)}},
\end{eqnarray}
and the free energy (\ref{eq:f_F}) as
\bea
f[\chi]
&=&-\lim_{n\to 0}\frac{1}{\beta n}
\Big\{
\log \Big(\sum_{\bsigma}\rme^{L(\bsigma)-c\alpha}\Big)
-\frac{\sum_\bsigma L(\bsigma)\rme^{L(\bsigma)}}{\sum_\bsigma \rme^{L(\bsigma)}}
\nonumber
\\
&&
+\alpha ~\Big\bra\log \Big(\sum_{\bM}\int_{-\pi}^{\pi}\!\frac{\rmd\bomega}{(2\pi)^n}~
\rme^{\rmi\bomega\cdot \bM+\sum_\alpha \beta \chi(M_\alpha,\psi)}Q(\bomega) \Big)\Big\ket_\psi
\Big\},
\label{eq:f_Q}
\eea
where we used $\alpha \int_{-\pi}^\pi\! \rmd\bomega\, P(\bomega) \log Q(\bomega)=\sum_\bsigma L(\bsigma)\rme^{L(\bsigma)}/\sum_\bsigma \rme^{L(\bsigma)}$.
Clearly $\int_{-\pi}^\pi\! \rmd\bomega~P(\bomega)=1$, $Q(\bomega)\in\R$,
$Q(-\bomega)=Q(\bomega)$, and $Q(\bnull)=\rme^c$.
We can now switch from the order parameter $Q(\bomega)$
to a new order parameter $\tilde Q(\bM)$, defined on
$\bM\in\Zset^n$, via the following one-to-one transformations:
\begin{eqnarray}
\tilde Q(\bM)=\int_{-\pi}^\pi\! \frac{\rmd\bomega}{(2\pi)^n}~Q(\bomega)\rme^{\rmi\bomega\cdot\bM},~~~~~~~~Q(\bomega)=\sum_{\bM\in\Zset^n}\tilde Q(\bM)\rme^{-\rmi\bomega\cdot\bM}.
\end{eqnarray}
The validity of these equations follows from the two identities
$(2\pi)^{-1}\int_{-\pi}^\pi\! \rmd\omega~\rme^{\rmi\omega m}=\delta_{m0}$ for $m\in\Zset$, and $(2\pi)^{-1}\sum_{M\in\Zset}\rme^{\rmi\omega M}=\delta(\omega)$ for $\omega\in[-\pi,\pi]$.
By construction we now have $\sum_{\bM}\tilde Q(\bM)=\rme^c$. Moreover,
since $Q(-\bomega)=Q(\bomega)$ we also know that
$\tilde Q(\bM)=(2\pi)^{-n}\int_{-\pi}^\pi\! \rmd\bomega~Q(\bomega)
\cos(\bomega\cdot\bM)\in\R$.
%
One can write the saddle point equations in terms of these order
functions (see \ref{app:L} for details):
\bea
\tilde Q(\bM)&=& \int_{-\pi}^\pi\! \rmd\bomega\, \cos(\bomega \cdot\bM)\exp\Big[c~\frac{\sum_\bsigma \cos(\bomega\cdot \bsigma)\rme^{L(\bsigma)}}{\sum_\bsigma \rme^{L(\bsigma)}}\Big],
\label{eq:Q_saddle}
\\
L(\bsigma)&=&
\alpha c~\rme^{\frac{\beta n}{2c}}\Big\bra\frac{\sum_\bM \tilde Q(\bM)
\rme^{\beta \sum_\alpha \chi(M_\alpha,\psi)} \cosh[\beta
(\frac{1}{c}\bM\cdot\bsigma+\psi \sum_\alpha \sigma^\alpha)]}
{\sum_\bM \tilde Q(\bM)
\rme^{\beta \sum_\alpha \chi(M_\alpha,\psi)} }
\Big\ket_\psi.
\label{eq:L_saddle}
\eea
and the free energy reads
\bea
\hspace*{-2mm}
f[\chi]
&=&-\lim_{n\to 0}\frac{1}{\beta n}
\Big\{
\log\sum_{\bsigma}\rme^{L(\bsigma)-c\alpha}
-\frac{\sum_\bsigma L(\bsigma)\rme^{L(\bsigma)}}{\sum_\bsigma \rme^{L(\bsigma)}}
+\alpha \Big\bra \!\log \Big[\sum_{\bM}
\rme^{\sum_\alpha \beta \chi(M_\alpha,\psi)}\tilde Q(\bM) \Big]\Big\ket_\psi
\Big\}.
\label{eq:f_tilde}
\nonumber\\
\eea
From (\ref{eq:PM_F}) we find that the distribution of overlaps can be written as
\bea
P(M|\psi)=\lim_{n\to 0}
\frac{\sum_\bM \Big(\frac{1}{n}\sum_{\gamma=1}^n \delta_{M,M_\gamma}\Big)
\rme^{\beta\sum_\alpha \chi(M_\alpha,\psi)}\tilde Q(\bM)}{\sum_\bM
\rme^{\beta\sum_\alpha \chi(M_\alpha,\psi)}\tilde Q(\bM)}\Bigg|_{\chi(M,\psi)=M^2/2c+M\psi}.
\label{eq:to_test}
\eea
In \ref{app:tests} we confirm the correctness of (\ref{eq:to_test}) in several special limits.

\subsection{The RS ansatz -- route II}

We now try to construct the RS solution of our new equations
(\ref{eq:L_saddle}, \ref{eq:Q_saddle}),
by applying the RS ansatz to the functions $L(\bsigma)$ and $\tilde Q(\bM)$:
\bea
L(\bsigma)&=&\alpha c \int\! \rmd h\, W(h)\prod_{\alpha=1}^n \rme^{\beta h \sigma^\alpha},
\quad\quad
\tilde Q(\bM)=\rme^c \int\! \{\rmd\pi\} W[\{\pi\}] \prod_\alpha \pi(M^\alpha),
\label{eq:newRSform}
\eea
with  $\int\! \rmd h\, W(h)=1,~W(h)=W(-h)$, and with a (normalised) functional measure $W[\pi]$ that is only non-zero for functions $\pi(M)$ that are themselves normalised according to $\sum_{M\in\Zset} \pi(M)=1$.
This ansatz meets the requirements $L(-\bsigma)=L(\bsigma)$,
$L({\bf 0})=\alpha c$ and $\sum_\bM \tilde Q(\bM)=\rme^c$, and is the most general form of the functions $L(\bsigma)$ and $\tilde Q(\bM)$ that is invariant under all replica permutations.
The advantage of this second formulation of the theory is that
it allows us to  work with a distribution $W(h)$ of effective fields,  instead of functional measures over distributions,
which have easier physical interpretations, and are more easy to solve numerically from self-consistent equations.

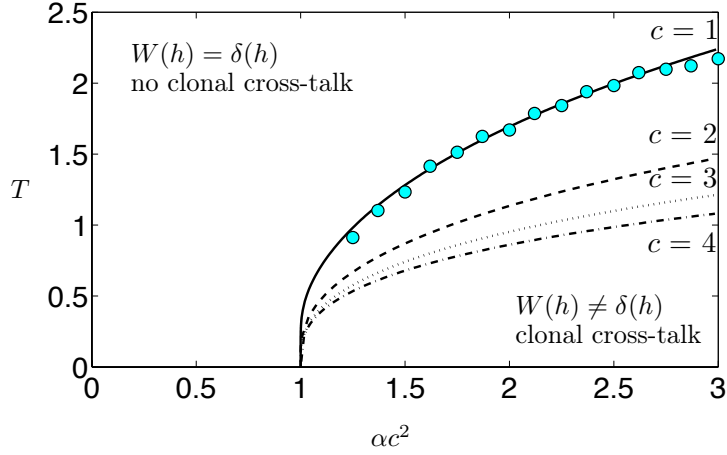
\begin{figure}[t]
\unitlength=0.55mm
\hspace*{26mm}
\begin{picture}(200,98)
\put(2,-5){\includegraphics[width=180\unitlength]{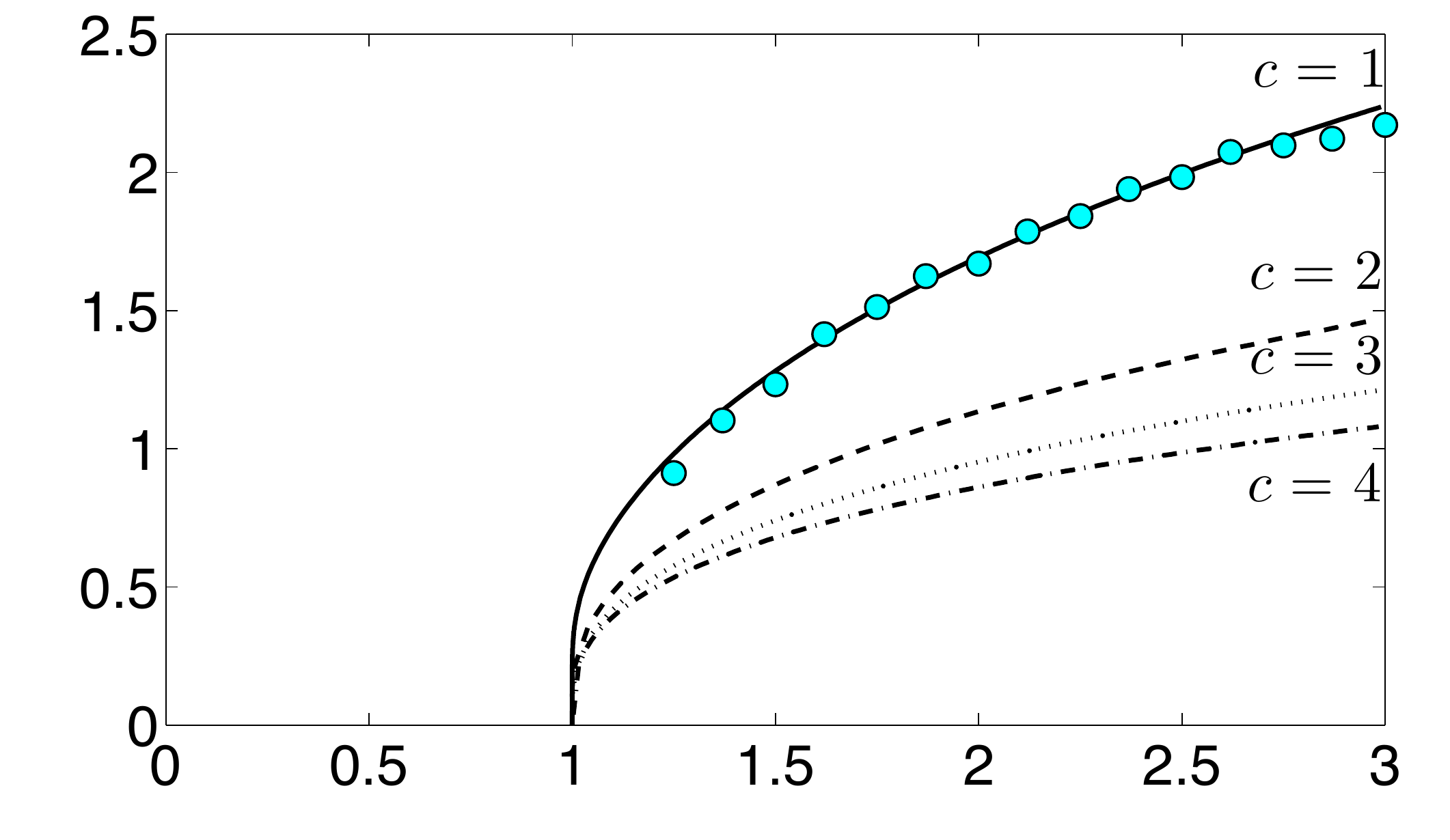}}
\put(90,-10){$\alpha c^2$}
\put(3,50){$T$}
\put(125,22){$W(h)\neq \delta(h)$}
\put(125,15){\rm clonal cross-talk}
\put(32,83){$W(h)=\delta(h)$}
\put(32,76){\rm no clonal cross-talk}
\end{picture}
\vspace*{8mm}
\caption{Transition lines (\ref{eqn:crit}) for $c=1,2,3,4$, in the $(\alpha c^2,T)$ plane, with $T=\beta^{-1}$. The distribution $W(h)$ represents the statistics of clonal cross-talk fields, which are caused by increased connectivity in the graph ${\mathcal G}$.  If $W(h)=\delta(h)$ the clones are controlled via signaling strategies that can act independently; we see that this is possible even above the percolation threshold if the temperature (i.e. the signalling noise) is nonzero. Circles: transition calculated via numerical solution of (\ref{eq:single_eqn}) for $c=1$ (see section \ref{sec:popdyn}).}
\label{fig:Tc}
\end{figure}

We relegate to \ref{app:derivation} all the details of the derivation of the RS equations, based on the form (\ref{eq:newRSform}), the results of which can be  summarised as follows.
The  RS  functional measure $W[\pi]$ and the field distribution $W(h)$ obey the following closed equations:
\begin{eqnarray}
\hspace*{-10mm}
W(h)&=&
\int\!\{\rmd\pi\}~W[\pi] \Big\bra\!\Big\bra\delta\Big[h\!-\tau\psi-\frac{1}{2\beta}\log\Big(\frac{\sum_M \pi(M)\rme^{\beta(M^2/2c+M(\psi+\tau/c)) }}
{\sum_M \pi(M)\rme^{\beta(M^2/2c+M(\psi-\tau/c) )}}\Big)\Big]
\Big\ket_{\!\psi}\Big\ket_{\!\tau=\pm 1},
\\
\hspace*{-10mm}
W[\pi]&=& e^{-c}\sum_{k\geq 0}\frac{c^k}{k!}e^{-\alpha c k}\sum_{r\geq 0}\frac{(\alpha c)^r}{r!}
\int_{-\infty}^\infty\!\rmd h_1\ldots \rmd h_r\Big[\prod_{s\leq r}W(h_s)\Big]\sum_{\ell_1\ldots \ell_r\leq k}
\nonumber
\\
\hspace*{-10mm}
&&
\hspace*{30mm}\times
\prod_M \delta\left[\pi(M)-
\frac{
\big\bra
 \rme^{\beta \sum_{s\leq r} h_s \sigma_{\ell_s}}
\delta_{M,\sum_{\ell\leq k}\sigma_\ell}
\big\ket_{\sigma_1\ldots\sigma_k}}
{
\big\bra
 \rme^{\beta \sum_{s\leq r} h_s \sigma_{\ell_s}}
\big\ket_{\sigma_1\ldots\sigma_k}}
\right]
\label{eq:WpiinWh}.
\end{eqnarray}
Both $W(h)$ and $W[\pi]$ are correctly normalised, $W(h)=W(-h)$, and $W[\pi]$ allows only for functions $\pi$ such that $\pi(M)=\pi(-M)$ and $\sum_M\pi(M)=1$.
We can substitute the second equation into the first and eliminate the functional measure $W[\pi]$, leaving us with a compact RS equation for the field distribution $W(h)$ only:
\begin{eqnarray}
\hspace*{-15mm}
W(h)&=&
\rme^{-c}\sum_{k\geq 0}\frac{c^k}{k!}\rme^{-\alpha c k}\sum_{r\geq 0}\frac{(\alpha c)^r}{r!}
\int_{-\infty}^\infty\!\rmd h_1\ldots \rmd h_r\Big[\prod_{s\leq r}W(h_s)\Big]\sum_{\ell_1\ldots \ell_r\leq k}
\label{eq:single_eqn}
\\
\hspace*{-10mm}
&&\hspace*{-4mm}\times
 \Big\bra\!\Big\bra
\delta\left[h-\tau\psi-\frac{1}{2\beta}\log\left(\frac{
\big\bra
\rme^{\beta(\sum_{\ell\leq k}\!\tau_{\ell})^2/2c+\beta(\sum_{\ell\leq k}\!\tau_{\ell})(\psi +\tau/c)+\beta \sum_{s\leq r} h_s \tau_{\ell_s}}\big\ket_{\tau_1\ldots\tau_k=\pm 1}}
{
\big\bra
\rme^{\beta(\sum_{\ell\leq k}\!\tau_{\ell})^2/2c+\beta(\sum_{\ell\leq k}\!\tau_{\ell})(\psi-\tau/c) +\beta \sum_{s\leq r} h_s \tau_{\ell_s}}\big\ket_{\tau_1\ldots\tau_k=\pm 1}}\right)\right]
\Big\ket_{\!\psi}\Big\ket_{\!\tau=\pm 1}.
\nonumber
\end{eqnarray}
We see that $W(h)=\delta(h)$ is a solution of
(\ref{eq:single_eqn}) for any temperature; one easily confirms that this is in fact the earlier state (\ref{eq:para}), recovered within the alternative formulation of the theory.
If we inspect continuous bifurcations of  new solutions with moments
$m_r=\int\!\rmd h~ h^r W(h)$ different from zero,
we find (see \ref{app:phase_transition}) a second order transition
along the critical
surface in the $(\alpha,\beta,c)$-space defined by
\begin{eqnarray}\label{eqn:crit}
1&=&
\alpha c^2
\sum_{k\geq 0}\rme^{-c}\frac{c^{k}}{k!}
\left\{
\frac{\int\!{\rm D}z~\tanh(z\sqrt{\beta/c}\!+\!\beta/c)\cosh^{k+1}(z\sqrt{\beta/c}\!+\!\beta/c)
}
{\int\! {\rm D}z~\cosh^{k+1}(z\sqrt{\beta/c}\!+\!\beta/c)
}
\right\}^2.
\label{eq:surface}
\end{eqnarray}
\noindent We note that the right-hand side obeys  $0\leq {\rm RHS}\leq \alpha c^2$, with $\lim_{\beta\to 0}{\rm RHS}=0$ and $\lim_{\beta\to \infty}{\rm RHS}=\alpha c^2$.
Hence a transition at finite temperature $T_c(\alpha,c)=\beta_c^{-1}(\alpha,c)>0$ exists to a new state with $W(h)\neq \delta(h)$ as soon as $\alpha c^2>1$. The critical temperature becomes zero when
$\alpha c^2=1$, consistent with the percolation
treshold (\ref{eq:percolation}) derived from the network analysis.
We show in \ref{app:equivalence}
that
the critical surface (\ref{eq:surface}) is indeed identical to the one found
in (\ref{eq:critical_surface}), within the approach involving functional distributions.
\vsp

Finally, within the new formulation of the theory,
the replica-symmetric field-conditioned overlap distribution is found to be
\begin{eqnarray}
P(M|\psi)&=&\lim_{n\to 0}
\frac{
\int\{\rmd\pi\}~W[\pi]~\Big(\sum_{M^\prime}
\pi(M^\prime)
\rme^{\beta (M^{\prime 2}/2c+\psi M^\prime)}\Big)^{\!n-1}\pi(M)
\rme^{\beta(M^{2}/2c+\psi M)}}
{
\int\{\rmd\pi\}~W[\pi]~\Big(\sum_{M^\prime}
\pi(M^\prime)
\rme^{\beta (M^{\prime 2}/2c+\psi M^\prime)}\Big)^{\!n}
}
\nonumber\\
&=&
\int\{\rmd\pi\}W[\pi]\Big\{
\frac{\pi(M)
\rme^{\beta (M^{2}/2c+\psi M)}}{\sum_{M^\prime}
\pi(M^\prime)
\rme^{\beta (M^{\prime 2}/2c+\psi M^\prime)}}
\Big\}.
\eea
Insertion of (\ref{eq:WpiinWh}) allows us to eliminate the functional measure in favour of effective field distributions:
\bea
P(M|\psi)
&=&
\rme^{-c}\sum_{k\geq 0}\frac{c^k}{k!}\rme^{-\alpha c k}\sum_{r\geq 0}\frac{(\alpha c)^r}{r!}
\int_{-\infty}^\infty\!\rmd h_1\ldots \rmd h_r\Big[\prod_{s\leq r}W(h_s)\Big]\sum_{\ell_1\ldots \ell_r\leq k}
\nonumber
\\
&&\times
\left\{
\frac{\big\bra
 \rme^{\beta \sum_{s\leq r} h_s \sigma_{\ell_s}}
\delta_{M,\sum_{\ell\leq k}\sigma_\ell}
\big\ket_{\!\sigma_1\ldots\sigma_k}
\rme^{\beta (M^{2}/2c+\psi M)}}{\sum_{M^\prime}
\big\bra
 \rme^{\beta \sum_{s\leq r} h_s \sigma_{\ell_s}}
\delta_{M^\prime,\sum_{\ell\leq k}\sigma_\ell}
\big\ket_{\!\sigma_1\ldots\sigma_k}
\rme^{\beta (M^{\prime 2}/2c+\psi M^\prime)}}
\right\}
\nonumber\\
&=&
\rme^{-c}\sum_{k\geq 0}\frac{c^k}{k!}\rme^{-\alpha c k}\sum_{r\geq 0}\frac{(\alpha c)^r}{r!}
\int_{-\infty}^\infty\!\rmd h_1\ldots \rmd h_r\Big[\prod_{s\leq r}W(h_s)\Big]\sum_{\ell_1\ldots \ell_r\leq k}
\nonumber
\\
&&\times\left\{
\frac{\big\bra
\delta_{M,\sum_{\ell\leq k}\tau_\ell} ~
\rme^{\beta (\sum_{\ell\leq k}\tau_\ell)^{2}/2c+\beta\psi \sum_{\ell\leq k}\tau_\ell+\beta \sum_{s\leq r} h_s \tau_{\ell_s}}
\big\ket_{\tau_1\ldots\tau_k=\pm 1}
}{
\big\bra
\rme^{\beta (\sum_{\ell\leq k}\tau_\ell)^{2}/2c+\beta\psi \sum_{\ell\leq k}\tau_\ell+\beta \sum_{s\leq r} h_s \tau_{\ell_s}}
\big\ket_{\tau_1\ldots\tau_k=\pm 1}}
\right\}.
\label{eq:PM_h}
\end{eqnarray}
Again, we can rewrite this result (\ref{eq:PM_h}) in the form  (\ref{eq:conditional_PM}), which is more useful to investigate the system's performance since it quantifies the statistics of overlaps relative to their maximum value $k$,  with
\bea
P(M|k,\psi)&=&\rme^{-\alpha c k}\sum_{r\geq 0}\frac{(\alpha c)^r}{r!}
\int_{-\infty}^\infty\!\rmd h_1\ldots \rmd h_r\Big[\prod_{s\leq r}W(h_s)\Big]\sum_{\ell_1\ldots \ell_r\leq k}
\nonumber
\\
&&\times\left\{
\frac{\big\bra
\delta_{M,\sum_{\ell\leq k}\tau_\ell}
\rme^{\beta (\sum_{\ell\leq k}\tau_\ell)^{2}/2c+\beta\psi \sum_{\ell\leq k}\tau_\ell+\beta \sum_{s\leq r} h_s \tau_{\ell_s}}
\big\ket_{\tau_1\ldots\tau_k=\pm 1}
}{
\big\bra
\rme^{\beta (\sum_{\ell\leq k}\tau_\ell)^{2}/2c+\beta\psi \sum_{\ell\leq k}\tau_\ell+\beta \sum_{s\leq r} h_s \tau_{\ell_s}}
\big\ket_{\tau_1\ldots\tau_k=\pm 1}}
\right\}.
\label{eq:PMk_h}
\end{eqnarray}
The latter formula shows very clearly that $h$ is to be interpreted as a clonal interference field, which is caused by overlapping signalling strategies in the bi-partite graph ${\mathcal B}$ and leads to clique interactions in the effective H-H graph ${\cal G}$. Biologically these interference fields can manifest themselves in unwanted clonal  expansions (in the absence if the required antigen), or unwanted clonal reductions (in the presence of the required antigen), due to accidental (frozen) random interactions between clones. Fortunately, we see in Figure \ref{fig:Tc} that even above the percolation threshold $\alpha c^2=1$ the system is able to suppress clonal cross-talk (i.e. have $W(h)=\delta(0)$), provided the noise level is nonzero, and that even in the cross-talk  phase the signalling performance of the system degrades only smoothly (see the section below).

\section{Numerical results: population dynamics and numerical simulations}
\label{sec:popdyn}

\subsection{Population dynamics calculation of the cross-talk field distribution}

We  solve numerically equation (\ref{eq:single_eqn}) for the clonal interference field distribution $W(h)$ with a population dynamics algorithm \cite{MezPar00},
which is based on interpreting  (\ref{eq:single_eqn}) as the fixed-point equation of a stochastic process and simulating this process numerically.
One observes that (\ref{eq:single_eqn})  has the structural form
\begin{eqnarray}
W(h)&=&  \Big\bra\!\Big\bra
\delta\left[h- h(k,r,\bh,\bl,\tau,\psi)\right]
\Big\ket\Big\ket_{\!k,r,\bh,\bl,\tau,\psi},
\label{eq:single_eqn2}
\end{eqnarray}
with the following set of random variables:
\begin{eqnarray*}
	k\sim {\rm Poisson}(c)\\
	r\sim {\rm Poisson}(\alpha ck)\\
	\bh=(h_1,\ldots,h_r):~~r~~{\rm i.i.d. ~random ~fields~with ~probability ~density~ }W(h)\\
	\bl=(\ell_1,\ldots,\ell_r):~~r~~{\rm i.i.d.~ discrete~ random ~variables,~distributed~ uniformly~ over}~\{1,\ldots,k\}\\
	\tau:~~{\rm  dichotomic~ random ~variable,~,~distributed~ uniformly~ over}~\{-1,1\}\\
	\psi:~~{\rm distributed~ according~ to~} P(\psi)
\end{eqnarray*}
and with
\begin{eqnarray}
\hspace*{-10mm}
h(k,r,\bh,\bl,\tau,\psi)&=& \tau\psi+\frac{1}{2\beta}\log\left(\frac{
\big\bra
\rme^{\beta(\sum_{\ell\leq k}\!\tau_{\ell})^2/2c+\beta(\sum_{\ell\leq k}\!\tau_{\ell})(\psi +\tau/c)+\beta \sum_{s\leq r} h_s \tau_{\ell_s}}\big\ket_{\tau_1\ldots\tau_k=\pm 1}}
{
\big\bra
\rme^{\beta(\sum_{\ell\leq k}\!\tau_{\ell})^2/2c+\beta(\sum_{\ell\leq k}\!\tau_{\ell})(\psi-\tau/c) +\beta \sum_{s\leq r} h_s \tau_{\ell_s}}\big\ket_{\tau_1\ldots\tau_k=\pm 1}}
\right).
\end{eqnarray}
We approximate $W(h)$ by the empirical field frequencies computed from a large number
 (i.e. a population)
of fields, which are made to evolve by repeated numerical iteration of a stochastic map.
We start by initialising $S$ fields $h_s\in\R$, with $s=1,\ldots,S$, randomly with uniform probabilities over the interval $[-h_{\rm max},h_{\rm max}]$. Their empirical distribution then represents the zero-step approximation $W_0(h)$ of $W(h)$. We then evolve the fields stochastically via the following Markovian process, giving at each  step $n$ an empirical distribution $W_n(h)$ which as $n$ increases given an increasingly precise approximation of the invariant measure $W(h)$:
\begin{itemize}
\item choose randomly the variables $k,r,\bl,\tau,\psi$ according to their (known) probability distributions
\item choose randomly $r$ fields $\bh=h_1,\ldots,h_r$ from the $S$ fields available, i.e. draw $r$ fields from the probability distribution
$W_{n-1}(h)$ of the previous step
\item compute $h(k,r,\bh,\bl,\tau,\psi)$
\item choose randomly one field from the set of the $M$ available, and set
its value to $h(k,r,\bh,\bl,\tau,\psi)$
\end{itemize}
In all population dynamics calculations in this paper we used populations of size $S=5000$.
\vsp

 We iterate the procedure until convergence, checking every $\mathcal{O}(S^2)$ steps the distance between different $W_n(h)$, and  speed up the computation of $h(k,r,\bh,\bl,\tau,\psi)$ by rewriting it as
\begin{eqnarray}
\hspace*{-10mm}
h(k,r,\bh,\bl,\tau,\psi)&=& \tau\psi+\frac{1}{2\beta}\log\left(\frac{
\int\!{\rm D}z~
\big\bra
\rme^{z\sqrt{\beta/c}\sum_{\ell\leq k}\!\tau_{\ell}+\beta(\sum_{\ell\leq k}\!\tau_{\ell})(\psi +\tau/c)+\beta \sum_{s\leq r} h_s \tau_{\ell_s}}\big\ket_{\tau_1\ldots\tau_k=\pm 1}}
{
\int\!{\rm D}z~
\big\bra
\rme^{z\sqrt{\beta/c}\sum_{\ell\leq k}\!\tau_{\ell}+\beta(\sum_{\ell\leq k}\!\tau_{\ell})(\psi-\tau/c) +\beta \sum_{s\leq r} h_s \tau_{\ell_s}}\big\ket_{\tau_1\ldots\tau_k=\pm 1}}
\right)
\nonumber
\\
&=&
\tau\psi+\frac{1}{2\beta}\log\left(\frac{
\int\!{\rm D}z~
\prod_{\ell\leq k}
\cosh[z\sqrt{\beta/c}+\beta(\psi \!+\!\tau/c)+\beta \sum_{s\leq r} h_s \delta_{\ell \ell_s}]}
{
\int\!{\rm D}z~
\prod_{\ell\leq k}
\cosh[z\sqrt{\beta/c}+\beta(\psi\!-\!\tau/c) +\beta \sum_{s\leq r} h_s \delta_{\ell \ell_s}]}
\right),
\label{eq:h_parameters}
\end{eqnarray}
which requires  Gaussian integration instead of the average over $\{\tau_1,\ldots,\tau_k\}$.
Having computed $W(h)$, we can build $P(M|\psi)$
using equation (\ref{eq:PM_h}). The latter can be rewritten as
\begin{eqnarray}
P(M|\psi)
&=&
\Big\bra\!\Big\bra
\frac{\big\bra
\delta_{M,\sum_{\ell\leq k}\tau_\ell}
\rme^{\beta (\sum_{\ell\leq k}\tau_\ell)^{2}/2c+\beta\psi \sum_{\ell\leq k}\tau_\ell+\beta \sum_{s\leq r} h_s \tau_{\ell_s}}
\big\ket_{\tau_1\ldots \tau_k=\pm 1}
}{
\big\bra
\rme^{\beta (\sum_{\ell\leq k}\tau_\ell)^{2}/2c+\beta\psi \sum_{\ell\leq k}\tau_\ell+\beta \sum_{s\leq r} h_s \tau_{\ell_s}}
\big\ket_{\tau_1\ldots \tau_k=\pm 1}}
\Big\ket\!\Big\ket_{k,r,\bh,\bl,\psi}
\nonumber
\\
&=&
\Big\bra\!\Big\bra
\frac{\big\bra\delta_{M,\sum_{\ell\leq k}\tau_\ell}
\rme^{\beta (\sum_{\ell\leq k}\tau_\ell)^{2}/2c+\beta\psi \sum_{\ell\leq k}\tau_\ell+\beta \sum_{s\leq r} h_s \tau_{\ell_s}}
\big\ket_{\tau_1\ldots\tau_k=\pm 1}}{Z(k,r,\bh,\bl,\psi)}
\Big\ket\!\Big\ket_{k,r,\bh,\bl,\psi},
\label{eqn:pm1}
\end{eqnarray}
with $Z(\ldots)=\int\!{\rm D}z~
\prod_{\ell\leq k}
\cosh[z\sqrt{\frac{\beta}{c}}\!+\!\beta(\psi\!-\!\tau/c) \!+\!\beta \sum_{s\leq r} h_s \delta_{\ell \ell_s}]$ as determined as in (\ref{eq:h_parameters}).
Hence we can carry out the
ensemble average
over the parameters $\{\btau,k,r,\bh,\bl,\psi\}$ in this last expression as
an arithmetic average over a large number $L$ of samples drawn from
their joint distribution, for which in this paper
we choose $L=\mathcal{O}(10^7)$.
The distribution  (\ref{eq:PMk_h}) is handled in the same way, and can be rewritten as
\be
P(M|k,\psi)=
\Big\bra\!\Big\bra
\frac{\big\bra\delta_{M,\sum_{\ell\leq k}\tau_\ell}
\rme^{\beta (\sum_{\ell\leq k}\tau_\ell)^{2}/2c+\beta\psi \sum_{\ell\leq k}\tau_\ell+\beta \sum_{s\leq r} h_s \tau_{\ell_s}}
\big\ket_{\tau_1\ldots\tau_k=\pm 1}}{Z(k,r,\bh,\bl,\psi)}
\Big\ket\!\Big\ket_{r,\bh,\bl,\psi},
\label{eqn:pmcond}
\ee
i.e. upon simply omitting the averaging over $k$.
\vsp

In the interest of  transparancy and an intuitive understanding, it helps to identify the physical meaning of the random variables involved in the above stochastic process.
Given a subsystem of $k$ spins linked to a particular cytokine pattern
(say pattern $\mu=1$, without loss of generality), we may ask
how many other patterns $\mu\neq 1$ interfere with it. This number is the cardinality of the set
\be
R=\left\{\xi^{\mu}_i,~ i\!=\!1,\ldots,N;~ \mu\!=\!2,\ldots,\alpha N\!:~ \xi^{\mu}_i\xi^1_i\neq 0 \right\}.
\ee
With each of the $k$ spins
(labelled by $i$, with $\xi^1_i\neq 0$) correspond $\alpha N\!-\!1$ cytokine variables
$\xi^{\mu}_i$ with $\mu>1$. Hence we have, for a set of $k$ spins,
$k (\alpha N\!-\!1)$ independent possibilities to generate interfering cytokine signals, each nonzero with probability
$c/N$. Thus, for $N\to\infty$
the number of possible interferences is a Poissonian random variable with mean $\alpha c k$, which is recognised to be the variable $r$.
For each value of $r$ we next ask on {\em which} of the $k$ spins each interference acts, i.e.  which are the $r$ indices $i$ such that
$\xi^{\mu}_i\xi^1_i\neq 0$ for some $\mu>1$.
Each $i$ refers to one of the $k$ spins selected,  so we can describe this situation by $r$ random variables $\ell_s$, with $s=1,\ldots,r$, each distributed uniformly in $\{1,\ldots,k\}$, with are recognised as the vector $\bl$.
The parameters $k$, $r$ and $\bl$ considered so far depend only on the (quenched) structure of the B-H network. By conditioning on these random variables we can write
\begin{eqnarray}
P(M|\psi)&=&\sum_{k=0}^{\infty}\rme^{-c}\frac{c^k}{k!}\sum_{r=0}^{\infty}\rme^{-\alpha ck}
\frac{(\alpha c k)^r}{r!}\sum_{\ell_1,\ldots,\ell_r=1}^k k^{-r} P(M|k,r,\bl,\psi)
\nonumber
\\
&=&\Big\bra\!\Big\bra\ \sum_{\bsigma} \delta_{M,\sum_{\ell=1}^k\xi^1_{\ell}\sigma_{\ell}} Z^{-1}(k,r,\bl,\psi)\rme^{-\beta H(\bsigma|k,r,\bl,\psi)}
\ \Big\ket\!\Big\ket_{\!k,r,\bl}.
  \label{eqn:pmeur2}
\end{eqnarray}
Inside the brackets we have the overlap $M$ of a single pattern ($\mu=1$) with $k$ non-null entries, whose correlation with the other patterns is specified uniquely by the parameters $(k,r,\bl)$. We can write the effective Hamiltonian governing this $k$-spin subsystem by
isolating in the Hamiltonian (\ref{eq:H})  $\mu=1$ contribution:
\be
H_{\rm eff}(\bsigma)=-M_1^2(\bsigma)/2c-\psi M_1(\bsigma)- \sum_{i=1}^k \sigma_{i}\sum_{\mu>2}\xi^{\mu}_i (M_{\mu}(\bsigma)/c+\psi_\mu).
\ee
Upon transforming $\tau_{\ell}=\xi^1_{\ell}\sigma_{\ell}$, and defining $h^{\mu}_{\ell}(\btau)=\xi^{1}_{\ell}\xi^{\mu}_{\ell} (M_{\mu}/c +\psi_\mu)$, and using the meaning of the parameters $r$ and $\ell_s$, we arrive at a description involving $r$ non zero fields $h_s(\btau)$,
each acting on a spin $\ell_s$:
\be
H_{\rm eff}(\tau_1,\ldots,\tau_k)=-(\sum_{\ell\leq k} \tau_{\ell})^2/2c-\psi \sum_{\ell\leq k} \tau_{\ell} - \sum_{s\leq r} h_s(\btau) \tau_{\ell_s}.
\ee
If we then regard each field  $h_s(\btau)$ as a independent random field (conditional on $(k,r,\bl)$), with probability distribution $W(h_s)$, we arrive at
\be
P(M)=\Big\bra\!\Big\bra \int\! \rmd\bh ~W(\bh)\Big\bra \delta_{M,\sum_{\ell=1}^k\tau_{\ell}}
\frac{\rme^{\beta(\sum_{\ell\leq k} \tau_{\ell})^2/2c+\beta\psi \sum_{\ell\leq k} \tau_{\ell} +\beta \sum_{s\leq r} h_s \tau_{\ell_s}} }{Z(k,r,\bl,\psi)} \Big\ket_{\!\btau}\ \Big\ket\!\Big\ket_{\!k,r,{\bl}}.
\label{eqn:pmeur}
\ee
This is exactly equation $(\ref{eq:PM_h})$ obtained within the RS ansatz. Hence
 the parameters $\bh$ in (\ref{eq:single_eqn2}) represent the effective fields induced by the cross-talk interference of cytokine patterns.
The only difference between the rigorous RS derivation and the above heuristic one is that in the
former we effectively find
$W(\bh)=\prod_{s\leq r}W(h_s)$, i.e. the random fields are
independent. This may not always be the case: if we recall the definition of the $r$ effective
fields, viz.
$h^{\mu}_{\ell}(\btau)=\xi^{1}_{\ell}\xi^{\mu}_{\ell} (M_{\mu}/c+\psi_\mu)$,
we see that as soon as different patterns have more then one spin in common, their interference fields will not be independent. One therefore expects that the RS equation is no longer exact if the bi-partite  B-H network is not-tree like but contains loops.

\begin{figure}[t]
\unitlength=0.65mm
\hspace*{35mm}
 \begin{picture}(200,80)
\put(0,0){\includegraphics[width=145\unitlength]{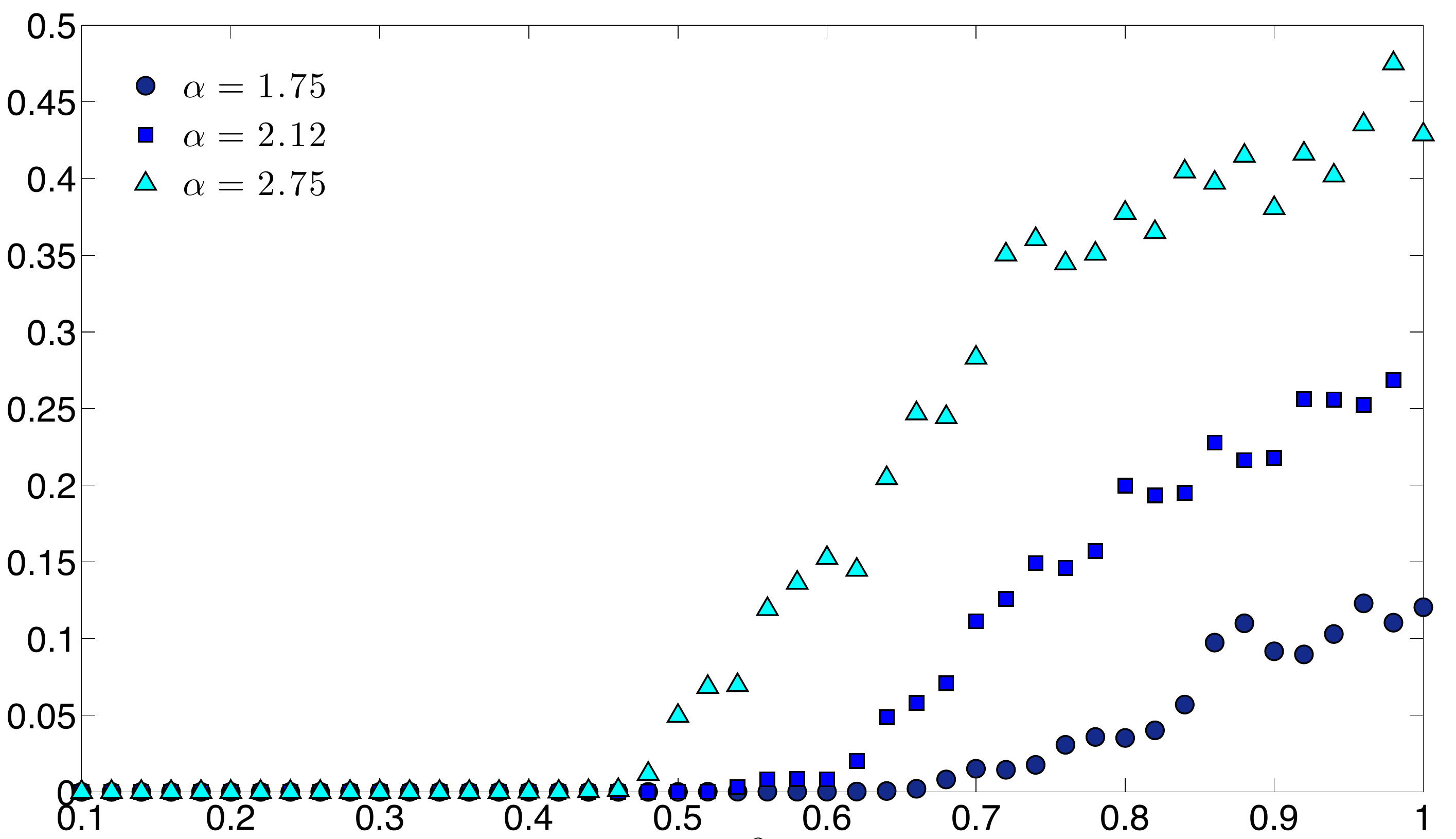}}
\put(70,-8){$\beta$}
\put(-11,45){$m_2$}
\end{picture}
\vspace*{2mm}

\caption{Widths (variances) $m_2=\int\!\rmd h~W(h)h^2$ of the distribution of clonal cross-talk fields, shown as markers versus the inverse temperature $\beta$ for different values of $\alpha$. In all cases $c=1$.
The values of $m_2$ are calculated from the population dynamics solution of (\ref{eq:single_eqn2}), and are (modulo finite size fluctuations in population dynamics algorithm) in excellent agreement
with (\ref{eqn:crit}). The latter predicts that for the $\alpha$-values considered and for $c=1$ the widths $m_2$ should become nonzero at:
$\beta_c=0.6634$ (for $\alpha=1.75$), $\beta_c= 0.5639$ (for $\alpha=2.12$), and $\beta_c=0.4707$ (for $\alpha=2.75$).
}
\label{fig:crit}
\end{figure}

\begin{figure}[t]

\unitlength=0.65mm
\hspace*{35mm}
 \begin{picture}(200,105)
\put(0,0){\includegraphics[width=160\unitlength]{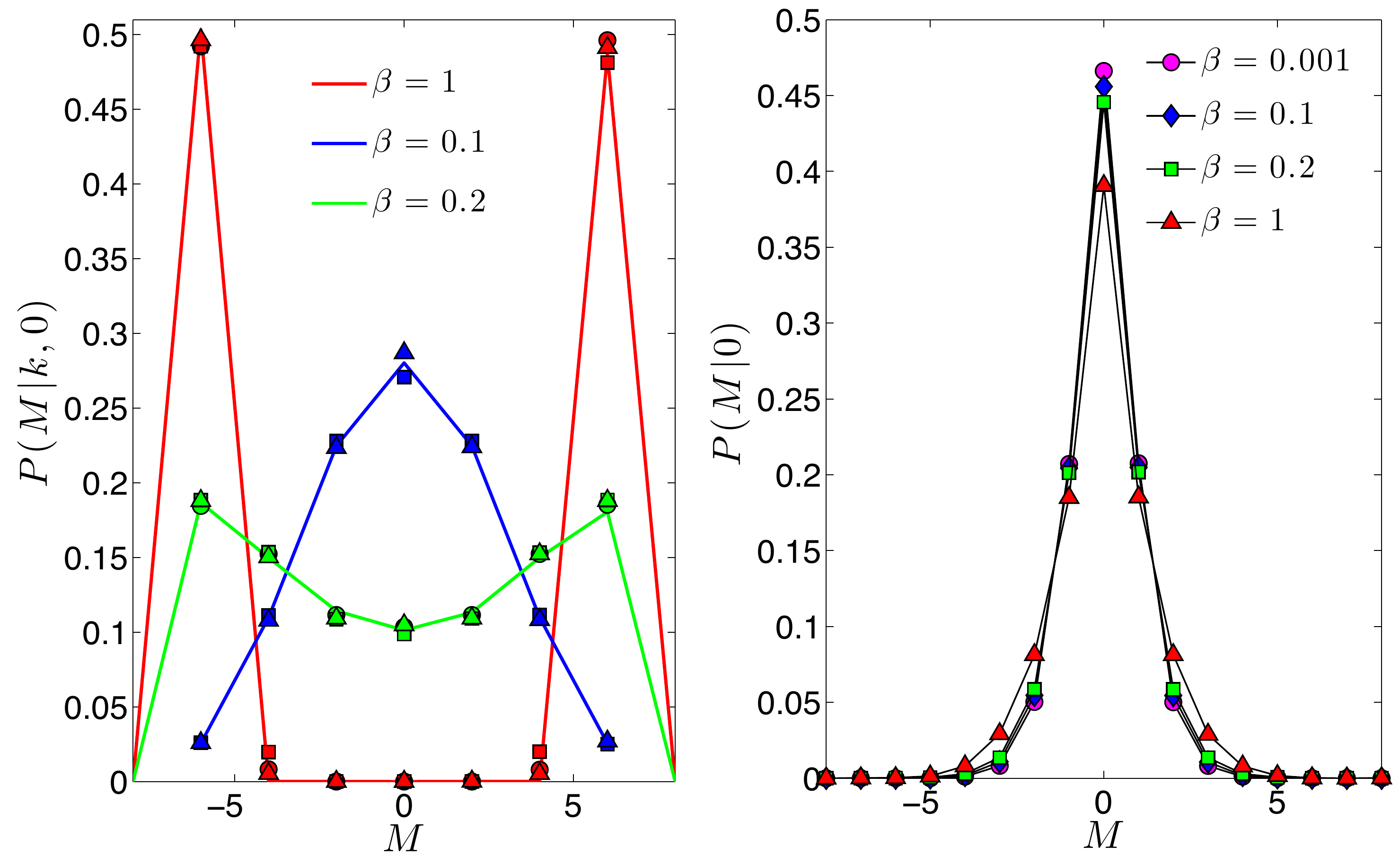}}
\end{picture}
\vspace*{-5mm}

\caption{
Left: degree-conditioned conditioned overlap distribution $P(M|k,0)$ in the under-percolated regime, for $k=6$, $c=1$, and different $\beta$ values (see legend), without external fields.
Solid lines: theoretical predictions. Markers: results of measuring the overlap statistics in  Monte-Carlo simulations of the spin system with Hamiltonian (\ref{eq:hopfield}), with $N=3.10^4$ H-cells.
Different symbols represent different values of $\alpha$, namely $\alpha = 0.005$ (bullets), $\alpha=0.008$ (squares) and $\alpha=0.011$ (triangles). The theory predicts that here $P(M|k,0)$ is independent of $\alpha$, which we find confirmed.
Right panel: overlap distribution $P(M|0)$ at zero field in the under-percolated regime, for $k=6$,  $c=1$ and $\alpha=0.5$,  and different temperatures (see legend). Note that $M\in\Zset$, so line segments are only guides to the eye.
}
\label{fig:under}
\end{figure}

\subsection{Critical line,  overlap distributions, and interference field distribution}

\begin{figure}[t]
\centering
\vspace*{-2mm}
\includegraphics[width=95mm]{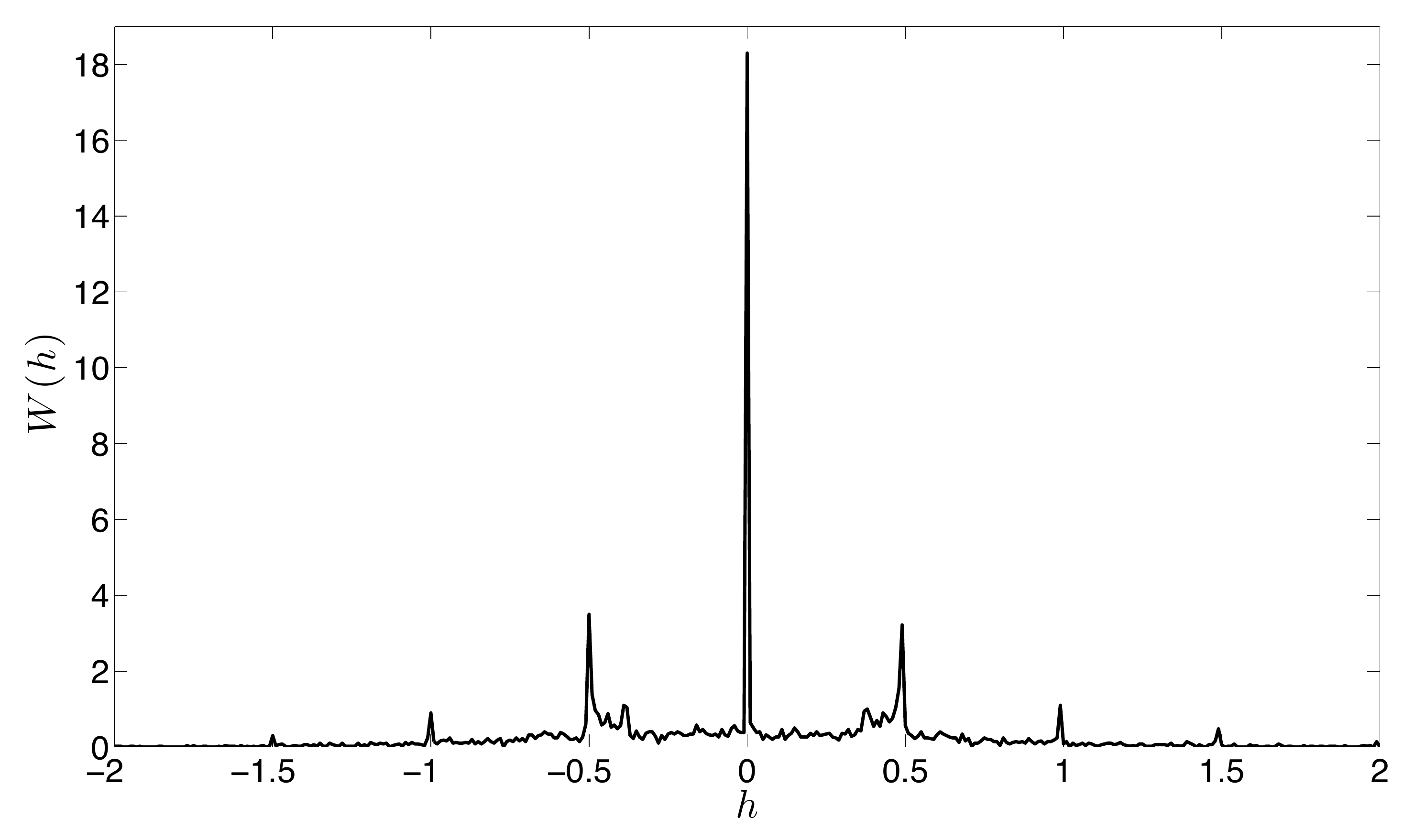}
\caption{The clonal cross-talk interference field distribution $W(h)$ below the critical temperature and in the absence of external fields, as calculated (approximately) via the population dynamics algorithm, for
$c=2$, $\alpha=2$ and
$\beta=6.2$. Note that the support of $W(h)$ is $\Zset/c$. One indeed observes the weight of $W(h)$ being concentrated on these points; due to the finite population size in the algorithm (here $S=5000$) one finds small nonzero values for $h\notin\Zset/c$ due to finite size fluctuations. }
\label{fig:Wh}
\end{figure}

First we use the population dynamics algorithm to validate the location of the
 critical  line (\ref{eqn:crit}). To do so we keep $\alpha$ fixed and compute $W(h)$ for different values of the inverse
temperature $\beta$. From the solution we compute $m_2=\int\!\rmd h~ h^2 W(h)$, and determine for which $\beta$-value  it becomes nonzero (starting from the high temperature phase), i.e. where clonal cross-talk sets in. The result is shown in Figure \ref{fig:crit}, which reveals excellent agreement between the predicted bifurcation temperatures (\ref{eqn:crit}) and those obtained from population dynamics. We also see that there is no evidence for discontinuous transitions. In Figure \ref{fig:Tc} we plotted the bifurcation temperatures obtained via population dynamics versus $\alpha c^2$ (markers), together with the full transition lines predicted by (\ref{eqn:crit}) and again see excellent agreement between the two.

In the under-percolated regime $\alpha c^2<1$, there is no possbility of a phase transition and the only solution of  (\ref{eq:single_eqn}) is
$W(h)=\delta(h)$. Both equations (\ref{eq:PM_h},\ref{eq:PMk_h}) then lose their
dependence on $\alpha$, and after some simple manipulations we recover our earlier results (\ref{eq:conditional_PM},\ref{eq:PM_beta}). In Figure \ref{fig:under} we test our predictions for the overlap statistics against the results of numerical (Monte-Carlo) simulations of the spin process defined by  Hamiltonian (\ref{eq:hopfield}), in the absence of external fields.
There is excellent agreement between theory and numerical experiment.  Comparison of $P(M|k,0)$ to $P(M|0)$ shows that the former
 changes shape as the inverse temperature $\beta$ is increased
from zero,dx
from a single peak at $M=0$ to two symmetric peaks,
showing that the system behaviour at high versus low noise levels is very different. In contrast,
$P(M|0)$ has always a maximum in $M=0$, due to the Poissonian distribution of $k$, and does not capture the two different behaviours. Hence $P(M|k,0)$ is the most useful
quantifier of retrieval behavior, which from now on we will simply denote in the absence of external fields as $P(M|k)$.

\begin{figure}[t]

\unitlength=0.65mm
\hspace*{35mm}
 \begin{picture}(200,102)
\put(0,0){\includegraphics[width=160\unitlength]{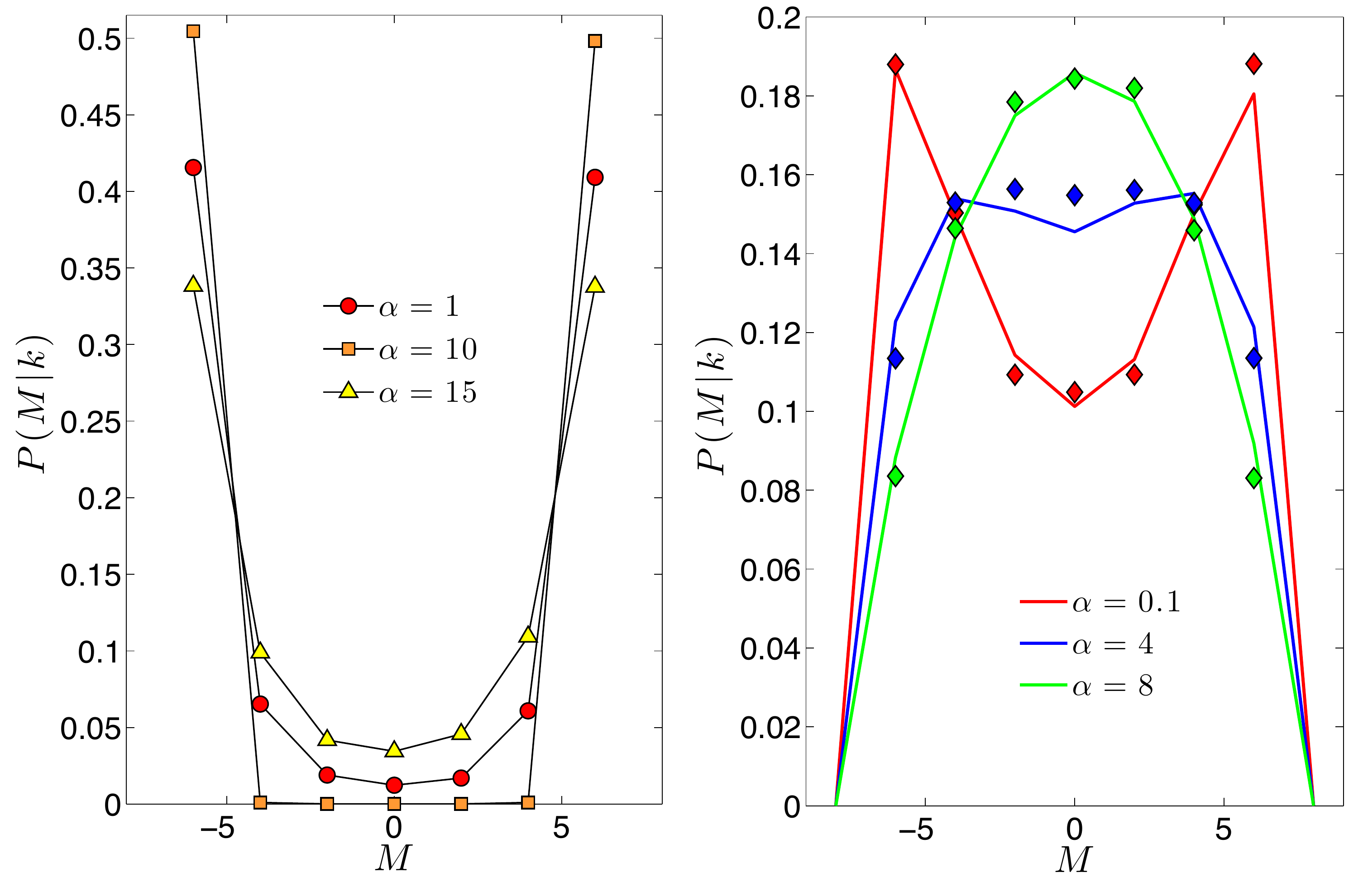}}
\end{picture}
\vspace*{-5mm}

\caption{
Left panel: overlap distribution $P(M|k)$ at zero field in the over-percolated regime, for $k=6$,  $c=1$ and $\beta=0.8$,  and different $\alpha$ values (see legend).
Right: the same distribution, but now  for $k=6$, $c=3$, and different $\alpha$ values (see legend). Note the different vertical axis scales of the two panels.
Solid lines: theoretical predictions, calculated via the population dynamics method. Markers: results of measuring the overlap statistics in  Monte-Carlo simulations of the spin system with Hamiltonian (\ref{eq:hopfield}), with $N=3.10^4$ H-cells.
 The theory predicts that here $P(M|k,0)$ is no longer independent of $\alpha$, which we find confirmed.
 Note that $M\in\Zset$, so line segments are only guides to the eye.
}
\label{fig:over}
\end{figure}

When $\alpha c^2>1$, and below the critical line defined by equation $(\ref{eqn:crit})$, the solution of equation (\ref{eq:single_eqn})
in the absence of external fields will exhibit $W(h)\neq \delta(h)$, see Figure \ref{fig:Wh}.
As a consequence, the effective Boltzmann factor governing the behavior of a set of $k$ spins, linked to a single pattern, acquires a term  $\beta \sum_{s\leq r} h_s \tau_{\ell_s}$ (see equation $(\ref{eqn:pmcond})$). This term means that each subsystem is no longer isolated as in the underpercolated regime, but feels the interference due to the other patterns in the form of effective random fields.
Numerical results for $P(M|k)$ in the overpercolated regime, including comparisons between population dynamics calculations and measurements taken in numerical similations (involving spin systems with $N=3.10^4$ H-cells) are shown in
Figure \ref{fig:over}. Again we observe excellent
agreement. Moreover, we see that in the regime of clonal cross-talk the system's signalling preformance
degrades only gracefully; provided $\alpha$ is not yet too large, the overlap distribution maintains its bimodal form.

\section{Conclusions}

The adaptive immune system consists of a large and diverse ensemble of cells and  chemical messengers, such as antibodies and
cytokines. Helper and suppressor T-lymphocytes (the coordinator branches)
control the activity of B-lymphocytes (the effector branches)
through a rich and continuous exchange of cytokines, which
elicit or suppress effector actions.
From a theoretical point of view, a fascinating feature of the immune system is the ability of T-lymphocytes to manage  multiple B-clones at once, which is vital in defending the host from simultaneous attacks by multiple pathogens.
We investigated this ability in the present study, as an emergent, collective, feature of a spin-glass model, that describes the adaptive response of by B-cells under the coordination of T-cells.

In particular, the focus of this paper is on the ability of the T-cells to coordinate very effectively an {\em extensive} number of B-soldiers, when a suitable degree of dilution in the B-T network is employed.
We assumed that the number $N_B$ of B-cells scales with the number $N_T$ of T-cells as $N_B = \alpha N_T$, with $\alpha>0$, and we modeled the interactions between B-cells and T-cells by means of a finitely connected bi-partite spin-glass, where each B-cell has a likelihood to be connected to a T-cell which scales
as $c/N_T$. This is in agreement with the biological picture of highly-selective touch-interactions among B- and T-cells. The system is thermodynamically equivalent to a diluted monopartite graph ${\mathcal G}$ that describes effective interactions beween T-cells, whose topological properties are shown to depend crucially on the parameters $\alpha$ and $c$. In particular, when $\alpha c^2<1$
the typical components in ${\mathcal G}$
are finite-sized, and
form cliques whose occurrence frequency decays exponentially with their size.
Each clique corresponds to a cytokine signalling pattern, and this kind of arrangement easily allows for the simultaneous recall of multiple patterns. On the other hand, when $\alpha c^2 >1$, the effective network can exhibit a giant component, which can
compromise the system's parallel processing ability.

We have analysed the operation of the system
as an effective equilibrated stochastic process of interacting T-cells,
by using  techniques from the statistical mechanics of finitely connected spin systems.
Within the replica-symmetric (RS) ansatz, we found a critical surface $T_{\rm c}(\alpha,c)$ that separates two distinct phases. For
$T>T_{\rm c}(\alpha,c)$, the system behaves as a extensively large set of independent
neural networks, each of finite size and
each storing a single undiluted
pattern. Here, the only source of
noise are the thermal fluctuations within each subsystem.
For high temperature, each subsystem behaves as a
paramagnet while, at low temperature each subsystem retrieves
one particular pattern (or its inverse), representing parallel
retrieval (perfectly at zero temperature) of an extensive number of finite-size cytokine patterns.
The regulators are able to activate and
inhibit independently the whole B repertoire, as they should.
In particular, each subsystem will oscillate between positive and negative
signalling (with a timescale which increases with its size
and only tends to infinity at zero temperature), because there is only
weak ergodicity breaking.
In the presence of noise (temperature),
no clone can be expanded forever and
there is no expansion without contraction,
unless there is a persistent external stimulus (field) pinning the system
into one particular strategy. This may be a key feature for the homeostatic
regulation of lymphocytes numbers, as cells that are not signaled in a given
time undergo anergy and apoptosis.
The critical temperature becomes zero when $\alpha c^2=1$, i.e.
$T_{\rm c}(\alpha,1/\sqrt{\alpha})=0 ~\forall ~\alpha \geq 0$,
so for $\alpha c^2<1$ no transition at finite temperature
away from this phase is possible.

When the load increases, i.e. when $\alpha$ becomes larger and we cross the transition line, overlaps among bit entries of the `cytokine patterns' to be recalled become more and more frequent, and this gives rise to a source of cross-clonal interference which acts as an effective random field on each node. This represents an additional source of noise for the
system at any finite temperature, and the only source of noise at zero
temperature, and is seen to diminish the parallel processing capabilities. However, the signalling
performance  is found to degrade only
smoothly as one enters further into the clonal cross-talk regime.

Remarkably, the high-temperature phase without clonal cross-talk is the one that gives the
desired emerging behaviour of parallel retrieval, in contrast with
traditional associative networks. This is due to the fact that the
distribution of overlaps, which is the order function of the model,
encodes both the thermal fluctuations of the overlap of the system with each
pattern, {\it and} the fluctuations of the overlap
across different patterns.
Below the percolation threshold, i.e. $\alpha<1/c^2$,
where the system consists of independent
subsystems, each dealing with one pattern,
fluctuations of the overlap across different patterns (i.e. subsystems) are
uncorrelated {\it even} at zero temperature (when all spins are frozen and
ergodicity is broken by each subsystem), so each replica evolves
independently.
Increasing the temperature restores ergodicty in each subsystem, and
the regime of $\alpha$-values without clonal cross-talk gets wider.
From physical arguments and interpretations of our formulae we expect that parallel retrieval
without cross-talk is replica-symmetric,
whereas sequential retrieval (or parallel retrieval in the presence of cross-talk) will not be.
Our predictions and results are tested against numerical simulations  wherever possible, and we consistently find perfect agreement.

Finally, we are tempted to add a last note on the solvability of this model. Despite
the graph ${\mathcal G}$ exhibiting many short loops, which are usually
an obstacle to statistical mechanical techniques, the present spin model on ${\mathcal G}$
is found to be solvable, due to the separable nature of the effective
interaction matrix. This separability allows  us to
unfold the effective network into a bi-partite network ${\mathcal B}$, where loops are
few or absent and factorization over sites can be achieved.
\vsp

\noindent{\bf Acknowledgements}
\\[1mm]
EA, AB and DT acknowledge the FIRB grant RBFR08EKEV and Sapienza Universit\'{a} di Roma for financial support.
ACCC is grateful for support from the Biotechnology and Biological Sciences Research Council (BBSRC) of the United Kingdom. DT would like to thank King's College London for hospitality.

\section*{References}


\appendix

\section{Simple limits}
\label{app:Simple_cases}

Here we work out the theory in some simple limits, which can be worked out independently, to test more complicated stages of our general calculation:
\begin{itemize}
\item The paramagnetic state at $\beta=0$:
\begin{eqnarray}
\lim_{\beta\to 0}\beta\overline{f}&=&   -\lim_{N\to\infty}\lim_{n\to 0}\frac{1}{Nn}\log \sum_{\bsigma^1\ldots\bsigma^n}1=
 -\log 2.
\end{eqnarray}
The conditioned overlap distribution at $\beta=0$ would be
\begin{eqnarray}
P(M|\psi)&=& \frac{1}{P(\psi)}\lim_{N\to\infty}
\frac{1}{\alpha N}\sum_{\mu=1}^{\alpha N}\delta(\psi-\psi_\mu)\int_{-\pi}^\pi\!\frac{\rmd\phi}{2\pi}\rme^{\rmi M\phi}~ 2^{-N}\sum_{\bsigma}\overline{\rme^{-\rmi\phi\sum_i \xi_i^\mu\sigma_i}}
\nonumber
\\
&=&
\lim_{N\to\infty}
\int_{-\pi}^\pi\!\frac{\rmd\phi}{2\pi}\rme^{\rmi M\phi}~\Big(1+\frac{c}{N}[\cos(\phi)\!-\!1]\Big)^N
\nonumber
\\
&=&
\int_{-\pi}^\pi\!\frac{\rmd\phi}{2\pi}\rme^{\rmi M\phi+c[\cos(\phi)-1]}
=  \rme^{-c}\sum_{k\geq 0}\frac{c^k}{k!}\int_{-\pi}^\pi\!\frac{\rmd\phi}{2\pi}\rme^{\rmi M\phi}\bra \rme^{\rmi\phi\sigma}\ket_{\sigma=\pm1}^k
\nonumber
\\
&=&
 \rme^{-c}\sum_{k\geq 0}\frac{c^k}{k!}\big\bra\delta_{M,\sum_{\ell\leq k}\sigma_k} \big\ket_{\sigma_1\ldots\sigma_k=\pm1}.
\end{eqnarray}

\item The case of external fields only:
\\
This simply corresponds to removing the $M_\mu^2$ terms, and gives
\begin{eqnarray}
\hspace*{-10mm}
\overline{f}&=&   -\lim_{N\to\infty}\lim_{n\to 0}\frac{1}{\beta Nn}\log \overline{\prod_{i\alpha}\Big( \sum_{\sigma}\rme^{\beta\sigma\sum_{\mu=1}^{\alpha N}\psi_\mu \xi_i^\mu }\Big)}
\nonumber
\\
\hspace*{-10mm}
&=&
-\frac{1}{\beta}\log 2
 -\lim_{N\to\infty}\lim_{n\to 0}\frac{1}{\beta n}\log~\overline{\cosh^{n\!}\Big(\beta\!\!\sum_{\mu\leq \alpha N}\psi_\mu \xi^\mu\Big) }
\nonumber
\\
\hspace*{-10mm}
&=&-\frac{1}{\beta}\log 2
 -\lim_{N\to\infty}\lim_{n\to 0}\frac{1}{\beta n}\log\int\!\frac{\rmd h\rmd\hat{h}}{2\pi}~\rme^{\rmi\hat{h}h}\cosh^n(\beta h)~\overline{\rme^{-\rmi\hat{h}\sum_{\mu\leq \alpha N}\psi_\mu \xi^\mu}}
\nonumber
\\
\hspace*{-10mm}
&=&-\frac{1}{\beta}\log 2
 -\lim_{N\to\infty}\lim_{n\to 0}\frac{1}{\beta n}\log\int\!\frac{\rmd h\rmd\hat{h}}{2\pi}~\rme^{\rmi\hat{h}h}\cosh^n(\beta h)\prod_{\mu=1}^{\alpha N}\Big(1\!+\!\frac{c}{N}[\cos(\hat{h}\psi_\mu)\!-\!1]\Big)
\nonumber
\\
\hspace*{-10mm}
&=&-\frac{1}{\beta}\log 2
 -\lim_{n\to 0}\frac{1}{\beta n}\log\int\!\frac{\rmd h\rmd\hat{h}}{2\pi}\rme^{\rmi\hat{h}h}\cosh^n(\beta h)\rme^{\alpha c\int\!\rmd\psi~P(\psi)[\cos(\hat{h}\psi)-1]}
\nonumber
\\
\hspace*{-10mm}
&=&-\frac{1}{\beta}\log 2
 -\lim_{n\to 0}\frac{1}{\beta n}\log\int\!\frac{\rmd h\rmd\hat{h}}{2\pi}\rme^{\rmi\hat{h}h+\alpha c\int\!\rmd\psi~P(\psi)[\cos(\hat{h}\psi)-1]}\Big\{
1\!+\!n\log\cosh(\beta h)\!+\!\order(n^2)\Big\}
\nonumber
\\
\hspace*{-10mm}
&=&-\frac{1}{\beta}\log 2
 -\frac{1}{\beta }\int\!dh~W(h)\log\cosh(\beta h),
\end{eqnarray}
with the effective field distribution
\begin{eqnarray}
W(h)&=&\int\!\frac{\rmd\hat{h}}{2\pi}~\rme^{\rmi\hat{h}h+\alpha c\int\!\rmd\psi~P(\psi)[\cos(\hat{h}\psi)-1]}
\nonumber
\\
&=&
\rme^{-\alpha c}\sum_{k\geq 0}\frac{(\alpha c)^k}{k!}
\int\Big[\prod_{\ell\leq k}P(\psi_\ell)\rmd\psi_\ell\Big]
\int\!\frac{\rmd\hat{h}}{2\pi}\rme^{\rmi\hat{h}h}\prod_{\ell\leq k}\cos(\hat{h}\psi_\ell)
\nonumber
\\
&=&
\rme^{-\alpha c}\sum_{k\geq 0}\frac{(\alpha c)^k}{k!}
\int\Big[\prod_{\ell\leq k}P(\psi_\ell)\rmd\psi_\ell\Big]
\Big\bra\int\!\frac{\rmd\hat{h}}{2\pi}\rme^{\rmi\hat{h}(h-\sum_{\ell\leq k}\psi_\ell\sigma_\ell)}\Big\ket_{\!\sigma_1\ldots\sigma_k=\pm 1}
\nonumber
\\
&=&
\sum_{k\geq 0}\rme^{-\alpha c}\frac{(\alpha c)^k}{k!}
\Big\bra\Big\bra\delta\Big[h-\sum_{\ell\leq k}\psi_\ell\sigma_\ell)\Big]\Big\ket_{\!\psi_1\ldots\psi_k}\Big\ket_{\!\sigma_1\ldots\sigma_k=\pm 1}.
\end{eqnarray}
\end{itemize}

\section{Normalization of $F(\bomega)$}
\label{app:normalization}

In this appendix we derive
equation (\ref{eq:normalization}). It follows from
\bea
\hspace*{-20mm}
\int_{-\pi}^{\pi}\! \rmd\bomega\,\cos(\bomega \cdot \bsigma)
&=&\int \!\frac{\rmd m \rmd\hat m}{2\pi}\, \rme^{\rmi m\hat m}\cos(m)
\int_{-\pi}^{\pi}\! \rmd\bomega\,\rme^{-\rmi \hat m \bomega\cdot \bsigma}
=\int \!\frac{\rmd m \rmd\hat m}{2\pi}\, \rme^{\rmi m\hat m}\cos(m)
\Big[
\frac{2c}{\hat m}\sin(\hat m \pi)
\Big]^n
\nonumber\\
&=&\int\! \rmd\hat m\, \frac{\delta(\hat m\!+\!1)+\delta(\hat m\!-\!1)}{2}\Big[
\frac{2c}{\hat m}\sin(\hat m \pi)
\Big]^n=0,
\eea
where we isolated $\bsigma \cdot \bomega$ via
$1=(2\pi)^{-1}\int\! \rmd m\rmd\hat m\,\rme^{\rmi m\hat m-i\hat m \bomega \cdot \bsigma}$ and
used
\bea
\int_{-\pi}^{\pi}\! \rmd\bomega\,\rme^{-\rmi \hat m \bomega\cdot \bsigma}
&=&\prod_{\alpha=1}^n \int_{-\pi}^{\pi}\!\rmd\omega_\alpha \,\rme^{-\rmi \hat m \omega_\alpha \sigma_\alpha}
=\prod_{\alpha=1}^n \Big(2\int_{0}^{\pi}\!\rmd\omega_\alpha \,\cos(\hat m \omega_\alpha \sigma_\alpha)\Big)
\nonumber\\
&=&\prod_{\alpha=1}^n \Big(\frac{2c\sigma^\alpha}{\hat m}\sin(\hat m \pi \sigma^\alpha)\Big)
=\Big[ \frac{2c}{\hat m}\sin(\hat m \pi)\Big]^n.
\eea

\section{Continuous RS phase transitions via route I}
\label{app:transition}

Here we derive the equation for the continuous phase transitions in the absence of external fields, i.e. for $P(\psi)=\delta(\psi)$, away from the solution (\ref{eq:para}).
At the transition, the  the function $D_0(\omega|\beta)$, which we will denote simply as $D(\omega|\beta)$, still satisfies (\ref{eq:symmetry_0}).
Continuous bifurcations away from (\ref{eq:para}) can be identified
via a Guzai (or functional moment) expansion \cite{Guzai}. We transform
\be
\pi(\omega)\to \cos(\omega)
+\Delta(\omega),
\label{eq:variation}
\ee
with $f_k(\{\pi_1,\ldots,\pi_\ell\})\to\tilde f_k(\{\Delta_1,\ldots,\Delta_k\})$,
$W[\{\pi\}]\to \tilde W[\{\Delta\}]$, and
$\tilde W[\{\Delta\}]=0$ as soon as $\int\! \rmd\omega~ \Delta(\omega)\neq 0$ (because $\int\! \rmd\omega\, \pi(\omega)=1$), and $\lambda(\theta|W)\to \tilde\lambda(\theta|\tilde W)$.
We  expand our equations in powers of the functional moments
$\varrho(\omega_1,\ldots,\omega_r)=
\int \{\rmd\Delta\}~\tilde W[\{\Delta\}] \Delta(\omega_1)\ldots \Delta(\omega_r)$.
One assumes that close to the transition there exists some small $\epsilon$
such that $\varrho(\omega_1,\ldots,\omega_r)=\order{(\epsilon^r)}$.
If the lowest bifurcating is of order $\epsilon^1$, we obtain, upon multiplying
(\ref{eq:W}) by $\Delta$ and subsequently integrating over $\Delta$:
\bea
\varrho(\omega)&=&\int\! \{\rmd\Delta\}\, \Delta(\omega)
\int \!\frac{\rmd\theta}{2\pi}~ \tilde\lambda(\theta|\tilde W) \prod_\omega \delta\Big[
\Delta(\omega)+\cos(\omega)-\cos(\omega\!-\!\theta)
\Big]
\nonumber\\
&=&\int \!\frac{\rmd\theta}{2\pi}\, \tilde\lambda(\theta|\tilde W)
[\cos(\omega\!-\!\theta)-\cos(\omega)]=\cos(\omega)\int \!\frac{\rmd\theta}{2\pi}\, \tilde\lambda(\theta|\tilde W)[\cos \theta\!-\!1],
\label{eq:psi}
\eea
where we used the invariance under $\theta\to -\theta$ of
\be
\tilde\lambda(\theta|\tilde W)=\sum_{m\in\Zset} \rme^{\rmi m\theta+c\alpha
\sum_{k\geq 0}\frac{c^k \rme^{-c}}{k!} \int\! \prod_{\ell=1}^k \big[\{\rmd\Delta_\ell\}
\tilde W[\{\Delta_\ell\}]\big] \{\cos[m \arctan \tilde f_k(\{\Delta_1,\ldots,\Delta_k\})]
-1\}}.
\label{eq:tlambda}
\ee
The solution of (\ref{eq:psi}) is clearly $\varrho(\omega)=\phi \cos(\omega)$,
with
\be
\phi=\int \!\frac{\rmd\theta}{2\pi}\, \tilde\lambda(\theta|\tilde W) [\cos(\theta)-1],
\label{eq:phi}
\ee
which we need to evaluate further by expanding $\tilde\lambda(\theta|\tilde W)$  for small $\epsilon$.
Conversely, if the lowest bifurcating order is $\epsilon^2$ one must focus on
\bea
\varrho(\omega_1,\omega_2)&=&
\int\! \{\rmd\Delta\}\, \Delta(\omega_1) \Delta(\omega_2)
\int\! \frac{\rmd\theta}{2\pi} ~\tilde\lambda(\theta|\tilde W) \prod_\omega \delta\Big[
\Delta(\omega)-\cos(\omega)-\cos(\omega\!-\!\theta)
\Big]\nonumber\\
&=&
\cos(\omega_1)\cos(\omega_2)
\int \!\frac{\rmd\theta}{2\pi}\, \tilde\lambda(\theta|\tilde W) [\cos(\theta)\!-\!1]^2
+
\sin(\omega_1)\sin(\omega_2)
\int\! \frac{\rmd\theta}{2\pi}\, \tilde\lambda(\theta|\tilde W) \sin^2(\theta).
\nonumber\\
\label{eq:psi_two}
\eea
We first inspect (\ref{eq:phi}).
Transforming each $\pi_\ell$ in (\ref{eq:fk}) according to (\ref{eq:variation}), we have
\bea
\prod_{\ell=1}^k \pi_\ell(\omega)&=&
\prod_{\ell=1}^k [\cos(\omega)\!+\!\Delta_\ell(\omega)]=\cos^k(\omega)\Big[1\!+\!\sum_{\ell=1}^k \frac{\Delta_\ell(\omega)} {\cos(\omega)}
\Big]+\order{(\Delta^2)}.
\eea
Inserting this result into (\ref{eq:fk}), and using the properties (\ref{eq:symmetry_0}), allows us to expand $\tilde f_k(\{\Delta_1,\ldots,\Delta_k\})$:
\be
\tilde f_k(\{\Delta_1,\ldots,\Delta_k\})=\frac{\sum_{\ell=1}^k \int_{-\pi}^\pi \rmd\omega\,
\sin(\omega)\cos^{k-1}(\omega) \Delta_\ell(\omega) \Db}
{\int_{-\pi}^\pi \rmd\omega\,
\cos^{k+1}(\omega) \Db}+\order(\Delta^2).
\ee
We substitute the above into (\ref{eq:tlambda}) and expand $\cos(m\arctan (x))=1-\frac{1}{2}m^2x^2+\order(x^4)$. Upon introducing
\bea
{\mathcal I}_k &=&\int \!\prod_{\ell=1}^k \big[\{\rmd\Delta_\ell\}
\tilde W[\{\Delta_\ell\}]\big]
\Big[
\sum_{s=1}^k \int_{-\pi}^\pi\! \rmd\omega\,
\sin(\omega)\cos^{k-1}(\omega) \Delta_s(\omega) \Db
\Big]^2,
\label{eq:I}
\\
A_k&=&\int_{-\pi}^{\pi}\! \rmd\omega\, \cos^{k+1}(\omega)\Db,
\eea
we see that ${\mathcal I}_k=\order(\epsilon^2)$, so we  can now expand  $\tilde\lambda(\theta|\tilde W)$ as
\bea
\tilde\lambda(\theta|\tilde W)&=&\sum_{m\in\Zset} \exp\Big[\rmi m\theta-\frac{c\alpha}{2}m^2
\sum_{k\geq 0}\frac{\rme^{-c}c^k}{k!} \frac{{\mathcal I}_k}{A_k^2}+\order(\epsilon^4)\Big]
\nonumber
\\
&=& \sum_{m\in\Zset} \rme^{\rmi m\theta}\Big[1-\frac{c\alpha}{2}m^2
\sum_{k\geq 0}\frac{\rme^{-c}c^k}{k!} \frac{{\mathcal I}_k}{A_k^2}+\order(\epsilon^4)\Big]
\nonumber
\\
&=& 2\pi\delta(\theta)+\pi\alpha c~\delta^\pprime(\theta)
\sum_{k\geq 0}\frac{\rme^{-c}c^k}{k!} \frac{{\mathcal I}_k}{A_k^2}+\order(\epsilon^4).
\label{eq:tlambda_I}
\eea
Next we need to work out the factors ${\mathcal I}_k$.
Using  the functional moment definition
$\varrho(\omega_1,\ldots,\omega_r)=
\int \{\rmd\Delta\}~\tilde W[\{\Delta\}] \Delta(\omega_1)\ldots \Delta(\omega_r)$, one may write
\bea
&& \hspace*{-25mm} \int\! \prod_{\ell=1}^k \big[\{\rmd\Delta_\ell\}
\tilde W[\{\Delta_\ell\}]\big]\sum_{r,s=1}^k \!\Delta_r(\omega^\prime) \Delta_s(\omega^\pprime)\nonumber\\[-1mm]
&&\hspace*{-7mm}=~\sum_r\int\! \{\rmd\Delta_r\} \,\tilde W[\{\Delta_r\}]\Delta_r(\omega^\prime) \Delta_r(\omega^\pprime)+
\sum_{r\neq s}\int \{\rmd\Delta_r\} \{\rmd\Delta_s\} \,\tilde W[\{\Delta_r\}]\tilde W[\{\Delta_s\}] \Delta_r(\omega^\prime) \Delta_s(\omega^\pprime)
\nonumber\\
&&\hspace*{-7mm}=~ k \varrho(\omega^\prime\!,\omega^\pprime)+k(k-1)\varrho(\omega^\prime)
\varrho(\omega^\pprime).
\label{eq:deltas}
\eea
This allows us to work out (\ref{eq:I}) further:
\bea
{\mathcal I}_k&=&k\int_{-\pi}^\pi\! \rmd\omega^\prime\rmd\omega^\pprime\, \sin(\omega^\prime)\cos^{k-1}(\omega^\prime)\Dbp \sin(\omega^\pprime)\cos^{k-1}(\omega^\pprime)\Dbd
\psi(\omega^\prime,\omega^\pprime)
\nonumber\\
&&+ k(k\!-\!1) \Big[\int_{-\pi}^\pi\! \rmd\omega^\prime\, \sin(\omega^\prime)\cos^{k-1}(\omega^\prime)\Dbp \psi(\omega^\prime)\Big]^2
\nonumber\\
&=& k\int_{-\pi}^\pi\! \rmd\omega^\prime\rmd\omega^\pprime\, \Dbp\Dbd \sin(\omega^\prime)\cos^{k-1}(\omega^\prime)
\psi(\omega^\prime\!,\omega^\pprime) \sin(\omega^\pprime)\cos^{k-1}(\omega^\pprime),
\eea
where in the last equality we have used the symmetry of $D(\omega|\beta)$ and
$\varrho(\omega)=\phi \cos(\omega)$.
Inserting this last expression in (\ref{eq:tlambda_I}) and shifting the summation index $k\to k+1$ then leads to
\bea
\hspace*{-10mm}
\tilde\lambda(\theta|\tilde W)&=&2\pi\delta(\theta)+\pi\alpha c^2\delta^\pprime(\theta) S(\{\varrho\})+\order(\epsilon^4),
\label{eq:lambda_psi}
\\
\hspace*{-10mm}
S(\{\varrho\})&=&
\sum_{k\geq 0}\frac{\rme^{-c}c^{k}}{k!} \frac{\int_{-\pi}^\pi\! \rmd\omega^\prime\rmd\omega^\pprime\, \Dbp\Dbd \sin(\omega^\prime)\cos^{k}(\omega^\prime)
\varrho(\omega^\prime\!,\omega^\pprime) \sin(\omega^\pprime)\cos^{k}(\omega^\pprime) }{\Big[\int_{-\pi}^{\pi}\! \rmd\omega\,\Db \cos^{k+2}(\omega)\Big]^2}.
\label{eq:Spsi}
\eea
To make further progress we need to calculate $\varrho(\omega^\prime,\omega^\pprime)$.
We can first simplify (\ref{eq:psi_two}) using (\ref{eq:phi}), giving
\bea
\varrho(\omega_1,\omega_2)&=&\phi^\prime
\sin(\omega_1)\sin(\omega_2)
-(2\phi+\phi^\prime)\cos(\omega_1)\cos(\omega_2),
\label{eq:psi_two_comb}
\eea
where we defined
\be
\phi^\prime=\int_{-\pi}^\pi \!\frac{\rmd\theta}{2\pi} ~\tilde\lambda(\theta|\tilde W)\,
\sin^2(\theta).
\label{eq:phip}
\ee
With this we can simplify (\ref{eq:Spsi}) to
\bea
S(\{\varrho\})&=& \phi^\prime
\sum_{k\geq 0}\frac{\rme^{-c}c^{k}}{k!} \frac{\Big[\int_{-\pi}^\pi\! \rmd\omega\, \Db \sin^2(\omega)\cos^{k}(\omega)
\Big]^2 }{\Big[\int_{-\pi}^{\pi}\! \rmd\omega\,\Db \cos^{k+2}(\omega)\Big]^2}.
\eea
Together with (\ref{eq:lambda_psi}), this allows us to established equations from which to solve the two amplitudes $\phi$ and $\phi^\prime$, by substitution into (\ref{eq:phi}) and (\ref{eq:phip}). This results in, after intergation by parts over $\theta$:
\bea
\phi&=&\frac{1}{2}\alpha c^2S(\{\varrho\}) \int_{-\pi}^\pi \!\rmd\theta~[\cos(\theta)\!-\!1]\delta^\pprime(\theta)+\order(\epsilon^4)
=
-\frac{1}{2}\alpha c^2S(\{\varrho\}) +\order(\epsilon^4)
\nonumber
\\
&=& -\frac{1}{2}\alpha c^2 \phi^\prime
\sum_{k\geq 0}\frac{\rme^{-c}c^{k}}{k!} \frac{\Big[\int_{-\pi}^\pi\! \rmd\omega\, \Db \sin^2(\omega)\cos^{k}(\omega)
\Big]^2 }{\Big[\int_{-\pi}^{\pi}\! \rmd\omega\,\Db \cos^{k+2}(\omega)\Big]^2}+\order(\epsilon^4)
\\
\phi^\prime&=&\frac{1}{2}\alpha c^2S(\{\varrho\})\int_{-\pi}^\pi \!\rmd\theta~
\sin^2(\theta) \delta^\pprime(\theta)+\order(\epsilon^4)
\nonumber
\\
&=&\alpha c^2 \phi^\prime
\sum_{k\geq 0}\frac{\rme^{-c}c^{k}}{k!} \frac{\Big[\int_{-\pi}^\pi\! \rmd\omega\, \Db \sin^2(\omega)\cos^{k}(\omega)
\Big]^2 }{\Big[\int_{-\pi}^{\pi}\! \rmd\omega\,\Db \cos^{k+2}(\omega)\Big]^2}+\order(\epsilon^4).
\eea
Since $\phi^\prime=0$ immediately implies that  $\phi=0$, the only possible continuous bifurcation must be the first instance where $\phi^\prime\neq 0$. According to the above equation this $\order(\epsilon^2)$ bifurcation happens when
\bea
1=\alpha c^2
\sum_{k\geq 0}\frac{\rme^{-c} c^k }{k!}
\left[\frac{\int_{-\pi}^\pi\! \rmd\omega \sin^2(\omega)\cos^k(\omega)\Db}{\int_{-\pi}^\pi\! \rmd\omega \cos^{k+2}(\omega)\Db}\right]^2,
\label{eq:critical_surface}
\eea
with $D(\omega|\beta)=(2\pi)^{-1}\sum_{m\in\Zset}\cos(m\omega)\rme^{\beta m^2/2c}$.
Equation (\ref{eq:critical_surface})  defines the transition point, where
the system will leave the state (\ref{eq:para}).
The right-hand side of (\ref{eq:critical_surface})
obeys $\lim_{\beta\to 0}{\rm RHS}=0$. In \ref{app:phase_transition} we show that
 $\lim_{\beta\to \infty}{\rm RHS}=\alpha c^2$,
so a transition at finite temperature $T_c=\beta^{-1}_c>0$ exists to a new state
with $W[\{\pi\}]\neq \prod_\omega \delta[\pi(\omega)-
\cos(\omega)]$ as soon as $\alpha c^2>1$.
The critical temperature becomes zero when $\alpha c^2=1$.

\section{Saddle point equations in terms of $L(\bsigma)$}
\label{app:L}

Here we derive equation (\ref{eq:L_saddle}), starting from the definition (\ref{eq:newOPs}) and relation
(\ref{eq:PinQ}):
\bea
L(\bsigma)&=&\alpha c~\Bigg\langle\frac{\int_{-\pi}^\pi\! \rmd\bomega \cos(\bomega\cdot \bsigma)
Q(\bomega)\sum_\bM \rme^{\rmi\bomega \cdot\bM+\sum_\alpha \chi(M_\alpha,\psi)}}
{\int_{-\pi}^\pi\! \rmd\bomega \,Q(\bomega)\sum_{\bM}
\rme^{\rmi\bomega\cdot\bM+\sum_\alpha \chi(M_\alpha,\psi)}}\Bigg\rangle_{\!\psi}
\nonumber\\
&=&\alpha c\Bigg\langle\frac{\int\! \rmd\bomega \cos(\bomega\cdot \bsigma)
\sum_{\bM'} \tilde Q(\bM')\sum_\bM \rme^{\rmi\bomega\cdot (\bM-\bM')+\sum_\alpha \chi(M_\alpha,\psi)}}
{\int\! \rmd\bomega\,\sum_{\bM'}\tilde Q(\bM')\sum_{\bM}
\rme^{\rmi\bomega\cdot(\bM-\bM')+\sum_\alpha \chi(M_\alpha,\psi)}}\Bigg\rangle_{\!\psi}.
\label{eq:L1}
\eea
We can then work out the integrals
\bea
\int_{-\pi}^\pi\! \rmd\bomega ~\cos(\bomega\cdot \bsigma)\rme^{\rmi\bomega \cdot(\bM-\bM')}&=&\half \int_{-\pi}^\pi\! \rmd\bomega~ ( \rme^{\rmi\bomega\cdot \bsigma}\!+\!\rme^{\rmi\bomega\cdot \bsigma})\rme^{i\bomega\cdot (\bM-\bM')}
\nonumber\\
&=&\pi(\delta_{\bM',\bM+\bsigma}+\delta_{\bM',\bM-\bsigma}),
\eea
and substituting into (\ref{eq:L1}) gives
\bea
L(\bsigma)&=&\half \alpha c~\Bigg\langle
\frac{\sum_{\bM} \left[\tilde Q(\bM\!+\!\bsigma)\rme^{\beta \sum_\alpha \chi(M_\alpha,\psi)}
+\tilde Q(\bM\!-\!\bsigma)\rme^{\beta \sum_\alpha \chi(M_\alpha,\psi)}
\right]}
{\sum_{\bM}\tilde Q(\bM)
\rme^{\sum_\alpha \chi(M_\alpha,\psi)}}\Bigg\rangle_{\!\psi}
\nonumber\\
&=&\half \alpha c~\Bigg\langle
\frac{\sum_{\bM} \tilde Q(\bM)\left[\rme^{\beta \sum_\alpha \chi(M_\alpha-\sigma^\alpha,\psi)}
+\rme^{\beta \sum_\alpha \chi(M_\alpha+\sigma^\alpha,\psi)}
\right]}
{\sum_{\bM}\tilde Q(\bM)
\rme^{\sum_\alpha \chi(M_\alpha,\psi)}}\Bigg\rangle_{\!\psi}
\nonumber\\
&=&c\alpha~\Bigg\langle\rme^{\frac{\beta n}{2c}}~\frac{\sum_\bM \tilde Q(\bM)
\rme^{\beta \sum_\alpha \chi(M_\alpha,\psi)} \cosh[\beta
(\frac{1}{c}\bM\cdot\bsigma+\psi \sum_\alpha \sigma^\alpha)]}
{\sum_\bM \tilde Q(\bM)
\rme^{\beta \sum_\alpha \chi(M_\alpha,\psi)} }\Bigg\rangle_{\!\psi}.
\label{eq:L2}
\eea

\section{Simple limits to test the replica  theory}
\label{app:tests}

Here we inspect several simple limits to test our results for the overlap distribution and the free energy.
\begin{itemize}
\item
Infinite temperature:\\[2mm]
Using $\lim_{\beta\to 0}L(\bsigma)=\alpha c$ and $\sum_\bM \tilde Q(\bM)=\rme^c$
in (\ref{eq:f_tilde})
we immediately find the correct free energy
\begin{eqnarray}
\lim_{\beta\to 0}\beta\overline{f}_{\rm RSB}=
-\lim_{n\to 0}\frac{1}{n}\Big\{
\log\sum_{\bsigma}1
\Big\}=-\log 2.
\end{eqnarray}
Moreover, from (\ref{eq:Q_saddle})  we can extract
\begin{eqnarray}
\lim_{\beta\to 0}\tilde Q(\bM)&=&
\int_{-\pi}^{\pi}\!\frac{\rmd\bomega}{(2\pi)^n}~\cos(\bomega\cdot\bM)~\rme^{
c ~2^{-n}\sum_{\bsigma}\cos(\bomega\cdot\bsigma)}
\nonumber
\\
&=& \sum_{k\geq 0}\frac{c^k}{k!}2^{-nk}\!\!\sum_{\bsigma^1\ldots\bsigma^k}
\int_{-\pi}^{\pi}\!\frac{\rmd\bomega}{(2\pi)^n}~\cos(\bomega\cdot\bM)~\prod_{\ell\leq k}\cos(\bomega\cdot\bsigma^\ell)
\nonumber
\\
&
=& \sum_{k\geq 0}\frac{c^k}{k!}~2^{-nk}\!\!\sum_{\bsigma^1\ldots\bsigma^k}\!
\delta_{\bM,\sum_{\ell\leq k}\bsigma^\ell}
= \sum_{k\geq 0}\frac{c^k}{k!} \prod_{\alpha=1}^n \bra
\delta_{M_\alpha,\sum_{\ell\leq k}\sigma_\ell}\ket_{\sigma_1\ldots\sigma_k=\pm 1}.
\end{eqnarray}
Hence, it now follows from (\ref{eq:to_test}) that
\begin{eqnarray}
\lim_{\beta\to 0}P(M|\psi)&=&\lim_{n\to 0}
\frac{1}{n}\sum_{\gamma=1}^n
\frac{
\sum_{\bM\in
\Zset^n} \sum_{k\geq 0}\frac{c^k}{k!} \prod_{\alpha=1}^n \bra
\delta_{M_\alpha,\sum_{\ell\leq k}\sigma_\ell}\ket_{\sigma_1\ldots\sigma_k=\pm 1}~\delta_{M,M_\gamma}
}
{
\sum_{\bM\in
\Zset^n} \sum_{k\geq 0}\frac{c^k}{k!} \prod_{\alpha=1}^n \bra
\delta_{M_\alpha,\sum_{\ell\leq k}\sigma_\ell}\ket_{\sigma_1\ldots\sigma_k=\pm 1}~
}
\nonumber
\\
&=&
 \rme^{-c}\sum_{k\geq 0}\frac{c^k}{k!}
 \bra
\delta_{M,\sum_{\ell\leq k}\sigma_\ell}\ket_{\sigma_1\ldots\sigma_k=\pm 1}.
\end{eqnarray}
This coincide with our RS expression,
as it should since at high temperature the RS ansatz is exact.
\vsp

\item External fields only:\\[2mm]
In the case of having only external fields we simply remove all terms that come from the interaction energy in (\ref{eq:L2}), obtaining
\bea
L(\bsigma)
&=&
\alpha c ~\bra
\cosh\big[\beta \psi \sum_\alpha\sigma_\alpha\big]
\ket_\psi.
\eea
Inserting this into (\ref{eq:Q_saddle}),
and introducing the normalised measure
\begin{eqnarray}
\lambda(\bsigma)&=& \frac{\rme^{\alpha c ~\bra
\cosh[\beta \psi \sum_\alpha\sigma_\alpha ]
\ket_\psi}}
{\sum_{\bsigma^\prime}\rme^{\alpha c ~\bra
\cosh[\beta \psi \sum_\alpha\sigma_\alpha^\prime ]
\ket_\psi}},
\end{eqnarray}
we get
\bea
\tilde Q(\bM)&=&
\int_{-\pi}^{\pi}\!\frac{\rmd\bomega}{(2\pi)^n}~\cos(\bomega\cdot\bM)~\rme^{
c
\sum_{\bsigma}\lambda(\bsigma)\cos(\bomega\cdot\bsigma)
}
\nonumber\\
&=&\sum_{k\geq 0}\frac{c^k}{k!}
\int_{-\pi}^{\pi}\!\frac{\rmd\bomega}{(2\pi)^n}~\rme^{\rmi\bomega\cdot\bM}
\Big(\sum_{\bsigma}\lambda(\bsigma)\rme^{-\rmi\bomega\cdot\bsigma}\Big)^k
\nonumber\\
&=&\sum_{k\geq 0}\frac{c^k}{k!}
\int_{-\pi}^{\pi}\!\frac{\rmd\bomega}{(2\pi)^n}~\rme^{\rmi\bomega\cdot\bM}\sum_{\bsigma_1\ldots\bsigma_k}
\Big[\prod_{\ell=1}^k \lambda(\bsigma_\ell)\Big]\rme^{-\rmi\bomega\cdot\sum_{\ell\leq k}\bsigma_\ell}
\nonumber\\
&=&
\sum_{k\geq 0}\frac{c^k}{k!}
\sum_{\bsigma_1\ldots\bsigma_k}
\Big[\prod_{\ell=1}^k \lambda(\bsigma_\ell)\Big]\delta_{\bM,\sum_{\ell\leq k}\bsigma_\ell}.
\end{eqnarray}
This then gives for the free energy, upon removing the interaction energy:
\begin{eqnarray}
\overline{f}_{\rm RSB}&=&
-\lim_{n\to 0}\frac{1}{\beta n}\Big\{
\alpha \Big\bra\log \sum_{\bM\in
\Zset^n}\!\! \tilde Q(\bM)~\rme^{\beta\psi\sum_\alpha M_\alpha}\Big\ket_\psi
\nonumber
\\
&&
+\log\sum_{\bsigma}\rme^{
\alpha c [\bra
\cosh[\beta \psi \sum_\alpha\sigma_\alpha]
\ket_\psi-1]}
-\alpha c
\sum_{\bsigma}\lambda(\bsigma)
\bra
\cosh[\beta \psi \sum_\alpha\sigma_\alpha]
\ket_\psi
\Big\}
\nonumber
\\
&=&
-\lim_{n\to 0}\frac{1}{\beta n}\Big\{
\alpha \Big\bra\log \Big[
\sum_{k\geq 0}\frac{c^k}{k!}
\sum_{\bsigma_1\ldots\bsigma_k}
\Big[\prod_{\ell=1}^k \lambda(\bsigma_\ell)\Big]\sum_{\bM\in
\Zset^n}\delta_{\bM,\sum_{\ell\leq k}\bsigma_\ell}\rme^{\beta\psi\sum_\alpha M_\alpha}
\Big]\Big\ket_\psi
\nonumber
\\
&&
+\log\sum_{\bsigma}\rme^{
\alpha c [\bra
\cosh[\beta \psi \sum_\alpha\sigma_\alpha]
\ket_\psi -1]}
-\alpha c
\sum_{\bsigma}\lambda(\bsigma)
\bra
\cosh[\beta \psi \sum_\alpha\sigma_\alpha]
\ket_\psi
\Big\}
\nonumber
\\
&=&
-\lim_{n\to 0}\frac{1}{\beta n}\Big\{
\alpha \Big\bra\log \Big[
\sum_{k\geq 0}\frac{c^k}{k!}
\Big(\sum_{\bsigma}\lambda(\bsigma)\rme^{\beta\psi\sum_\alpha\sigma_{\alpha}}\Big)^k
\Big]\Big\ket_\psi
\nonumber
\\
&&
+\log\sum_{\bsigma}\rme^{
\alpha c [\bra
\cosh[\beta \psi \sum_\alpha\sigma_\alpha]
\ket_\psi-1]}
-\alpha c
\sum_{\bsigma}\lambda(\bsigma)
\bra
\cosh[\beta \psi \sum_\alpha\sigma_\alpha]
\ket_\psi
\Big\}
\nonumber
\\
&=&
-\lim_{n\to 0}\frac{1}{\beta n}\Big\{
\alpha
c\Big\bra\sum_{\bsigma}\lambda(\bsigma)\rme^{\beta\psi\sum_\alpha\sigma_{\alpha}}\Big\ket_\psi-\alpha c
\nonumber
\\
&&
+\log\sum_{\bsigma}\rme^{
\alpha c [\bra
\cosh[\beta \psi \sum_\alpha\sigma_\alpha]
\ket_\psi-1]}
-\alpha c
\sum_{\bsigma}\lambda(\bsigma)
\bra
\cosh[\beta \psi \sum_\alpha\sigma_\alpha]
\ket_\psi
\Big\}
\nonumber
\\
&=&
-\lim_{n\to 0}\frac{1}{\beta n}\Big\{
\alpha
c\sum_{\bsigma}\lambda(\bsigma)\bra\cosh[\beta\psi\sum_\alpha\sigma_{\alpha}]\ket_\psi-\alpha c
\nonumber
\\
&&
+\log\sum_{\bsigma}\rme^{
\alpha c [\bra
\cosh[\beta \psi \sum_\alpha\sigma_\alpha]
\ket_\psi-1]}
-\alpha c
\sum_{\bsigma}\lambda(\bsigma)
\bra
\cosh[\beta \psi \sum_\alpha\sigma_\alpha]
\ket_\psi
\Big\}
\nonumber
\\
&=&
-\lim_{n\to 0}\frac{1}{\beta n}\Big\{
\log\sum_{\bsigma}\rme^{
\alpha c ~[\bra
\cosh[\beta \psi \sum_\alpha\sigma_\alpha]
\ket_\psi-1]}
\Big\},
\end{eqnarray}
where in the penultimate step we used $\lambda(\bsigma)=\lambda(-\bsigma)$.
We next use the following replica identity, which is proved via Taylor expansion of even non-negative analytical functions $F(x)$ that have $F(0)=1$:
\begin{eqnarray}
\lim_{n\to 0} n^{-1}\log\big\bra F(\sum_{\alpha=1}^n\sigma_\alpha)\big\ket_{\sigma_1\ldots\sigma_n=\pm 1}
=\sum_{k>0}\frac{F^{(k)}(0)}{k!}\Big(\frac{\rmd^k}{\rmd x^k}\log\cosh(x)\Big)\Big|_{x=0}.
\end{eqnarray}
Application to the function $F(z)=\exp[\alpha c \bra
\cosh[\beta \psi z]
\ket_\psi-\alpha c]$ gives
\begin{eqnarray}
\hspace*{-25mm}
\overline{f}_{\rm RSB}&=&
-\frac{1}{\beta}\log 2
-\frac{\rme^{-\alpha c}}{\beta}\lim_{x,z\to 0}\sum_{k>0}\frac{1}{k!}\Big(\frac{\rmd^k}{\rmd x^k}\log \cosh(x)\Big)
\frac{\rmd^k}{\rmd z^k}\rme^{\alpha c\bra \cosh(\beta\psi z)\ket_\psi}
\nonumber
\\
\hspace*{-25mm}
&=&
-\frac{1}{\beta}\log 2
-\frac{\rme^{-\alpha c}}{\beta}\sum_{\ell\geq 0}\frac{(\alpha c)^\ell}{\ell!}
\!\lim_{x,z\to 0}\sum_{k>0}\frac{1}{k!}\Big(\frac{\rmd^k}{\rmd x^k}\log \cosh(x)\Big)
\frac{\rmd^k}{\rmd z^k}\bra \cosh(\beta\psi z)\ket^\ell_\psi
\nonumber
\\
\hspace*{-25mm}
&=&
-\frac{1}{\beta}\log 2
-\frac{\rme^{-\alpha c}}{\beta}\sum_{\ell\geq 0}\frac{(\alpha c)^\ell}{\ell!}
\!\lim_{x,z\to 0}\sum_{k>0}\frac{1}{k!}\Big(\frac{\rmd^k}{\rmd x^k}\log \cosh(x)\Big)
\frac{\rmd^k}{\rmd z^k}\bra\bra e^{\beta\psi \sum_{r\leq \ell}\sigma_r z_r}\ket_{\psi_1\ldots\psi_\ell}\ket_{\sigma_1\ldots\sigma_\ell=\pm 1}
\nonumber
\\
\hspace*{-25mm}
&=&
-\frac{1}{\beta}\log 2
-\frac{\rme^{-\alpha c}}{\beta}\sum_{\ell\geq 0}\frac{(\alpha c)^\ell}{\ell!}\Big\bra\Big\bra
\sum_{k>0}\frac{1}{k!}\Big(\lim_{x\to 0}\frac{\rmd^k}{\rmd x^k}\log \cosh(x)\Big)
\Big(\beta \sum_{r\leq \ell}\sigma_r \psi_r\Big)^k\Big\ket_{\psi_1\ldots\psi_\ell}\Big\ket_{\sigma_1\ldots\sigma_\ell=\pm 1}
\nonumber
\\
\hspace*{-25mm}
&=&
-\frac{1}{\beta}\log 2
-\frac{\rme^{-\alpha c}}{\beta}\sum_{\ell\geq 0}\frac{(\alpha c)^\ell}{\ell!}\Big\bra\Big\bra
\log \cosh\Big(\beta \sum_{r\leq \ell}\sigma_r \psi_r\Big)
\Big\ket_{\psi_1\ldots\psi_\ell}\Big\ket_{\sigma_1\ldots\sigma_\ell=\pm 1}
\nonumber
\\
\hspace*{-25mm}
&=&
-\frac{1}{\beta}\log 2-\frac{1}{\beta}\int\!\rmd h~W(h)\log\cosh(\beta h),
\end{eqnarray}
with
\begin{eqnarray}
W(h)&=&
\sum_{k\geq 0}e^{-\alpha c}\frac{(\alpha c)^k}{k!}
\Big\bra\Big\bra\delta\Big[h-\sum_{\ell\leq k}\psi_\ell\sigma_\ell)\Big]\Big\ket_{\psi_1\ldots\psi_k}\Big\ket_{\sigma_1\ldots\sigma_k=\pm 1}.
\end{eqnarray}
This recovers correctly the solution of external fields only.
\end{itemize}

\section{Derivation of RS equations via route II}
\label{app:derivation}

The RS ansatz converts the saddle point equation (\ref{eq:L_saddle}) into
\begin{eqnarray}
\hspace*{-10mm}
\int \!\rmd h~W(h)\rme^{\beta h\sum_\alpha\sigma_\alpha}
&=&
\rme^{\beta n/2c}
 \Big\bra\!\Big\bra
\int\{\rmd\pi\}W[\pi]\prod_{\alpha}\Big(\!\sum_M \pi(M)\rme^{\beta(M^2/2c+\psi M+\tau (\psi+M/c) \sigma_\alpha)}\Big)
\Big\ket_\psi\Big\ket_{\!\tau=\pm 1}
\nonumber
\\
&=&
\rme^{\beta n/2c}
 \Big\bra\!\Big\bra
\int\!\{\rmd\pi\}W[\pi]
\Big(\!\sum_M \pi(M)\rme^{\beta(M^2/2c+\psi M+\tau (\psi+M/c) )}\Big)^{\frac{1}{2}n+\frac{1}{2}\sum_\alpha\sigma_\alpha}
\nonumber
\\
&&\hspace*{20mm}\times
\Big(\!\sum_M \pi(M)\rme^{\beta(M^2/2c+\psi M-\tau (\psi+M/c))}\Big)^{\frac{1}{2}n-\frac{1}{2}\sum_\alpha\sigma_\alpha}
\Big\ket_\psi\Big\ket_{\!\tau=\pm 1}
\nonumber
\\
&=&
\rme^{\beta n/2c}
 \Big\bra\!\Big\bra
\int\!\{d\pi\}W[\pi]
\Big(\frac{\sum_M \pi(M)\rme^{\beta(M^2/2c+\psi M+\tau (\psi+M/c) )}}
{\sum_M \pi(M)\rme^{\beta(M^2/2c+\psi M-\tau (\psi+M/c) )}}\Big)^{\!\frac{1}{2}\sum_\alpha\sigma_\alpha}
\Big\ket_\psi\Big\ket_{\!\tau=\pm 1}
\nonumber
\\
&&\hspace*{-38mm}=
\rme^{\beta n/2c}\!
\int\!\rmd h~\rme^{\beta h\sum_\alpha\sigma_\alpha}
 \Big\bra\!\Big\bra
\int\!\{\rmd\pi\}W[\pi]\delta\Big[h\!-\!\frac{1}{2\beta}\log\Big(\frac{\sum_M \pi(M)\rme^{\beta(M^2/2c+\psi M+\tau (\psi+M/c) )}}
{\sum_M \pi(M)\rme^{\beta(M^2/2c+\psi M-\tau (\psi+M/c) )}}\Big)\Big]
\Big\ket_{\!\psi}\Big\ket_{\!\tau=\pm 1}.
\nonumber\\
\end{eqnarray}
We conclude after sending $n\to 0$ that
\begin{eqnarray}
W(h)&=& \Big\bra\!\Big\bra
\int\!\{\rmd\pi\}W[\pi]~\delta\Big[h-\frac{1}{2\beta}\log\Big(\frac{\sum_M \pi(M)\rme^{\beta(M^2/2c+\psi M+\tau (\psi+M/c) )}}
{\sum_M \pi(M)\rme^{\beta(M^2/2c+\psi M-\tau (\psi+M/c) )}}\Big)\Big]
\Big\ket_\psi\Big\ket_{\!\tau=\pm 1}.
\end{eqnarray}
$W(h)$ is indeed symmetric.
Next we turn to equation (\ref{eq:Q_saddle}),
where we require quantities of the form
\begin{eqnarray}
\varrho(\bomega)&=&\sum_{\bsigma}\cos(\bomega\cdot\bsigma) \rme^{L(\bsigma)}=
\sum_{\bsigma}\cos(\bomega\cdot\bsigma)\rme^{\alpha c\int \!\rmd h~W(h)  \rme^{\beta h\sum_\alpha\sigma_\alpha}}.
\end{eqnarray}
In fact we will need only the ratio $\varrho(\bomega)/\varrho(\bnull)$.
We note that
\begin{eqnarray}
\varrho(\bnull)&=&
2^n \sum_{k\geq 0}\frac{(\alpha c)^k}{k!} \int\!\rmd h_1\ldots \rmd h_k\Big[\prod_{\ell\leq k}W(h_k)\Big]
\cosh^n\!\Big(\beta\sum_{\ell\leq k}h_\ell \Big)=
\rme^{\alpha c+\order(n)}.
\end{eqnarray}
We can hence write the RS version of our first saddle-point equation as follows, using $W(h)=W(-h)$:
\begin{eqnarray}
\hspace*{-25mm}
\int\!\{\rmd\pi\}W[\pi]\prod_{\alpha=1}^n\pi(M_\alpha)&=&
\rme^{-c}\int_{-\pi}^\pi\!\frac{\rmd\bomega}{(2\pi)^n}\cos(\bomega\cdot\bM)
\rme^{c\rme^{-\alpha c +\order(n)}\varrho(\bomega)}
\nonumber\\
&=& \rme^{-c+\order(n)}\sum_{k\geq 0}\frac{c^k}{k!}\int_{-\pi}^\pi\!\frac{\rmd\bomega}{(2\pi)^n}\cos(\bomega\cdot\bM)
\Big\bra\! \cos(\bomega\cdot\bsigma)\rme^{\alpha c\int \!\rmd h~W(h) [
 \rme^{\beta h\sum_\alpha\!\sigma_\alpha}\!-1]}\Big\ket_{\bsigma}^k
\nonumber\\
&&\hspace*{-30mm}= \rme^{-c+\order(n)}\sum_{k\geq 0}\frac{c^k}{k!}\Big\bra\!\Big\bra\int_{-\pi}^\pi\!\frac{\rmd\bomega}{(2\pi)^n}\rme^{\rmi\bomega\cdot(\tau\bM-\sum_{\ell\leq k}\tau_\ell\bsigma^\ell)}
\rme^{\alpha c\sum_{\ell\leq k}\int \!\rmd h~W(h) [
 \rme^{\beta h\sum_\alpha\!\sigma^\ell_\alpha}\!-1]}
\Big\ket_{\bsigma^1\ldots\bsigma^k}\Big\ket_{\tau,\tau_1\ldots\tau_k=\pm 1}
\nonumber
\\
&=&
\rme^{-c+\order(n)}\sum_{k\geq 0}\frac{c^k}{k!}\rme^{-\alpha c k}\Big\bra
\rme^{\alpha c\sum_{\ell\leq k}\int \!\rmd h~W(h)
 \rme^{\beta h \sum_\alpha \sigma_\alpha^\ell}}
\delta_{\bM,\sum_{\ell\leq k}\bsigma^\ell}
\Big\ket_{\bsigma^1\ldots\bsigma^k}
\nonumber
\\
&=&
\rme^{-c+\order(n)}\sum_{k\geq 0}\frac{c^k}{k!}\rme^{-\alpha c k}\sum_{r\geq 0}\frac{(\alpha c)^r}{r!}
\Big\bra
\Big(\int \!\rmd h~W(h)
\sum_{\ell\leq k} \rme^{\beta h \sum_\alpha \sigma_\alpha^\ell}\Big)^r
\delta_{\bM,\sum_{\ell\leq k}\bsigma^\ell}
\Big\ket_{\bsigma^1\ldots\bsigma^k}
\nonumber
\\\
&&\hspace*{-30mm}=
\rme^{-c+\order(n)}\sum_{k\geq 0}\frac{c^k}{k!}\rme^{-\alpha c k}\sum_{r\geq 0}\frac{(\alpha c)^r}{r!}
\int\!\rmd h_1\ldots \rmd h_r\Big[\prod_{s\leq r}W(h_s)\Big]\!\!\sum_{\ell_1\ldots \ell_r\leq k}\!\!
\prod_\alpha
\Big\bra
 \rme^{\beta \sum_{s\leq r} h_s \sigma_{\ell_s}}
\delta_{M_\alpha,\sum_{\ell\leq k}\sigma_\ell}
\Big\ket_{\sigma_1\ldots\sigma_k}
\nonumber
\\
&&\hspace*{-30mm}=
\rme^{-c+\order(n)}\sum_{k\geq 0}\frac{c^k}{k!}\rme^{-\alpha c k}\sum_{r\geq 0}\frac{(\alpha c)^r}{r!}
\!\int\!\rmd h_1\ldots \rmd h_r\Big[\prod_{s\leq r}W(h_s)\Big]\!\!\sum_{\ell_1\ldots \ell_r\leq k}\!\!
\prod_\alpha \left\{
\frac{
\big\bra
 \rme^{\beta \sum_{s\leq r} h_s \sigma_{\ell_s}}
\delta_{M_\alpha,\sum_{\ell\leq k}\sigma_\ell}
\big\ket_{\sigma_1\ldots\sigma_k}}
{
\big\bra
 \rme^{\beta \sum_{s\leq r} h_s \sigma_{\ell_s}}
\big\ket_{\sigma_1\ldots\sigma_k}}
\right\}
\nonumber\\
&&\hspace*{-30mm}=
\int\!\{\rmd\pi\} \Big(\prod_\alpha \pi(M_\alpha)\Big)
\rme^{-c+\order(n)}\sum_{k\geq 0}\frac{c^k}{k!}\rme^{-\alpha c k}\sum_{r\geq 0}\frac{(\alpha c)^r}{r!}
\int\!\rmd h_1\ldots \rmd h_r\Big[\prod_{s\leq r}W(h_s)\Big]\!\!\sum_{\ell_1\ldots \ell_r\leq k}
\nonumber
\\
&&\hspace*{20mm}
\times
\prod_M \delta\left[\pi(M)-
\frac{
\big\bra
 \rme^{\beta \sum_{s\leq r} h_s \sigma_{\ell_s}}
\delta_{M,\sum_{\ell\leq k}\sigma_\ell}
\big\ket_{\sigma_1\ldots\sigma_k}}
{
\big\bra
 \rme^{\beta \sum_{s\leq r} h_s \sigma_{\ell_s}}
\big\ket_{\sigma_1\ldots\sigma_k}}
\right].
\end{eqnarray}
We thus conclude that for $n\to 0$ the following equation for $W[\pi]$ solves our saddle-point problem:
\begin{eqnarray}
W[\pi]&=& \rme^{-c}\sum_{k\geq 0}\frac{c^k}{k!}\rme^{-\alpha c k}\sum_{r\geq 0}\frac{(\alpha c)^r}{r!}
\int_{-\infty}^\infty\!\rmd h_1\ldots \rmd h_r\Big[\prod_{s\leq r}W(h_s)\Big]\sum_{\ell_1\ldots \ell_r\leq k}
\nonumber
\\
&&
\hspace*{30mm}\times
\prod_M \delta\left[\pi(M)-
\frac{
\big\bra
 \rme^{\beta \sum_{s\leq r} h_s \sigma_{\ell_s}}
\delta_{M,\sum_{\ell\leq k}\sigma_\ell}
\big\ket_{\sigma_1\ldots\sigma_k}}
{
\big\bra
 \rme^{\beta \sum_{s\leq r} h_s \sigma_{\ell_s}}
\big\ket_{\sigma_1\ldots\sigma_k}}
\right].
\end{eqnarray}
Everything is properly normalised, and if $W(h)=W(-h)$ the measure $W[\pi]$ is seen to permit only real-valued distributions $\pi(M)$ such that $\pi(M)\in[0,\infty)$ and $\pi(-M)=\pi(M)$ for all $M\in\Zset$.

\section{Continuous RS phase transitions via route II}
\label{app:phase_transition}

Here we work with the order parameter equation that is written in terms of $W(h)$ only, i.e. (\ref{eq:single_eqn}), and look for phase transitions in the absence of external fields. For $P(\psi)=\delta(\psi)$ we must solve $W(h)$ from
\begin{eqnarray}
W(h)&=&
\rme^{-c}\sum_{k\geq 0}\frac{c^k}{k!}\rme^{-\alpha c k}\sum_{r\geq 0}\frac{(\alpha c)^r}{r!}
\int_{-\infty}^\infty\!\rmd h_1\ldots \rmd h_r\Big[\prod_{s\leq r}W(h_s)\Big]\sum_{\ell_1\ldots \ell_r\leq k}
\nonumber
\\
&&\hspace*{-5mm}\times
 \Big\bra
\delta\left[h-\frac{1}{2\beta}\log\left(\frac{
\big\bra
\rme^{\beta(\sum_{\ell\leq k}\!\tau_{\ell})^2/2c+\beta(\sum_{\ell\leq k}\!\tau_{\ell})\tau/c+\beta \sum_{s\leq r} h_s \tau_{\ell_s}}\big\ket_{\tau_1\ldots\tau_k=\pm 1}}
{
\big\bra
\rme^{\beta(\sum_{\ell\leq k}\!\tau_{\ell})^2/2c-\beta(\sum_{\ell\leq k}\!\tau_{\ell})\tau/c +\beta \sum_{s\leq r} h_s \tau_{\ell_s}}\big\ket_{\tau_1\ldots\tau_k=\pm 1}}\right)\!\right]
\Big\ket_{\!\tau=\pm 1}.~
\end{eqnarray}
Clearly $W(h)=\delta(h)$ solves this equation for any temperature.
Due to $W(h)=W(-h)$, we will always have $\int\!\rmd h~W(h)h=0$, so the first bifurcation away from $W(h)=\delta(h)$ is expected  to be in the second moment.
 We write $h=\epsilon y$, with $0<\epsilon\ll 1$, and expand in powers of $\epsilon$. Upon setting
$W(h)=\epsilon^{-1}\tilde{W}(h/\epsilon)$
we have
\begin{eqnarray}
\tilde{W}(y) &=&
\rme^{-c}\sum_{k\geq 0}\frac{c^k}{k!}\rme^{-\alpha c k}\sum_{r\geq 0}\frac{(\alpha c)^r}{r!}
\int_{-\infty}^\infty\!\rmd y_1\ldots \rmd y_r\Big[\prod_{s\leq r}\tilde{W}(y_s)\Big]\sum_{\ell_1\ldots \ell_r\leq k}
\nonumber
\\
&&\hspace*{-5mm}\times
 \Big\bra
\delta\left[y-\frac{1}{2\beta\epsilon}\log\left(\frac{
\big\bra
\rme^{\beta(\sum_{\ell\leq k}\!\tau_{\ell})^2/2c+\beta(\sum_{\ell\leq k}\!\tau_{\ell})\tau/c+\beta \epsilon\sum_{s\leq r} y_s \tau_{\ell_s}}\big\ket_{\tau_1\ldots\tau_k=\pm 1}}
{
\big\bra
\rme^{\beta(\sum_{\ell\leq k}\!\tau_{\ell})^2/2c-\beta(\sum_{\ell\leq k}\!\tau_{\ell})\tau/c +\beta\epsilon \sum_{s\leq r} y_s \tau_{\ell_s}}\big\ket_{\tau_1\ldots\tau_k=\pm 1}}\right)\!\right]
\Big\ket_{\!\tau=\pm 1}.
\label{eq:Weqnew}
\end{eqnarray}
Next we expand the logarithm in the last line. To leading orders in $\epsilon$ we obtain
\begin{eqnarray}
\frac{1}{2\beta\epsilon}\log\Big(\!\ldots\!\Big)
&=&
\frac{1}{2\beta\epsilon}\log\left(\frac{
\big\bra
\rme^{\beta(\sum_{\ell\leq k}\!\tau_{\ell})^2/2c+\beta(\sum_{\ell\leq k}\!\tau_{\ell})\tau/c}
\big[1+\beta \epsilon\sum_{s\leq r} y_s \tau_{\ell_s}\big]
\big\ket_{\tau_1\ldots\tau_k=\pm 1}}
{
\big\bra
\rme^{\beta(\sum_{\ell\leq k}\!\tau_{\ell})^2/2c-\beta(\sum_{\ell\leq k}\!\tau_{\ell})\tau/c }\big[1+\beta \epsilon\sum_{s\leq r} y_s \tau_{\ell_s}
\big]\big\ket_{\tau_1\ldots\tau_k=\pm 1}}\right)
\nonumber
\\
&=&
\frac{1}{2\beta\epsilon}\log\left(
\frac{1+\beta \epsilon \sum_{s\leq r} y_s
\frac{\big\bra  \tau_{\ell_s}
\rme^{\beta(\sum_{\ell\leq k}\!\tau_{\ell})^2/2c+\beta(\sum_{\ell\leq k}\!\tau_{\ell})\tau/c}\big\ket_{\tau_1\ldots\tau_k=\pm 1}}
{\big\bra
\rme^{\beta(\sum_{\ell\leq k}\!\tau_{\ell})^2/2c+\beta(\sum_{\ell\leq k}\!\tau_{\ell})\tau/c}\big\ket_{\tau_1\ldots\tau_k=\pm 1}
}
}
{1+\beta \epsilon \sum_{s\leq r} y_s
\frac{\big\bra  \tau_{\ell_s}
\rme^{\beta(\sum_{\ell\leq k}\!\tau_{\ell})^2/2c-\beta(\sum_{\ell\leq k}\!\tau_{\ell})\tau/c}\big\ket_{\tau_1\ldots\tau_k=\pm 1}}
{\big\bra
\rme^{\beta(\sum_{\ell\leq k}\!\tau_{\ell})^2/2c-\beta(\sum_{\ell\leq k}\!\tau_{\ell})\tau/c}\big\ket_{\tau_1\ldots\tau_k=\pm 1}
}
}
\right)
\nonumber
\\
&&
\hspace*{-30mm}=
\frac{1}{2}
 \sum_{s\leq r} y_s\left\{
\frac{\big\bra  \tau_{\ell_s}
\rme^{\beta(\sum_{\ell\leq k}\!\tau_{\ell})^2/2c+\beta(\sum_{\ell\leq k}\!\tau_{\ell})\tau/c}\big\ket_{\tau_1\ldots\tau_k}}
{\big\bra
\rme^{\beta(\sum_{\ell\leq k}\!\tau_{\ell})^2/2c+\beta(\sum_{\ell\leq k}\!\tau_{\ell})\tau/c}\big\ket_{\tau_1\ldots\tau_k}
}
-
\frac{\big\bra  \tau_{\ell_s}
\rme^{\beta(\sum_{\ell\leq k}\!\tau_{\ell})^2/2c-\beta(\sum_{\ell\leq k}\!\tau_{\ell})\tau/c}\big\ket_{\tau_1\ldots\tau_k}}
{\big\bra
\rme^{\beta(\sum_{\ell\leq k}\!\tau_{\ell})^2/2c-\beta(\sum_{\ell\leq k}\!\tau_{\ell})\tau/c}\big\ket_{\tau_1\ldots\tau_k}
}
\right\}
\nonumber
\\
&=&
\tau
 \sum_{s\leq r} y_s\left\{
\frac{\int\!{\rm Dz}~\tanh(z\sqrt{\beta/c}\!+\!\beta/c)\cosh^k(z\sqrt{\beta/c}\!+\!\beta/c)}
{\int\!{\rm Dz}~\cosh^k(z\sqrt{\beta/c}\!+\!\beta/c)
}
\right\}.
\end{eqnarray}
Hence our order parameter equation (\ref{eq:Weqnew}) becomes
\begin{eqnarray}
\tilde{W}(y) &=&
\rme^{-c}\sum_{k\geq 0}\frac{c^k}{k!}\rme^{-\alpha c k}\sum_{r\geq 0}\frac{(\alpha c)^r}{r!}
\int_{-\infty}^\infty\!\rmd y_1\ldots \rmd y_r\Big[\prod_{s\leq r}\tilde{W}(y_s)\Big]\sum_{\ell_1\ldots \ell_r\leq k}
\nonumber
\\
&&\hspace*{-5mm}\times
 \Big\bra
\delta\left[y-
\tau
 \sum_{s\leq r} y_s\left\{
\frac{\int\!{\rm Dz}~\tanh(z\sqrt{\beta/c}\!+\!\beta/c)\cosh^k(z\sqrt{\beta/c}\!+\!\beta/c)}
{\int\!{\rm Dz}~\cosh^k(z\sqrt{\beta/c}\!+\!\beta/c)
}
\right\}
\!\right]
\Big\ket_{\!\tau=\pm 1}.
\label{eq:Weqnewnew}
\end{eqnarray}
The first potential type of bifurcation away from $W(h)=\delta(h)$ would have $\int\!\rmd h~W(h)h=\epsilon\int\!\rmd y~\tilde{W}(y)y\equiv \epsilon m_1\neq 0$. However, we see that mutiplying both sides of (\ref{eq:Weqnewnew}) by $y$, followed by integration, immediately gives $m_1=0$.
Thus, as expected,  a bifurcation leading to a function $W(h)$ with $\int\!\rmd h~W(h)h\neq 0$ is impossible.

Any continous bifurcation will consequently have $\int\!\rmd h~W(h)h=0$ and $\int\!\rmd h~W(h)h^2=\epsilon^2\int\!\rmd y~\tilde{W}(y)y^2\equiv \epsilon^2 m_2\neq 0$. Multiplication of equation (\ref{eq:Weqnewnew})  by $y^2$, followed by integration over $y$ gives
\begin{eqnarray}
m_2&=&
\rme^{-c}\sum_{k\geq 0}\frac{c^k}{k!}\rme^{-\alpha c k}\sum_{r\geq 0}\frac{(\alpha c)^r}{r!}
\int_{-\infty}^\infty\!\rmd y_1\ldots \rmd y_r\Big[\prod_{s\leq r}\tilde{W}(y_s)\Big]\sum_{\ell_1\ldots \ell_r\leq k}
\nonumber
\\
&&\hspace*{5mm}\times
 \sum_{s\leq r} y^2_s
 \Big\bra
\left\{
\frac{\int\!{\rm Dz}~\tanh(z\sqrt{\beta/c}\!+\!\beta/c)\cosh^k(z\sqrt{\beta/c}\!+\!\beta/c)}
{\int\!{\rm Dz}~\cosh^k(z\sqrt{\beta/c}\!+\!\beta/c)
}
\right\}^2
\Big\ket_{\!\tau=\pm 1}.
\end{eqnarray}
So now we get a bifurcation when
\begin{eqnarray}
1&=&
\alpha c^2
\sum_{k\geq 0}e^{-c}\frac{c^{k}}{k!}
\left\{
\frac{\int\!Dz~\tanh(z\sqrt{\beta/c}\!+\!\beta/c)\cosh^{k+1}(z\sqrt{\beta/c}\!+\!\beta/c)
}
{\int\! Dz~\cosh^{k+1}(z\sqrt{\beta/c}\!+\!\beta/c)
}
\right\}^2.
\label{eq:transition}
\end{eqnarray}
We note that the right-hand side of (\ref{eq:transition}) obeys $\lim_{\beta\to 0}{\rm RHS}=0$ and $\lim_{\beta\to \infty}{\rm RHS}=\alpha c^2$.
Hence a transition at finite temperature $T_c(\alpha,c)>0$ exists to a new state with $W(h)\neq \delta(h)$ as soon as $\alpha c^2>1$. The critical temperature becomes zero when $\alpha c^2=1$, so $T_c(\alpha,1/\sqrt{\alpha})=0$ for all $\alpha\geq 0$.  For large $c$, using $\tanh x = x+\order{(x^3)}$
and $\cosh x=1+x^2/2$, valid for small $x$, we have $T_c=\sqrt{\alpha}$.

\section{Coincidence of the two formulae for the transition line}

\label{app:equivalence}
In order to prove that the two expressions (\ref{eq:transition}) and (\ref{eq:critical_surface}) for the RS transition line
are identical, as they should be, we show that
\bea
\hspace*{-5mm}
\left\{
\frac{\int_{-\pi}^\pi\! \rmd\omega ~\sin^2(\omega)\cos^k(\omega)\Db}{\int_{-\pi}^\pi\! \rmd\omega~ \cos^{k+2}(\omega)\Db} \right\}^{\!2}
=
\left\{
\frac{\int\!{\rm D}z~\tanh(z\sqrt{\beta/c}\!+\!\beta/c)\cosh^{k+1}(z\sqrt{\beta/c}\!+\!\beta/c)
}
{\int\! {\rm D}z~\cosh^{k+1}(z\sqrt{\beta/c}\!+\!\beta/c)
}
\right\}^{\!2},~~
\label{eq:equality}
\end{eqnarray}
where ${\rm D}z=(2\pi)^{-1/2}\rme^{-z^2/2}\, dz$.
We can rewrite the argument of the curly brackets on the right-hand side, which we will denote as $A$, as
\bea
A&=&\frac{\int\!{\rm D}z~\sinh(z\sqrt{\beta/c}\!+\!\beta/c)\cosh^k(z\sqrt{\beta/c}\!+\!\beta/c)
}
{\int\! {\rm D}z~\cosh^{k+1}(z\sqrt{\beta/c}\!+\!\beta/c)
}
\nonumber\\
&=&\frac{\int\!{\rm D}z~\bra \tau_{k+1}
e^{(z\sqrt{\beta/c}\!+\!\beta/c)\sum_{\ell\leq k+1}
\tau_\ell}
\ket_{\tau_1\ldots \tau_{k+1}=\pm 1}
}
{\int\!{\rm D}z~\bra \rme^{(z\sqrt{\beta/c}\!+\!\beta/c)\sum_{\ell\leq k+1}\tau_\ell}
\ket_{\tau_1\ldots \tau_{k+1}=\pm 1}
}
\nonumber\\
&=&\frac{\bra \tau_{k+1}
\rme^{(\beta/2c)(\sum_{\ell\leq k+1}\tau_\ell)^2+(\beta/c)\sum_{\ell\leq k+1}\tau_\ell}
\ket_{\tau_1\ldots \tau_{k+1}=\pm 1}
}
{\bra
\rme^{(\beta/2c)(\sum_{\ell\leq k+1}\tau_\ell)^2+(\beta/c)\sum_{\ell\leq k+1}\tau_\ell}
\ket_{\tau_1\ldots \tau_{k+1}=\pm 1}
},
\end{eqnarray}
where we have carried out the Gaussian integrations. Next we insert
$1=\sum_{M\in\Zset} \delta_{M,\sum_{\ell\leq k+1} \tau_\ell}$, and write the Kronecker delta in integral form. This gives
\bea
A&=&
\frac{\sum_{M\in\Zset} \rme^{(\beta/2c)M^2+(\beta/c)M}
 \int_{-\pi}^\pi\! \rmd\omega\, \rme^{\rmi\omega M} \bra \tau_{k+1} \rme^{-\rmi\omega
\sum_{\ell\leq k+1}\tau_\ell}
\ket_{\tau_1\ldots \tau_{k+1}=\pm 1}
}
{
\sum_{M\in\Zset}
\rme^{(\beta/2c)M^2+(\beta/c)M}
\int_{-\pi}^\pi\! \rmd\omega ~\rme^{\rmi\omega M}
\bra
\rme^{-\rmi\omega \sum_{\ell\leq k+1}\tau_\ell}
\ket_{\tau_1\ldots \tau_{k+1}=\pm 1}
}
\nonumber\\
&=&
-\rmi~\frac{\sum_{M\in\Zset} \rme^{(\beta/2c)M^2+(\beta/c)M}
 \int_{-\pi}^\pi\! \rmd\omega~ \rme^{\rmi\omega M}
\cos^k(\omega)\sin(\omega)
}
{
\sum_{M\in\Zset}
\rme^{(\beta/2c)M^2+(\beta/c)M}
\int_{-\pi}^\pi\! \rmd\omega ~\rme^{\rmi\omega M}
\cos^{k+1}(\omega)
}.
\eea
By
completing the square,
$\sum_M
\rme^{(\beta/2c)M^2+(\beta/c)M}=\rme^{-\beta/(2c)}\sum_M e^{(\beta/2c)(M+1)^2}$,
shifting the summation index $M\to M-1$, and
using the symmetry properties (\ref{eq:symmetry_0}) of $D(\omega|\beta)$ at zero fields, we finally get
\bea
A
&=&
-\rmi\frac{\sum_{M\in\Zset} \rme^{(\beta/2c)M^2}
 \int_{-\pi}^\pi\! \rmd\omega~\rme^{\rmi\omega (M-1)}
\cos^k(\omega)\sin(\omega)
}
{
\sum_{M\in\Zset}
\rme^{(\beta/2c)M^2}
\int_{-\pi}^\pi\! \rmd\omega~ \rme^{\rmi\omega (M-1)}
\cos^{k+1}(\omega)
}
\nonumber
\\
&=&
-\rmi\frac{
 \int_{-\pi}^\pi\! \rmd\omega~D(\omega|\beta)
\cos^k(\omega)\sin(\omega)[\cos(\omega)\!-\!\rmi\sin(\omega)]
}
{
\int_{-\pi}^\pi\! \rmd\omega~ D(\omega|\beta)
\cos^{k+1}(\omega)[\cos(\omega)\!-\!\rmi\sin(\omega)]
}
\nonumber
\\
&=&
-\frac{
\int _{-\pi}^\pi\! \rmd\omega~ D(\omega|\beta)
\cos^k(\omega)\sin^2(\omega)
}
{
\int_{-\pi}^\pi\! \rmd\omega~ D(\omega|\beta)
\cos^{k+2}(\omega)
},
\eea
which proves (\ref{eq:equality}).

\end{document}